\documentclass{article}
\usepackage[letterpaper, portrait, margin=1in]{geometry}
\usepackage[utf8]{inputenc}
\usepackage[english]{babel}
\usepackage{amsmath}
\usepackage{amssymb}
\usepackage{graphicx}
\usepackage{caption}
\usepackage{subcaption}
\usepackage[dvipsnames]{xcolor}
\usepackage{soul}
\usepackage{comment}

\title{A systematic approach for modeling a nonlocal eddy diffusivity}

\author{Jessie Liu\thanks{jeliu@stanford.edu}, Hannah H. Williams, and Ali Mani}
\date{\textit{Department of Mechanical Engineering, \\ 
              Stanford University, CA, 94305, USA}}

\begin{document}

\maketitle

\begin{abstract}
This study considers advective and diffusive transport of passive scalar fields by spatially-varying incompressible flows. Prior studies have shown that the eddy diffusivities governing the mean field transport in such systems can generally be nonlocal in space and time. While for many flows nonlocal eddy diffusivities are more accurate than commonly-used Boussinesq eddy diffusivities, nonlocal eddy diffusivities are often computationally cost-prohibitive to obtain and difficult to implement in practice. We develop a systematic and more cost-effective approach for modeling nonlocal eddy diffusivities using \textit{matched moment inverse} (MMI) operators. These operators are constructed using only a few leading-order moments of the exact nonlocal eddy diffusivity kernel, which can be easily computed using the inverse macroscopic forcing method (IMFM) (Mani and Park (2021)). The resulting reduced-order models for the mean fields that incorporate the modeled eddy diffusivities often improve Boussinesq-limit models since they capture leading-order nonlocal effects. But more importantly, these models can be expressed as partial differential equations that are readily solvable using existing computational fluid dynamics capabilities rather than as integro-partial differential equations. 

\end{abstract}

\newpage 
\section{Introduction}
Scalar transport phenomena are critical to a broad range of everyday applications such as in engineering \cite{su2004simultaneous}\cite{muppidi2008direct}\cite{li2009dns} and geophysics \cite{uchiyama2017eddy}\cite{katul1998active}. One commonly-used approach for modeling mean scalar transport by a turbulent flow is to Reynolds average the governing equations \cite{reynolds1895iv}. This approach leads to an unclosed scalar flux term that must be specified. If, 1) the time and length scales over which the mean scalar field varies are much greater than the mixing times and mixing lengths of the underlying turbulent fluctuations, and 2) the mixing by the underlying turbulent fluctuations is assumed to be isotropic, then the Boussinesq approximation \cite{boussinesq1877essai} is valid. Under this approximation, one can write the scalar flux as the product of an eddy diffusivity and the gradient of the mean scalar. The scalar flux at a given point depends only on the gradient of the mean scalar at that one point, i.e. the Boussinesq approximation is a purely local (and isotropic) approximation. 

However, in many realistic flows, the Boussinesq locality approximation is not valid \cite{corrsin1975limitations}. A modification introduced by Berkowicz and Prahm \cite{berkowicz_prahm_1980} allows for the mean scalar flux to depend on the gradient of the mean scalar at all points in space rather than a single point. The eddy diffusivity becomes a \textit{nonlocal} eddy diffusivity, capturing the spatial dependence of the scalar flux. Moreover, a nonlocal eddy diffusivity may also be modified to capture the temporal dependence of the scalar flux on the history of the gradient of the mean scalar \cite{romanof1985application}\cite{kraichnan1987eddy}\cite{hamba1995analysis}\cite{hamba2004nonlocal}. These ideas and definitions will be made concrete in the problem formulation in Section \ref{problem formulation}. 

A nonlocal eddy diffusivity can give the precise description of the mean scalar flux (as long as the scalar source fields and boundary conditions are precise in the mean space \cite{mani2021macroscopic}). Moreover, it can reveal information fundamental to the understanding and prediction of the evolution of the passive scalar field, such as the sensitivity of the scalar flux to the mean scalar gradient in certain regions. The practical issue, however, is that for a given flow the nonlocal eddy diffusivity may be very computationally expensive to obtain. One method, introduced by Kraichnan \cite{kraichnan1987eddy} and later modified by Hamba for directly computing the nonlocal eddy diffusivity \cite{hamba1995analysis} \cite{hamba2004nonlocal}, is to use the Green's function solution to a passive scalar equation. The nonlocal eddy diffusivity, representing the projection of the scalar flux into the mean space, is then written exactly in terms of the Green’s function and velocity fluctuation. Another method, which we use in this work, is the macroscopic forcing method (MFM), a numerical technique introduced by Mani and Park \cite{mani2021macroscopic} in which one probes the closure operator by applying an appropriate forcing (not necessarily a Dirac delta function) to the governing equations and measures the averaged response. Through this input-output analysis, one can determine the exact nonlocal eddy diffusivity corresponding to the unclosed term. However, because the full nonlocal eddy diffusivity captures the dependency of the scalar flux on the mean scalar gradient everywhere in the averaged space, either of these brute force approaches would require as many direct numerical simulations (DNS) as degrees of freedom in the averaged space.

Recent work by Bryngelson and Sch{\"a}fer et al. \cite{bryngelson2023fast} leverages hidden sparsity in the discretized nonlocal eddy diffusivity to substantially reduce the number of DNS required to obtain the nonlocal eddy diffusivity. 

However, even once obtained, a nonlocal eddy diffusivity may still be expensive to implement in a model due to the resulting integro-differential equation accounting for the effect of mean scalar gradients everywhere in space (and time). A spatially nonlocal eddy diffusivity would raise the computational cost from $\mathcal{O}(N)$ to $\mathcal{O}(N^2)$; a temporally nonlocal eddy diffusivity would require keeping the history of the mean scalar gradient stored in memory. 

Previous works have attempted to address the cost of implementing a nonlocal eddy diffusivity by suggesting various approaches for modeling a nonlocal eddy diffusivity using a partial differential equation rather than an integro-differential equation. Georgopoulos and Seinfeld \cite{georgopoulos1989nonlocal} assumed an exponential kernel shape for a temporally nonlocal (but spatially local) eddy diffusivity and arrived at a hyperbolic telegrapher's equation for the mean scalar. Yoshizawa \cite{yoshizawa1985statistical} expanded the nonlocal eddy diffusivity using the two-scale direct interaction approximation (TSDIA) and used higher-order terms involving products of the mean scalar gradient and mean velocity gradient as corrections to the local eddy diffusivity. Hamba \cite{hamba1995analysis}\cite{hamba2004nonlocal} expanded the nonlocal eddy diffusivity using a Taylor series expansion (also known as a Kramers-Moyal expansion \cite{kampen1961power}) and used higher-order terms of the Taylor series as corrections to the local eddy diffusivity. Such an expansion can have convergence issues as we discuss in Section \ref{modeling nonlocal eddy diffusivity}. Hamba \cite{hamba2004nonlocal} also suggested another model form based on a partial differential equation for the scalar flux; however, the coefficients of this model require knowledge of the full nonlocal eddy diffusivity.
More recently, Hamba \cite{hamba2022analysis} modeled the nonlocal eddy diffusivity for decaying homogeneous isotropic turbulence in Fourier space by using the energy spectrum.

To alleviate the computational cost while keeping the accuracy of the nonlocal eddy diffusivity, we introduce a systematic technique for modeling nonlocal eddy diffusivities using what we call matched moment inverse (MMI) operators. Determining these operators does not require obtaining the full nonlocal eddy diffusivity---as the name suggests, they require only a few \textit{moments} of the nonlocal eddy diffusivity, which can be cost-effectively obtained using inverse MFM (IMFM) \cite{mani2021macroscopic}.

The remainder of this article is organized as follows: In Section \ref{problem formulation}, we define the passive scalar transport problem and nonlocal eddy diffusivity. In Section \ref{intro to MFM}, we introduce IMFM for obtaining moments of the nonlocal eddy diffusivity. In Section \ref{modeling nonlocal eddy diffusivity}, we discuss modeling approaches for nonlocal eddy diffusivities, culminating in the development of MMI. Then, in Section \ref{homogeneous flows}, we demonstrate the use of MMI for a simple homogeneous problem and illustrate the importance of including nonlocal effects in a model. Lastly, in Section \ref{inhomogeneous flows}, we demonstrate the use of MMI for inhomogeneous flows and address some of the challenges with MMI for inhomogeneous wall-bounded flows. 
  
\subsection{Problem formulation} \label{problem formulation}
Consider a passive scalar, $c(\mathbf{x},t)$, being transported by a flow with velocity, $\mathbf{u}(\mathbf{x},t)$. The governing equation is
\begin{equation} \label{eq:scalar transport}
\frac{\partial c}{\partial t} + \frac{\partial}{\partial x_j}(u_j c) = D_M \frac{\partial^2 c}{\partial x_j \partial x_j},
\end{equation}
where $D_M$ is the molecular diffusivity. In many applications, instead of the full solution, $c$, one may be only interested in the average of the solution, $\bar{c}$. For example, the average may be taken over ensembles, in time if the flow is statistically stationary, or over homogeneous spatial directions, but its definition is system-dependent and varies between problems. One can derive an equation for $\bar{c}$ by applying the Reynolds decomposition \cite{reynolds1895iv}:
\begin{subequations}
\begin{align}
\label{eq:Reynolds decomp c}
c &= \bar{c} + c^\prime \\ 
\label{eq:Reynolds decomp uj}
u_j &= \bar{u}_j + u_j^\prime
\end{align}
\end{subequations}
where $\overline{()}$ denotes a mean quantity and $()^\prime$ denotes fluctuations about the mean quantity. Substituting (\ref{eq:Reynolds decomp c}) and (\ref{eq:Reynolds decomp uj}) into Equation (\ref{eq:scalar transport}) and then averaging the resulting equation leads to the mean scalar transport equation,
\begin{equation} \label{eq:mean scalar transport}
\frac{\partial \bar{c}}{\partial t} + \frac{\partial}{\partial x_j} (\bar{u}_j\bar{c}) = D_M \frac{\partial^2 \bar c}{\partial x_j \partial x_j} - \frac{\partial}{\partial x_j}(\overline{u_j^\prime c^\prime}).
\end{equation}
The scalar flux, $\overline{u_j^\prime c^\prime}$, is unknown and further attempts to analytically develop governing equations for this term would result in more unknown quantities, i.e. this term is unclosed. A commonly-used closure model, introduced by Boussinesq \cite{boussinesq1877essai}, approximates $-\overline{u_j^\prime c^\prime}$ as a diffusive flux: 
\begin{equation} \label{eq:Boussinesq eddy diffusivity}
    -\overline{u_j^\prime c^\prime}(\mathbf{x}) = D \frac{\partial \bar{c}}{\partial x_j}\big|_\mathbf{x},
\end{equation}
where $D$ is commonly referred to as the eddy diffusivity. Equation (\ref{eq:Boussinesq eddy diffusivity}) relies on two simplifying approximations. The first approximation is isotropy of the underlying mixing process, resulting in a scalar eddy diffusivity. More critically relevant to our study, the second approximation is that the mean scalar, $\bar{c}$, varies over a time and length scale much larger than that of the fluctuations, $c^\prime$. In other words, the fluctuations mix very quickly and \textit{very locally} due to the underlying flow. In this limit, one can draw an analogy to kinetic theory, where molecular mixing happens very quickly and locally due to Brownian motion, but the average motion represented at the continuum scale, which is much larger than the Brownian mean free path, can be approximated via a local diffusive flux. Reliant on a separation of scales, the Boussinesq approximation is a purely local approximation: $-\overline{u_j^\prime c^\prime}$ at a given location, $\mathbf{x}$, is only dependent on the gradient of $\bar{c}$ at the same location, $\mathbf{x}$. 

\begin{figure}[t]
     \centering
     \begin{subfigure}[t]{0.45\textwidth}
         \centering
         \includegraphics[width=\textwidth]{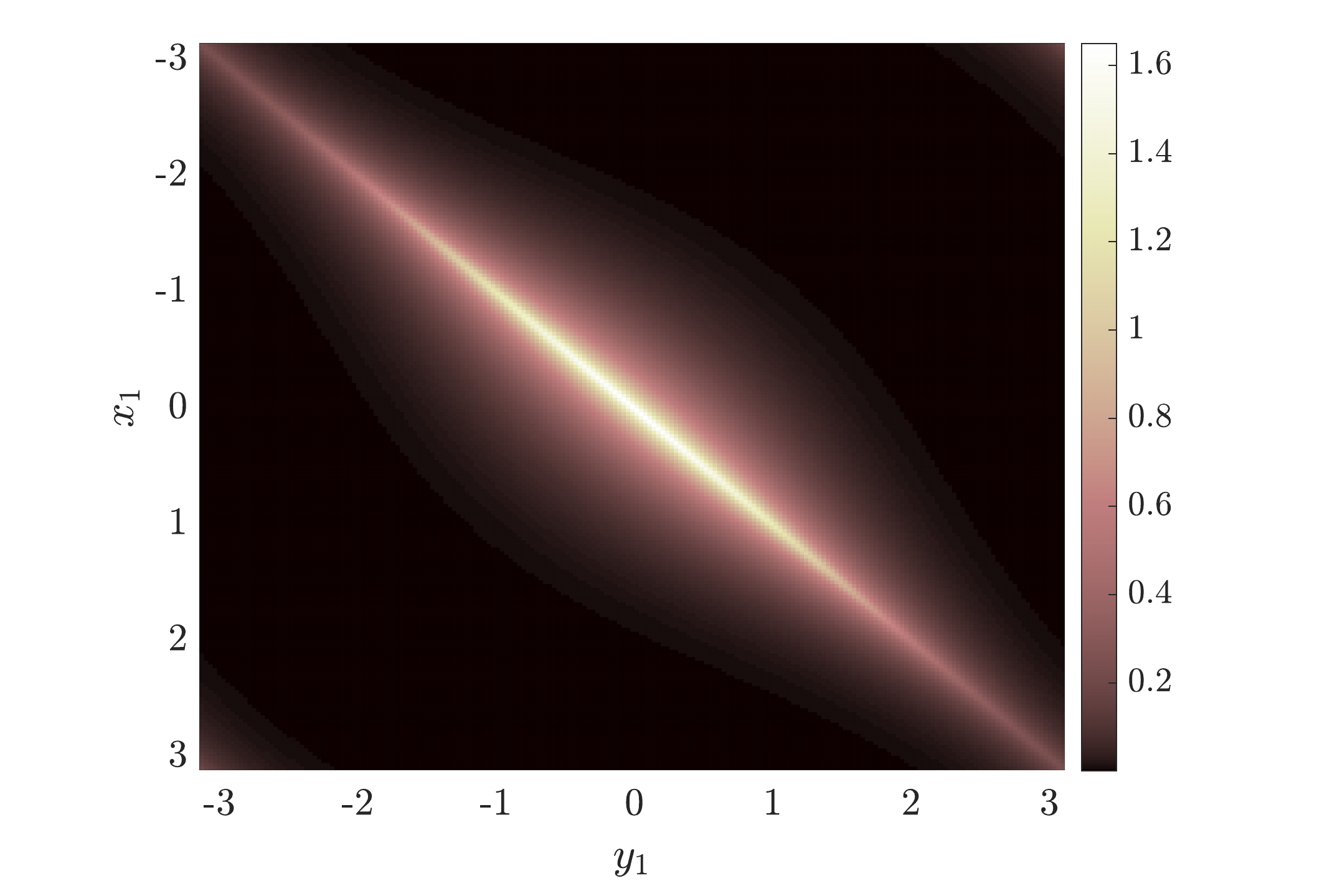}       \caption{}
         \label{fig:example_full_nonlocal_eddy_diffusivity}
     \end{subfigure}
     \begin{subfigure}[t]{0.45\textwidth}
         \centering
         \includegraphics[width=\textwidth]{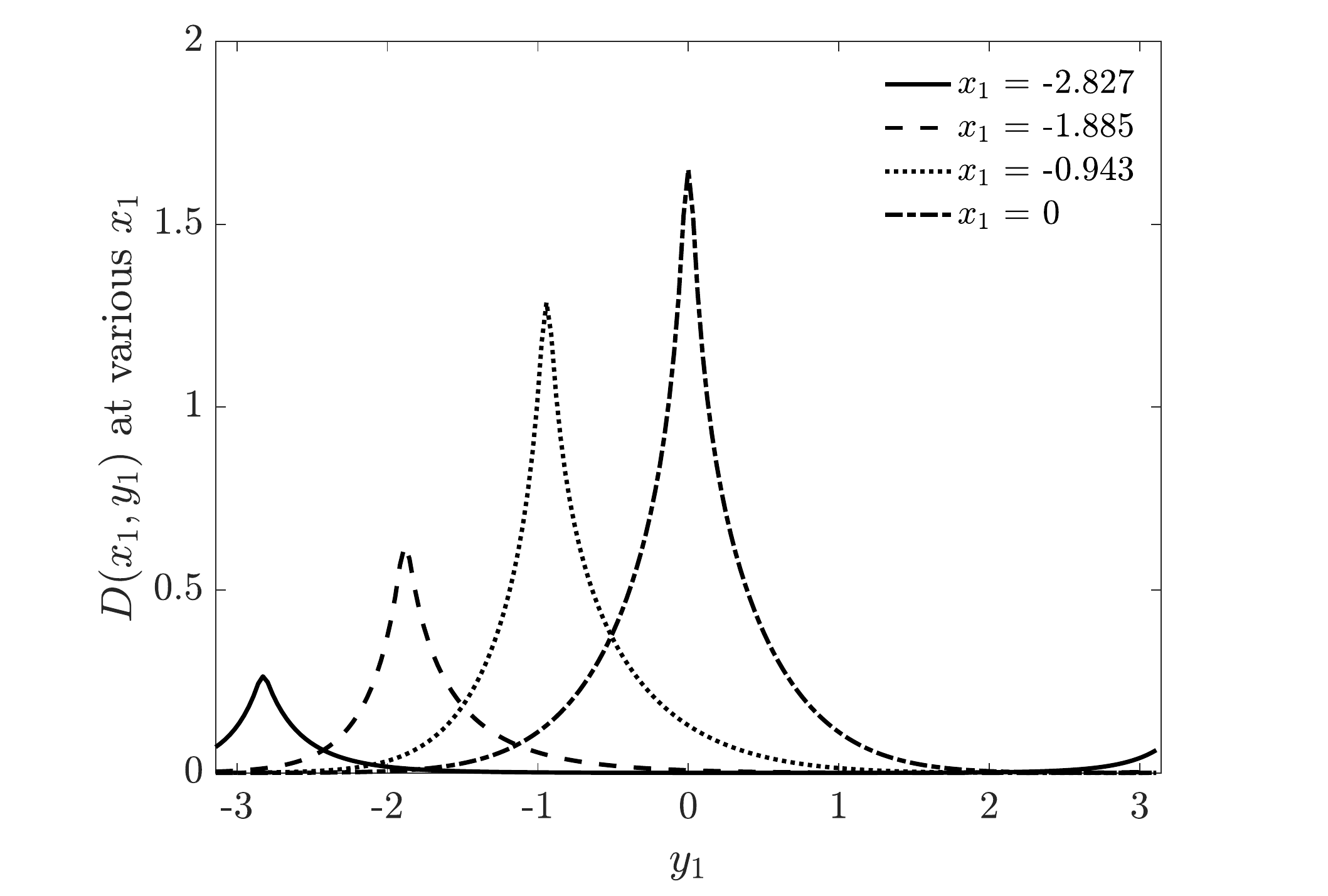}
         \caption{}
         \label{fig:example_slices_nonlocal_eddy_diffusivity}
     \end{subfigure}
     \caption{Example of a nonlocal eddy diffusivity. (a) Full nonlocal eddy diffusivity kernel, $D(x_1,y_1)$, and (b) cross sections of $D(x_1,y_1)$ at various $x_1$-locations.}
     \label{fig:example_nonlocal_eddy_diffusivity}
\end{figure}

However, such isotropy and idealized separation of scales between mean fields and fluctuations of continuum passive scalar fields often does not exist in turbulent flows \cite{corrsin1975limitations}. When the Boussinesq approximation breaks down, a more general form of the eddy diffusivity is introduced by Berkowicz and Prahm \cite{berkowicz_prahm_1980}:
\begin{equation} \label{eq:general spatial eddy diffusivity}
    -\overline{u_j^\prime c^\prime}(\mathbf{x}) = \int_{\mathbf{y}} D_{ji}(\mathbf{x},\mathbf{y})\frac{\partial \bar{c}}{\partial x_i}\big|_{\mathbf{y}} \mathrm{d}\mathbf{y},
\end{equation}
where $-\overline{u_j^\prime c^\prime}$ at a given location, $\mathbf{x}$, may now be influenced by the gradient, $\partial \bar{c}/\partial x_i$, at another location, $\mathbf{y}$. The eddy diffusivity is now tensorial and can capture anisotropy of the underlying mixing process by allowing the mean scalar flux to depend on different directions of the mean scalar gradient. Moreover, there is no requirement of length scale separation, and fluctuations are not assumed to mix locally. $D_{ji}(\mathbf{x}, \mathbf{y})$ is a \textit{nonlocal} eddy diffusivity kernel and captures how the scalar flux depends on gradients at other locations. Figure \ref{fig:example_nonlocal_eddy_diffusivity} shows an example of a nonlocal eddy diffusivity kernel (corresponding to the inhomogeneous model problem with periodic boundary conditions in Section \ref{inhomogeneous periodic model problem} in which only $D_{11}(x_1, y_1)$ is active). If one considers the eddy diffusivity as a discretized matrix, then for a purely local eddy diffusivity, only the diagonal values would be nonzero. However, the exact nonlocal eddy diffusivity in Figure \ref{fig:example_full_nonlocal_eddy_diffusivity} shows some spread around the diagonal, indicating the presence of nonlocality. Figure \ref{fig:example_slices_nonlocal_eddy_diffusivity} shows cross sections of the nonlocal eddy diffusivity at various $x_1$-locations. 

A nonlocal eddy diffusivity can also take into consideration temporal effects \cite{romanof1985application}\cite{kraichnan1987eddy}\cite{hamba1995analysis}\cite{hamba2004nonlocal}: 
\begin{equation} \label{eq:general eddy diffusivity}
    -\overline{u_j^\prime c^\prime}(\mathbf{x}, t) = \int_{\mathbf{y}, \tau} D_{ji}(\mathbf{x},\mathbf{y}, t, \tau)\frac{\partial \bar{c}}{\partial x_i}\big|_{\mathbf{y}, \tau} \mathrm{d}\mathbf{y} \mathrm{d}\tau,
\end{equation}
where $-\overline{u_j^\prime c^\prime}$ at a given time, $t$, may now also depend on the time history, $\tau$, of the gradient of $\bar{c}$. 

If the underlying flow is statistically stationary and homogeneous, the nonlocal eddy diffusivity expression in (\ref{eq:general eddy diffusivity}) has a simplified form: 
\begin{equation} \label{eq:homogeneous eddy diffusivity}
    -\overline{u_j^\prime c^\prime}(\mathbf{x}, t) = \int_{\mathbf{y}, \tau} D_{ji}(\mathbf{y}-\mathbf{x}, \tau-t)\frac{\partial \bar{c}}{\partial x_i}\big|_{\mathbf{y}, \tau} \mathrm{d}\mathbf{y} \mathrm{d}\tau,
\end{equation}
where the nonlocal eddy diffusivity does not depend on the specific point, $(\mathbf{x}, t)$, but rather only the distance, $(\mathbf{y}-\mathbf{x}, \tau-t)$. We consider nonlocal eddy diffusivities for homogeneous flows in Section \ref{homogeneous flows} before considering inhomogeneous flows in Section \ref{inhomogeneous flows}.

The moments of the nonlocal eddy diffusivity are related to the full nonlocal eddy diffusivity by considering the Taylor series expansion of Equation (\ref{eq:general eddy diffusivity}) locally about $\mathbf{y} = \mathbf{x}$ and $\tau = t$ (also known as a Kramers-Moyal expansion \cite{kampen1961power}) as done by Kraichnan \cite{kraichnan1987eddy} and Hamba \cite{hamba1995analysis}\cite{hamba2004nonlocal}:
\begin{equation} \label{eq:Kramers-Moyal expansion full}
    -\overline{u_j^\prime c^\prime}(\mathbf{x},t) = \int_{\mathbf{y},\tau}D_{ji}(\mathbf{x},\mathbf{y},t,\tau) \left[1 + (y_k - x_k)\frac{\partial}{\partial x_k} + \dots + (\tau - t)\frac{\partial}{\partial t} + \dots \right]\frac{\partial \bar{c}}{\partial x_i}\big|_{\mathbf{x},t}\mathrm{d}\mathbf{y}\mathrm{d}\tau.
\end{equation}
The terms involving $\bar{c}$ are now no longer a function of $\mathbf{y}$ and $\tau$ and can be taken out of the integral such that
\begin{equation}
    \label{eq:Kramers-Moyal expansion simplified}
    -\overline{u_j^\prime c^\prime}(\mathbf{x},t) = \left[ D^0_{ji}(\mathbf{x},t) + D^{1_s}_{jik}(\mathbf{x},t)\frac{\partial}{\partial x_k} + \dots + D^{1_t}_{ji}(\mathbf{x},t) \frac{\partial}{\partial t} + \cdots\right]\frac{\partial \bar{c}}{\partial x_{i}},
\end{equation}
where $D^0_{ji}$ is the zeroth-order spatial and temporal moment:
\begin{equation}
    \label{eq:zeroth moment}
    D_{ji}^0 (\mathbf{x},t) = \int_{\mathbf{y},\tau} D_{ji}(\mathbf{x},\mathbf{y}, t,\tau) \mathrm{d}\mathbf{y}\mathrm{d}\tau, 
\end{equation}
$D^{1_s}_{jik}$ is the first-order spatial moment (superscript $s$ for spatial):
\begin{equation}
    \label{eq:1st moment in space}
    D_{jik}^{1_s} (\mathbf{x}, t) = \int_{{y_k},\tau} (y_k - x_k) D_{jik}(\mathbf{x}, \mathbf{y}, t,\tau)\mathrm{d}y_k \mathrm{d}\tau,
\end{equation}
$D^{1_t}_{ji}$ is the first-order temporal moment (superscript $t$ for temporal):
\begin{equation}
    \label{eq:1st moment in time}
    D_{ji}^{1_t} (\mathbf{x}, t) = \int_{\mathbf{y},\tau} (\tau - t) D_{ji}(\mathbf{x}, \mathbf{y}, t,\tau)\mathrm{d}\mathbf{y} \mathrm{d}\tau,
\end{equation}
and so forth. The zeroth-order spatial and temporal moment, $D_{ji}^0$, is the local and anisotropic eddy diffusivity, and higher-order moments characterize the nonlocality of the eddy diffusivity.

\subsection{Methods for obtaining moments of the nonlocal eddy diffusivity} \label{intro to MFM}
While the moments of the nonlocal eddy diffusivity may be obtained by computing the full nonlocal eddy diffusivity and then integrating using the definitions in Equations (\ref{eq:zeroth moment})-(\ref{eq:1st moment in time}), Mani and Park \cite{mani2021macroscopic} use IMFM to compute the moments more directly. We first review methods for obtaining the full nonlocal eddy diffusivity before introducing direct methods for obtaining eddy diffusivity moments. 

The passive scalar transport equation in (\ref{eq:scalar transport}) can be written in operator form as
\begin{equation}
    [\mathcal{L}][c]=0,
\end{equation}
where [$\mathcal{L}$] is a matrix representing the discretized advection-diffusion operator. The desired equation governing the mean scalar field is 
\begin{equation}
    [\overline{\mathcal{L}}][\bar{c}]=0,
\end{equation}
where $[\overline{\mathcal{L}}]$ is the averaged operator containing both the closed advection-diffusion operator and the closure operator for the scalar flux in Equation (\ref{eq:mean scalar transport}). Let averaging be defined by
\begin{equation}
    [\bar{c}] = [P][c],
\end{equation}
where $[P]$ is the projection operator. Similarly, $[E]$ is an extension operator such that $[P][E] = [\mathcal{I}]$ where $[\mathcal{I}]$ is the identity operator. Mani and Park \cite{mani2021macroscopic} show that averaged operator can be found by using
\begin{equation}
    \label{eq:linalg Lbar}
    [\overline{\mathcal{L}}]=([P][\mathcal{L}]^{-1}[E])^{-1}.
\end{equation}
Once the averaged operator $[\bar{\mathcal{L}}]$ is obtained, one can subtract out the closed portion of the Reynolds-averaged advection-diffusion operator to find the closure operator, $[\bar{\mathcal{L^\prime}}]$. For example, if averages are taken over all directions except $x_1$ and the unclosed term is simply $\bar{\mathcal{L}}^\prime \bar{c} = \partial/\partial x_1 (\overline{u_1^\prime c^\prime})$, after obtaining $[\bar{\mathcal{L}}^\prime]$ one can then write the closure operator as 
\begin{equation}
    [\bar{\mathcal{L}}^\prime] = -[\partial/\partial x_1][D][\partial/\partial x_1].
\end{equation}
By removing the appropriate $[\partial/\partial x_1]$ matrices, one can recover the eddy diffusivity, $[D]$, in discretized form. If $[D]$ is a purely diagonal matrix, then the eddy diffusivity is purely local. If instead, there are nonzero off-diagonal entries in $[D]$, which then multiply a spread of corresponding entries in $[\partial \bar{c}/\partial x_1]$, then the eddy diffusivity is nonlocal. Matrix multiplication can be expressed as a convolution, and in continuous form, generalizes to the nonlocal eddy diffusivity formulation of Berkowicz and Prahm \cite{berkowicz_prahm_1980} in Equation (\ref{eq:general spatial eddy diffusivity}). 

We use this method for obtaining full nonlocal eddy diffusivities for the simple problems with low degrees of freedom in this work. However, inversion of $[\mathcal{L}]$, which is size-dependent on the number of degrees of freedom, can become prohibitively expensive for complex problems. Moreover, for general problems, $[\mathcal{L}']$ may include multiple unclosed terms from which it may not be possible to extract a nonlocal eddy diffusivity for each term separately using this approach.

Forcing methods, such as MFM of Mani and Park \cite{mani2021macroscopic} or use of the Green's function by Hamba \cite{hamba1995analysis}\cite{hamba2004nonlocal}, can take advantage of computational fluid dynamics solvers and probe the nonlocal eddy diffusivity directly. MFM is a more flexible technique that allows 1) explicit specification of the forcing and postprocessing of $\bar{c}$ to arrive at the closure operator, 2) specification of the gradient of $\bar{c}$ as a Dirac delta function and postprocessing of the scalar flux to obtain the nonlocal eddy diffusivity, which is consistent with the Green's function approach of Hamba \cite{hamba1995analysis}\cite{hamba2004nonlocal} as shown in Appendix \ref{Hamba comparison}, or 3) specification of $\bar{c}$ as polynomials and postprocessing of scalar fluxes to obtain moments of the nonlocal eddy diffusivity. Moreover, this linear-algebra-based forcing technique does not require the governing (microscopic) equations as analytical partial differential equations but only as operators in discretized form. 

Rather than obtaining the full nonlocal eddy diffusivity, we introduce IMFM and its usage for directly obtaining moments of the eddy diffusivity. In IMFM, one adds a forcing in order to maintain a pre-specified mean field and measures the scalar flux response. For example, a forcing, $s$, is added to the passive scalar transport equation in Equation (\ref{eq:scalar transport}):
\begin{equation}
    \frac{\partial c}{\partial t} + \frac{\partial}{\partial x_j}(u_j c) - D_M \frac{\partial^2 c}{\partial x_j \partial x_j} = s.
\end{equation}
At each time step, $\bar{c}$ is constrained to its pre-specified value due to $s$ while $c$ is free to evolve. In this procedure, at each time step, one can solve for $c$ without the forcing, and then shift $c$ appropriately (still observing the property $s=\bar{s}$) such that the average matches the pre-specified $\bar{c}$. Specifying the mean field as polynomials leads to moments of the eddy diffusivity. For example, specifying $\bar{c}=x_\alpha$ where $\alpha = 1, 2, \text{or } 3$ and substituting into the expansion in Equation (\ref{eq:Kramers-Moyal expansion simplified}) leads to 
\begin{equation}
    -\overline{u_j'c'}(\mathbf{x},t)|_{\bar{c}=x_\alpha} = D_{j\alpha}^0(\mathbf{x},t).
\end{equation}
Postprocessing the scalar flux gives the $i=\alpha$ component of the zeroth-order spatial and temporal moment of the eddy diffusivity, $D_{j\alpha}^0$. Specifying $\bar{c}$ as higher-order polynomials leads to higher-order moments of the eddy diffusivity. For the first-order spatial moment, specifying $\bar{c}=1/2x_\alpha^2$ leads to 
\begin{equation}
    -\overline{u_j'c'}(\mathbf{x},t)|_{\bar{c}=1/2x_\alpha^2} = x_\alpha D_{j\alpha}^0(\mathbf{x},t)+D_{j\alpha \alpha}^{1_s}(\mathbf{x},t),
\end{equation}
with no summation over $\alpha$ implied. Postprocessing the scalar flux and then subtracting out the contribution from the zeroth-order moment leads to $D_{j\alpha \alpha}^{1_s}$. Cross components of the first-order spatial moment may be obtained by by specifying $\bar{c}=x_\alpha x_\beta$ where $\alpha, \beta = 1, 2, \text{or } 3$ and $\alpha \neq \beta$. For the first-order temporal moment, specifying $\bar{c}=x_\alpha t$ leads to
\begin{equation}
    -\overline{u_j'c'}(\mathbf{x},t)|_{\bar{c}=x_\alpha t} = tD_{j\alpha}^0(\mathbf{x},t)+D_{j\alpha}^{1_t}(\mathbf{x},t).
\end{equation} 
Postprocessing the scalar flux and then subtracting out the contribution from the zeroth-moment leads to $D_{j\alpha}^{1t}$. Generally, calculation of higher-order moments depends on having previously obtained lower-order moments. IMFM has been used to investigate the nonlocal eddy diffusivity in turbulent flows such as channel flow \cite{park2021channel}, a separated boundary layer \cite{park2022direct} \cite{park2023direct}, and Rayleigh-Taylor instability \cite{lavacot2023RTI}.

\section{Approaches for modeling a nonlocal eddy diffusivity}
\label{modeling nonlocal eddy diffusivity}

Given a spatiotemporally nonlocal eddy diffusivity as in Equation (\ref{eq:general eddy diffusivity}), one may want to model the nonlocal eddy diffusivity in order to express the governing equation for the mean scalar as a partial differential equation rather than an integro-differential equation. For simplicity, consider a problem where averaging is taken over all directions except $x_1$, and there is only one component of the scalar flux, $\overline{u'_1c'}$. The scalar flux can be expressed exactly as
\begin{equation} 
    -\overline{u_1^\prime c^\prime}(x_1,t) = \int_{y_1, \tau} D(x_1, y_1,t,\tau)\frac{\partial \bar{c}}{\partial x_1}\big|_{y_1, \tau} \mathrm{d}y_1 \mathrm{d}\tau.
\end{equation}

\subsubsection*{Boussinesq model}
In the Boussinesq limit, where mixing is purely local, $D(x_1, y_1, t, \tau) = D^0 (x_1, t) \delta(y_1 - x_1)\delta(\tau - t)$. The purely local model is given by
\begin{equation} 
    \label{eq:Boussinesq eddy diffusivity model}
    -\overline{u_1^\prime c^\prime}(x_1, t) = D^0(x_1, t)\frac{\partial \overline{c}}{\partial x_1},
\end{equation}
where $D^0$ is the zeroth-order spatial and temporal moment. This is also sometimes called the $K$-model where the local eddy diffusivity, $K$, may be generalized to be a tensor to account for anisotropy \cite{kraichnan1987eddy}\cite{georgopoulos1989nonlocal}\cite{yoshizawa1985statistical}. The purely local model is the first term of the Taylor series expansion in Equation (\ref{eq:Kramers-Moyal expansion simplified}). Figure \ref{fig:example_boussinesq_eddy_diffusivity} shows an example of the eddy diffusivity kernel shape.

\subsubsection*{Explicit model}
\begin{figure}[t]
     \centering
     \begin{subfigure}[t]{0.45\textwidth}
         \centering
         \includegraphics[width=\textwidth]{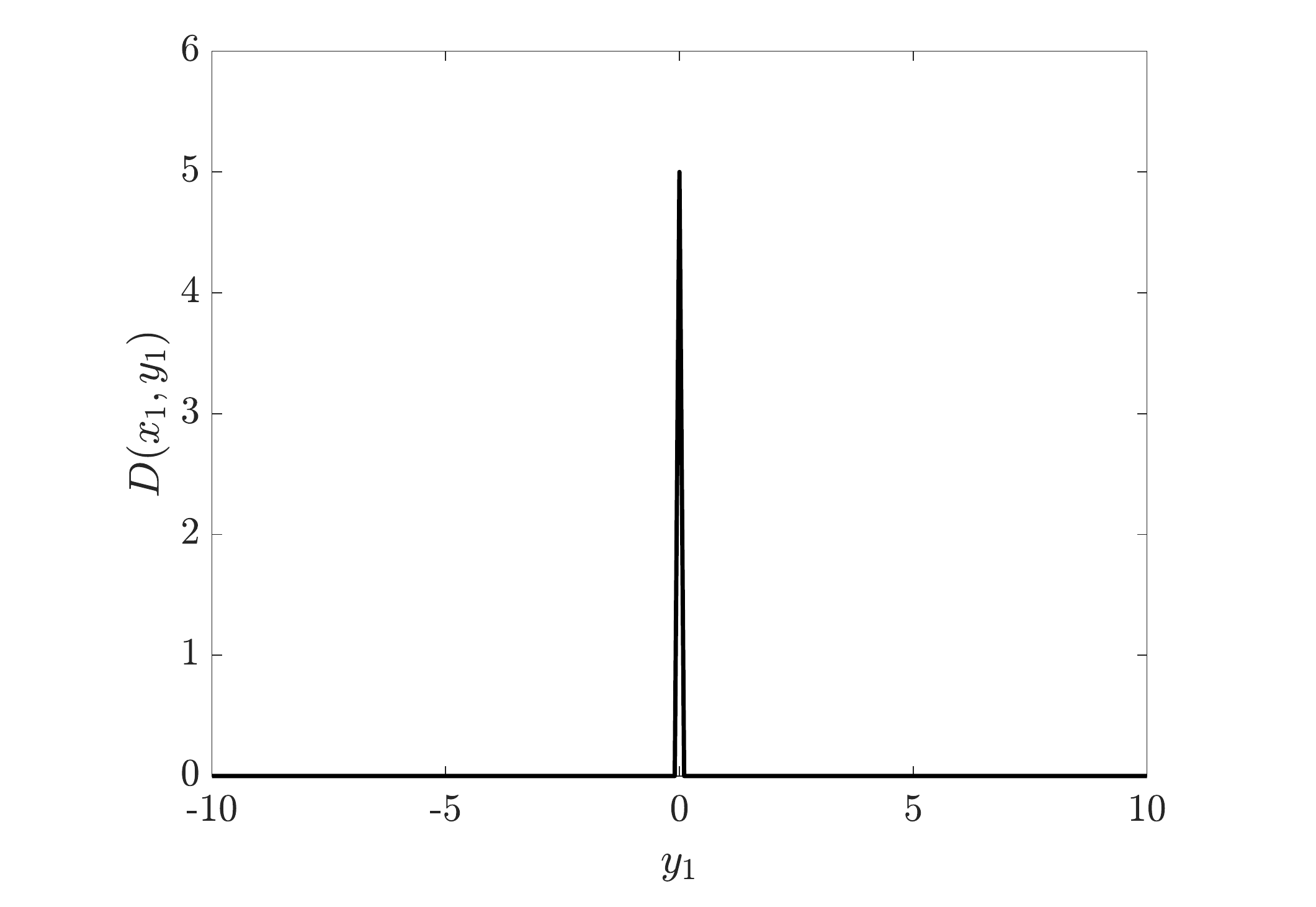}       \caption{}
         \label{fig:example_boussinesq_eddy_diffusivity}
     \end{subfigure}
     \begin{subfigure}[t]{0.45\textwidth}
         \centering
         \includegraphics[width=\textwidth]{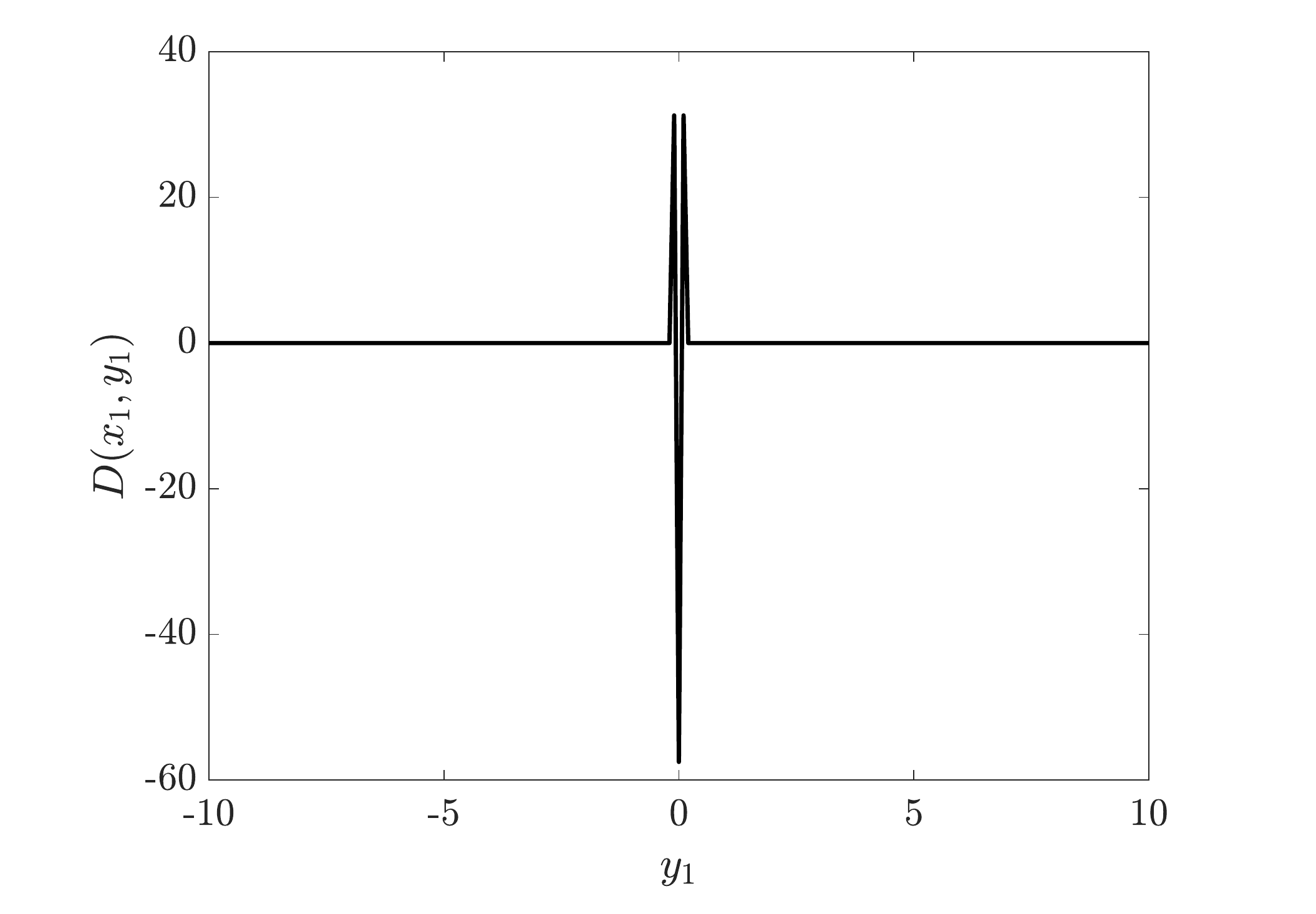}
         \caption{}
         \label{fig:example_explicit_eddy_diffusivity}
     \end{subfigure}
     \caption{(a) Example eddy diffusivity kernel corresponding to the Boussinesq model in Equation (\ref{eq:Boussinesq eddy diffusivity model}). (b) Example eddy diffusivity kernel corresponding to the explicit model in Equation (\ref{eq:explicit model}) with $D^{0} = 1/2$, $D^{1_s} = 0$, and $D^{2_s} = 1/32$ using a second-order central difference scheme and $\Delta y_1 = 0.1$.}
     \label{fig:example_boussinesq_explicit_eddy_diffusivity}
\end{figure}

If the Boussinesq locality approximation is not valid, one may consider higher-order terms of the expansion as nonlocal corrections to the local model. For example, including the first- and second-order spatial moments and first-order temporal moment of the nonlocal eddy diffusivity results in
\begin{equation}
    \label{eq:explicit model}
    -\overline{u_1^\prime c^\prime}(x_1, t) = D^0(x_1, t)\frac{\partial \overline{c}}{\partial x_1} + D^{1_s}(x_1, t)\frac{\partial^2 \overline{c}}{\partial x_1^2} + D^{2_s}(x_1, t)\frac{\partial^3 \overline{c}}{\partial x_1^3} + D^{1_t}(x_1,t)\frac{\partial^2\bar{c}}{\partial t \partial x_1}.
\end{equation}
This type of model form has been investigated by Hamba \cite{hamba1995analysis}\cite{hamba2004nonlocal}. In general, addition of these higher-order terms may not guarantee improvement of the model as demonstrated in Section \ref{model comparison homogeneous} and by Mani and Park \cite{mani2021macroscopic}. Non-convergence of the Kramers-Moyal expansion has also been previously shown in the context molecular dynamics \cite{pawula1967approximation}. This lack of convergence may be explained by examining the eddy diffusivity kernel shape as shown in Figure \ref{fig:example_explicit_eddy_diffusivity}. The leading term in Equation (\ref{eq:explicit model}) implies a Dirac delta function as the eddy diffusivity kernel, and adding higher-order corrections is equivalent to adding higher-order derivatives of Dirac delta functions; these corrections are still highly local and may not adequately capture the smooth shape of the nonlocal eddy diffusivity. 

\subsubsection*{A temporally nonlocal model}
If the flow is statistically stationary and homogeneous, a simple eddy diffusivity that is local in space but nonlocal in time is
\begin{equation}
    \label{eq:temporally nonlocal eddy diffusivity}
    D(y_1-x_1, \tau-t) = \delta(y_1 - x_1) \frac{\beta}{\alpha} \exp^{-(t-\tau)/\alpha},
\end{equation}
where $\alpha$ and $\beta$ are constants and the history effect of the eddy diffusivity decays exponentially backward in time. Figure \ref{fig:example_temporal_eddy_diffusivity} shows an example of the nonlocal eddy diffusivity kernel. The scalar flux is governed by
\begin{equation}
    \label{eq:temporal MMI eqn}
    \left[\alpha\frac{\partial}{\partial t}+1\right](-\overline{u_1'c'}) = \beta \frac{\partial \bar{c}}{\partial x_1},
\end{equation}
and as shown by Georgopoulos and Seinfeld \cite{georgopoulos1989nonlocal}, the mean scalar is governed by the hyperbolic telegrapher's equation:
\begin{equation}
    \alpha\frac{\partial^2 \bar{c}}{\partial t^2}  + \frac{\partial \bar{c}}{\partial t} = \beta \frac{\partial^2 \bar{c}}{\partial x_1^2},
\end{equation}
which was first derived by Goldstein \cite{goldstein1951diffusion} for a one-dimensional correlated random walk.

\subsubsection*{A spatially nonlocal model}
\begin{figure}[t]
     \centering
     \begin{subfigure}[t]{0.45\textwidth}
         \centering
         \includegraphics[width=\textwidth]{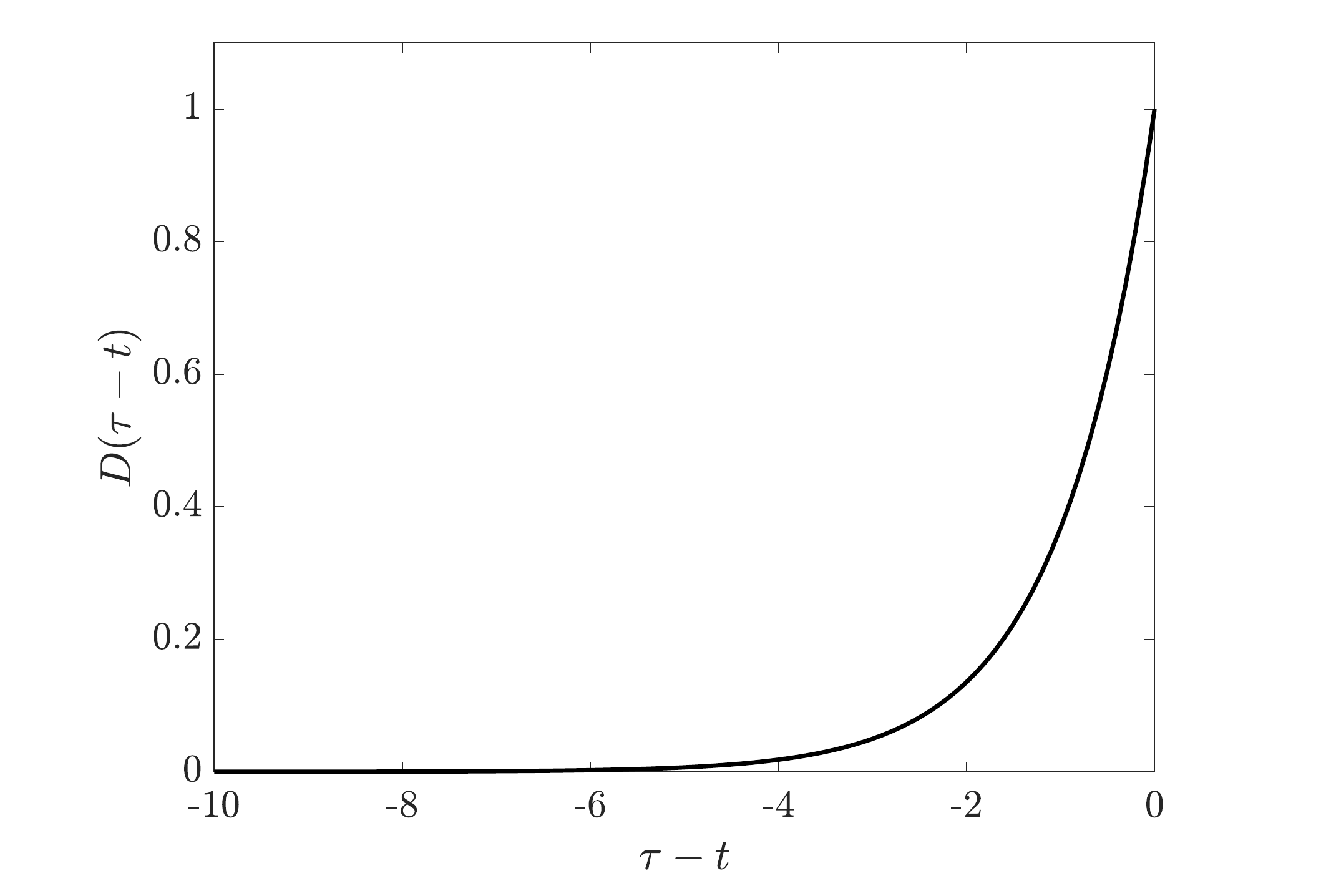}       
         \caption{}
         \label{fig:example_temporal_eddy_diffusivity}
     \end{subfigure}
     \begin{subfigure}[t]{0.45\textwidth}
         \centering
         \includegraphics[width=\textwidth]{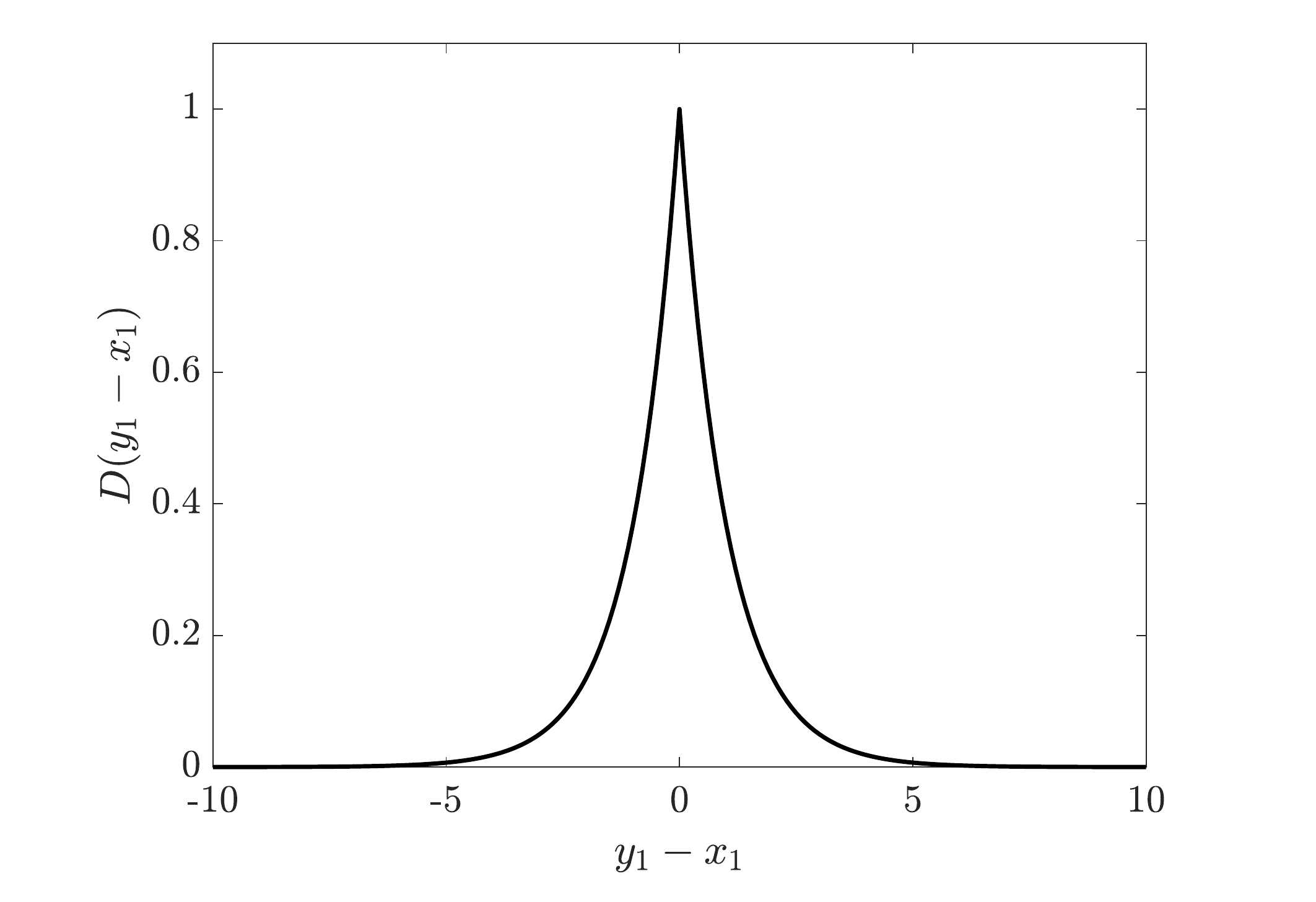}
         \caption{}
         \label{fig:example_spatial_eddy_diffusivity}
     \end{subfigure}
     \caption{(a) Example temporally nonlocal eddy diffusivity kernel in Equation (\ref{eq:temporally nonlocal eddy diffusivity}) with $\alpha = \beta = 1$. (b) Example spatially nonlocal eddy diffusivity kernel in Equation (\ref{eq:spatially nonlocal eddy diffusivity}) with $\alpha = \beta = 1$.}
     \label{fig:example_temporal_spatial_eddy_diffusivity}
\end{figure}

If the flow is statistically stationary and homogeneous, an example eddy diffusivity that is nonlocal in space but local in time is 
\begin{equation}
    \label{eq:spatially nonlocal eddy diffusivity}
    D(y_1 - x_1, \tau - t) = \frac{\beta}{\alpha}e^{-|y_1-x_1|/
    \alpha}\delta(\tau-t),
\end{equation}
where $\alpha$ and $\beta$ are constants and the spatial nonlocality of the eddy diffusivity is captured as a double-sided exponential. Figure \ref{fig:example_spatial_eddy_diffusivity} shows an example of the nonlocal eddy diffusivty kernel. As shown by Hamba \cite{hamba2004nonlocal}, the scalar flux is governed by 
\begin{equation}
    \label{eq:spatial MMI eqn}
    \left[1 - \alpha^2 \frac{\partial^2}{\partial x_1^2}\right](-\overline{u_1'c'}) = 2\beta\frac{\partial \bar{c}}{\partial x_1}.
\end{equation}

\subsubsection*{Matched moment inverse (MMI) operators}
For many flows, the measured nonlocal eddy diffusivity may not be exactly an exponential in time or space \cite{hamba2004nonlocal}\cite{hamba2005nonlocal}\cite{park2021channel}. However, it may still be appropriate to model the shape of the nonlocal eddy diffusivity as exponential functions or combinations of exponential functions. For such flows, the key question we seek to address is how to appropriately determine the coefficients. For this purpose, we introduce \textit{matched moment inverse} (MMI) operators which systematically allow determination of the coefficients using, as the name suggests, only measured moments of the nonlocal eddy diffusivity rather than the full measured nonlocal eddy diffusivity. For a statistically stationary and homogeneous flow, consider a model for a spatiotemporally nonlocal eddy diffusivity formed by combining Equation (\ref{eq:temporal MMI eqn}) and (\ref{eq:spatial MMI eqn}) (and more generally in which, higher-order terms may be included):
\begin{equation}
    \label{eq:MMI eqn}
    \left[a_3 \frac{\partial}{\partial t} + \left(1 + a_1 \frac{\partial}{\partial x_1} + a_2 \frac{\partial^2}{\partial x_1^2}\right)\right](-\overline{u_1'c'}) = a_0 \frac{\partial \bar{c}}{\partial x_1}.
\end{equation}
The coefficients will be determined by matching up to the second spatial moment and first temporal moment of the nonlocal eddy diffusivity. Rearranging,
\begin{equation}
    -\overline{u_1^\prime c^\prime} = \left[ 1 + a_1\frac{\partial}{\partial x_1} + a_2 \frac{\partial^2}{\partial x_1^2} + a_3\frac{\partial }{\partial t} \right]^{-1}a_0\frac{\partial \bar{c}}{\partial x_1},
\end{equation}
and Taylor series expanding the inverse operator on the right-hand-side leads to
\begin{equation} 
\label{eq:MMI eqn expanded}
    -\overline{u_1^\prime c^\prime} = \left[ 1 - a_1\frac{\partial}{\partial x_1} - a_2 \frac{\partial^2}{\partial x_1^2} + a_1\frac{\partial}{\partial x_1}\left(a_1 \frac{\partial}{\partial x_1}\right) + \dots - a_3\frac{\partial }{\partial t} + \dots \right] a_0 \frac{\partial \bar{c}}{\partial x_1}.
\end{equation}
Because the flow is statistically stationary and homogeneous, the coefficients, $a_i$, are constants. To match the first few moments of the modeled nonlocal eddy diffusivity with the exact measured moments, compare the expansion in (\ref{eq:MMI eqn expanded}) with the Taylor series expansion of the nonlocal eddy diffusivity in (\ref{eq:Kramers-Moyal expansion simplified}). This leads to the coefficients: $a_0 = D^0$, $a_1 = -D^{1_s}/D^0$, $a_2 = -D^{2_s}/D^0 + (D^{1_s}/D^0)^2$, and $a_3 = -D^{1_t}/D^0$.  After determination of the coefficients, the model form in Equation (\ref{eq:MMI eqn}) is used for the scalar flux. The bracketed operator on the left-hand-side of (\ref{eq:MMI eqn}) acting on the scalar flux is the MMI operator and may be generalized to include higher-order derivatives and mixed derivatives.

\begin{figure}[tb]
    \centering
    \begin{subfigure}[t]{0.45\textwidth}
         \centering
         \includegraphics[width=\textwidth]{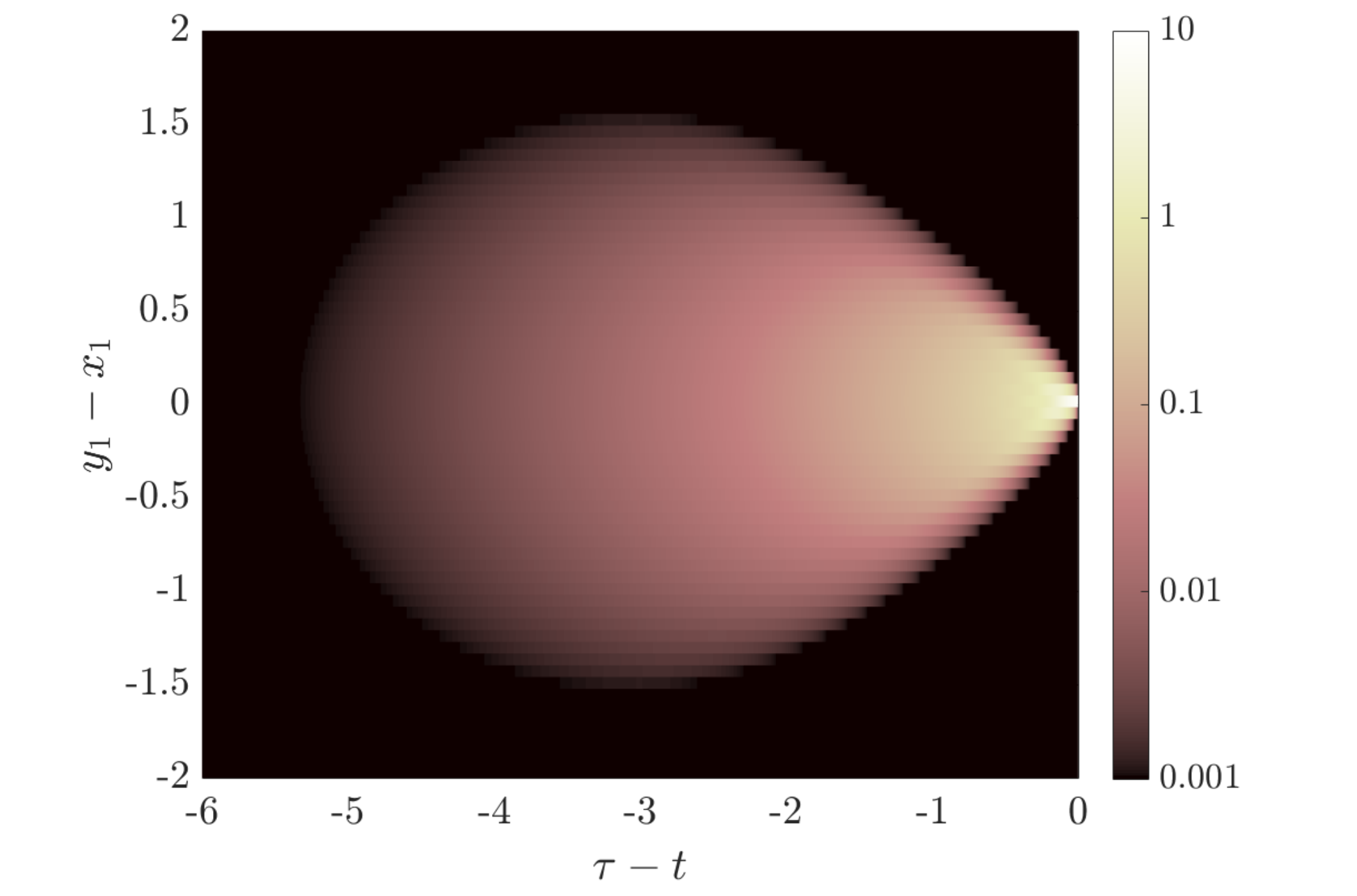}
        \caption{}
        \label{fig:full_kernel_homogeneous}
    \end{subfigure}
    \begin{subfigure}[t]{0.45\textwidth}
         \centering
         \includegraphics[width=\textwidth]{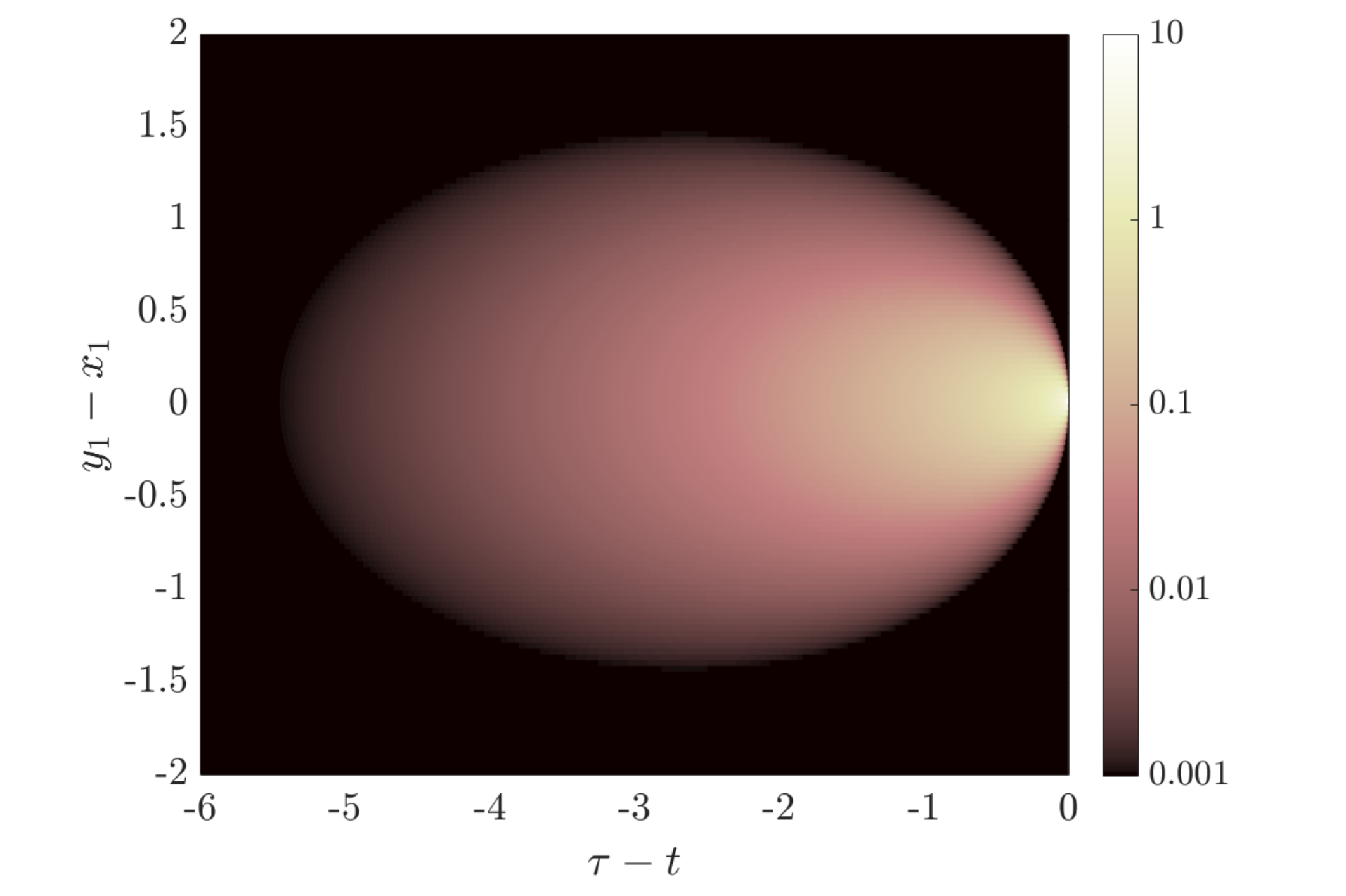}
         \caption{}
         \label{fig:MMI_kernel_homogeneous}
    \end{subfigure}
    \caption{(a) The exact spatiotemporally nonlocal eddy diffusivity, $D(y_1-x_1,\tau-t)$, for the homogeneous model problem in Section \ref{homogeneous model problem}. (b) The modeled eddy diffusivity using MMI.}
\end{figure}

Note that while the explicit model in Equation (\ref{eq:explicit model}) also matches the low-order measured moments of the eddy diffusivity, the shape of the nonlocal eddy diffusivity is not properly captured and leads to convergence issues. By keeping an infinite Taylor series in Equation (\ref{eq:MMI eqn expanded}) that smoothly approximates the nonlocal eddy diffusivity as an exponential function in time and a double-sided exponential function in space, MMI operators better approximate the true shape of the nonlocal eddy diffusivity. For example, Figure \ref{fig:full_kernel_homogeneous} shows the measured spatiotemporally nonlocal eddy diffusivity for the homogeneous model problem described in Section \ref{homogeneous flows}, and Figure \ref{fig:MMI_kernel_homogeneous} shows the MMI-constructed spatiotemporally nonlocal eddy diffusivity. The two nonlocal eddy diffusivities have the same zeroth-, first-, and second-order spatial and first-order temporal moments. 

For inhomogeneous flows, the moments of the eddy diffusivity  are functions of space, and correspondingly the coefficients of the MMI model will also be functions of space. In this case, the MMI coefficients cannot analytically be matched with the eddy diffusivity moments since an infinite number of higher-order derivatives of the unknown coefficients appear in the Taylor series expansion of the MMI operator. We present a modified numerical procedure for determining the MMI coefficients in order to match the low-order moments of the eddy diffusivity.

\begin{figure}[t]
     \centering
     \begin{subfigure}[t]{0.45\textwidth}
         \centering
         \includegraphics[width=\textwidth]{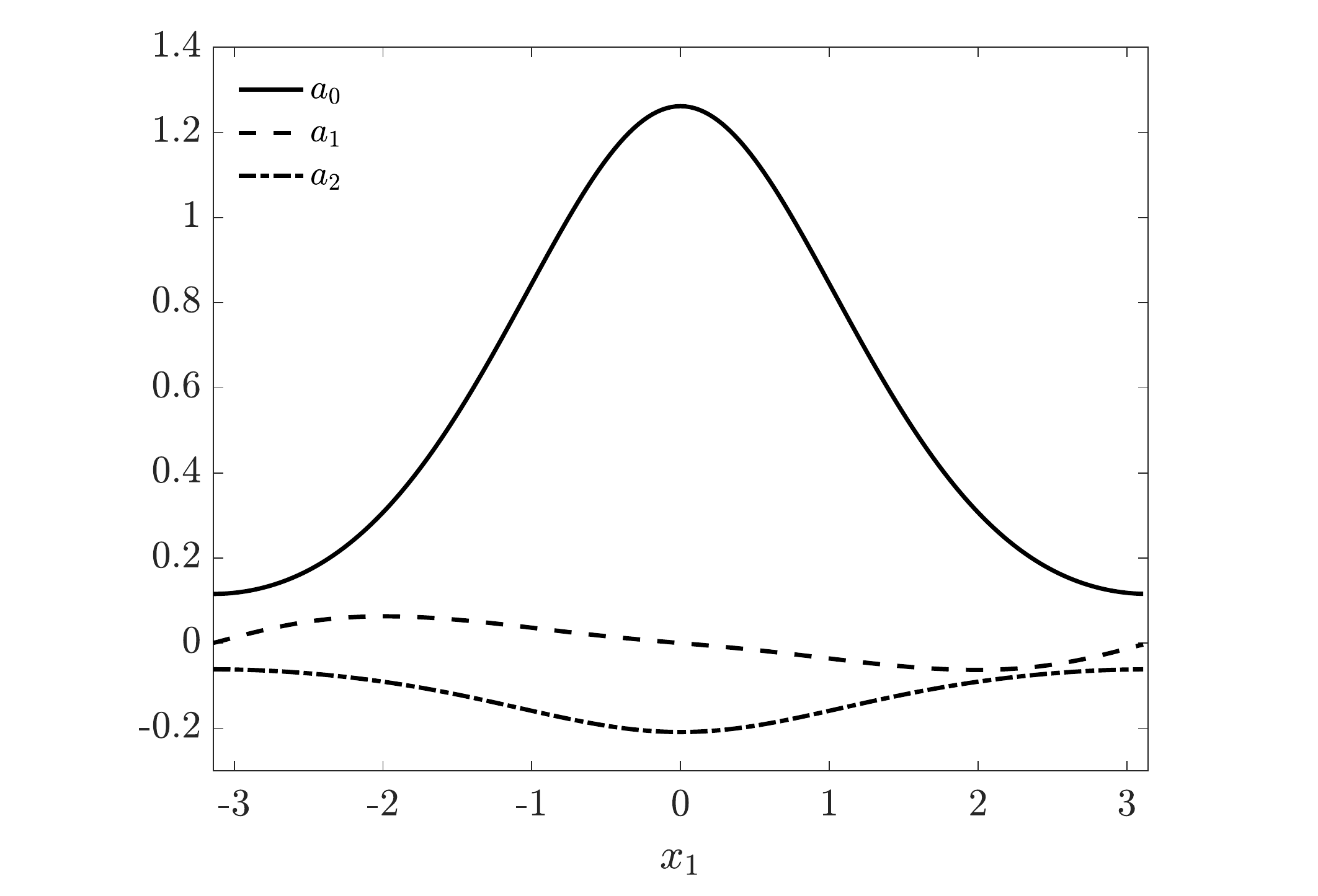}
        \caption{}
        \label{fig:MMI_coeff_inhomogeneous_periodic}
     \end{subfigure}
     \begin{subfigure}[t]{0.45\textwidth}
         \centering
         \includegraphics[width=\textwidth]{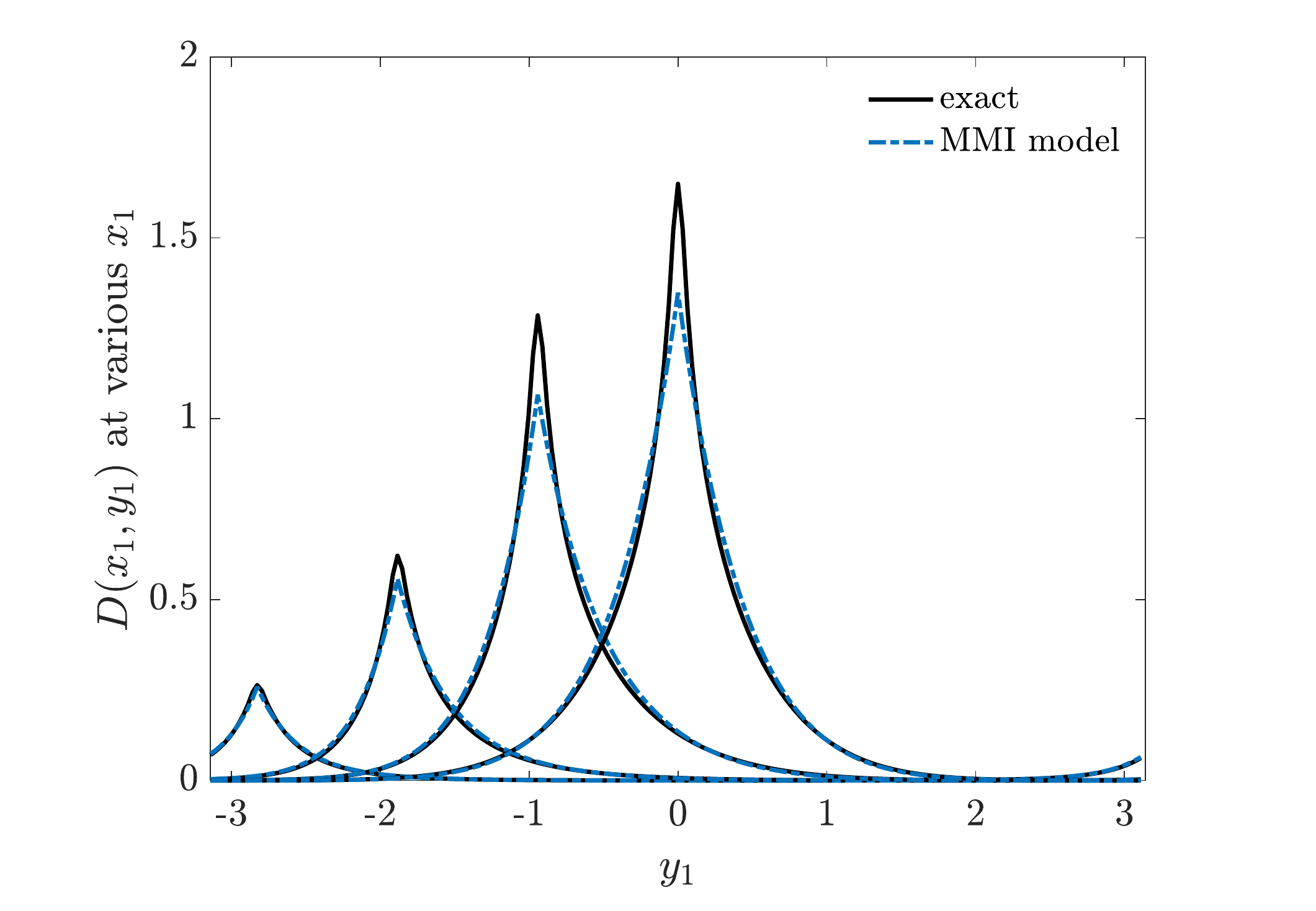}
         \caption{}
         \label{fig:kernel_slices_inhomogeneous_periodic}
     \end{subfigure}
     \caption{For the inhomogeneous model problem with periodic boundaries in Section \ref{inhomogeneous periodic model problem}: (a) MMI coefficients for Equation (\ref{eq:inhomogeneous MMI}). (b) The eddy diffusivity kernel from the MMI model closely approximates the exact eddy diffusivity kernel as shown for various $x_1$-locations. }
\end{figure}

For illustration, consider an inhomogeneous flow in which the eddy diffusivity is spatially nonlocal:
\begin{equation} 
    -\overline{u_1^\prime c^\prime}(x_1) = \int_{y_1} D(x_1,y_1)\frac{d \bar{c}}{d x_1}\big|_{y_1} dy_1.
\end{equation}
The MMI model including up to the second-order spatial moment is
\begin{equation} \label{eq:inhomogeneous MMI}
    \left[1 + a_1(x_1)\frac{d}{d x_1} + a_2(x_1)\frac{d^2}{d x_1^2} \right](-\overline{u_1^\prime c'}) = a_0(x_1)\frac{d \bar{c}}{d x_1}, 
\end{equation}
where the coefficients, $a_i$, are now also functions of $x_1$ and yet to be determined. The Taylor series expansion of the nonlocal eddy diffusivity is
\begin{equation} \label{eq:kernel Taylor series}
    -\overline{u_1^\prime c^\prime}(x_1) = \left[ D^0(x_1) + D^{1_s}(x_1)\frac{d}{d x_1} + D^{2_s}(x_1)\frac{d^2}{d x_1^2} + \dots \right]\frac{d\bar{c}}{d x_1},  
\end{equation}
where the moments of the eddy diffusivity are given by
\begin{subequations}
    \begin{align}
        D^0 &= \int_{y_1} D(x_1, y_1) dy_1, \\
        D^{1_s} &= \int_{y_1} (y_1-x_1) D(x_1, y_1) dy_1, \\
        D^{2_s} &= \int_{y_1} \frac{1}{2}(y_1-x_1)^2 D(x_1, y_1) dy_1.
    \end{align}
\end{subequations}
To obtain the low-order moments of the eddy diffusivity using IMFM, the forcing is applied such that $\bar{c} = x_1$, $\bar{c} = x_1^2/2$, $\bar{c} = x_1^3/6$, etc. Substituting these $\bar{c}(x_1)$ into Equation (\ref{eq:kernel Taylor series}) and post-processing the corresponding $-\overline{u_1^\prime c^\prime}|_{\bar{c}}$ leads to:
\begin{subequations}
    \begin{align}
        \label{eq:alpha0}
        -\overline{u_1^\prime c^\prime}|_{\bar{c} = x_1} &= D^0(x_1), \\
        \label{eq:alpha1}
        -\overline{u_1^\prime c^\prime}|_{\bar{c} = x_1^2/2} &= x_1 D^0(x_1)  + D^{1_s}(x_1), \\
        \label{eq:alpha2}
        -\overline{u_1^\prime c^\prime}|_{\bar{c} = x_1^3/6} &= \frac{x_1^2}{2} D^0(x_1) + x_1 D^{1_s}(x_1) + D^{2_s}(x_1).
    \end{align}
\end{subequations}
If $-\overline{u_1^\prime c^\prime}|_{\bar{c}}$ are directly available from IMFM, then one should use them directly. Otherwise, if only the moments are available, one should form $-\overline{u_1^\prime c^\prime}|_{\bar{c}}$ using the expressions above in Equations (\ref{eq:alpha0})-(\ref{eq:alpha2}). $-\overline{u_1^\prime c^\prime}|_{\bar{c}}$ contains exact information about the low-order moments of the true eddy diffusivity that one can now incorporate into the MMI model in Equation (\ref{eq:inhomogeneous MMI}). One can form three equations for $a_0(x_1)$, $a_1(x_1)$, and $a_2(x_1)$ by substituting $\bar{c} = x_1$, $\bar{c} = x_1^2/2$, and $\bar{c} = x_1^3/6$ and the corresponding $-\overline{u_1^\prime c^\prime}|_{\bar{c}}$ from Equations (\ref{eq:alpha0})-(\ref{eq:alpha2}) into Equation (\ref{eq:inhomogeneous MMI}):
\begin{subequations}
    \begin{align}
        \label{eq:alpha0 eqn}
        &-\overline{u_1^\prime c^\prime}|_{\bar{c}=x_1} + a_1 \frac{d}{dx_1}(-\overline{u_1^\prime c^\prime}|_{\bar{c}=x_1}) + a_2 \frac{d^2}{d x_1^2} (-\overline{u_1^\prime c^\prime}|_{\bar{c}=x_1}) = a_0, \\
        \label{eq:alpha1 eqn}
        &-\overline{u_1^\prime c^\prime}|_{\bar{c}=x_1^2/2} + a_1 \frac{d}{dx_1}(-\overline{u_1^\prime c^\prime}|_{\bar{c}=x_1^2/2}) + a_2 \frac{d^2}{d x_1^2} (-\overline{u_1^\prime c^\prime}|_{\bar{c}=x_1^2/2}) = a_0 x_1, \\
        \label{eq:alpha2 eqn}
        &-\overline{u_1^\prime c^\prime}|_{\bar{c}=x_1^3/6} + a_1 \frac{d}{dx_1}(-\overline{u_1^\prime c^\prime}|_{\bar{c}=x_1^3/6}) + a_2 \frac{d^2}{d x_1^2} (-\overline{u_1^\prime c^\prime}|_{\bar{c}=x_1^3/6}) = a_0 \frac{x_1^2}{2}.
    \end{align}
\end{subequations}
This linear system of equations for $a_0$, $a_1$, and $a_2$ is solved pointwise to obtain the MMI coefficients at each $x_1$-location. For example, Figure \ref{fig:MMI_coeff_inhomogeneous_periodic} shows the coefficients of the MMI model for the inhomogeneous model problem with periodic boundaries in Section \ref{inhomogeneous periodic model problem}. Figure \ref{fig:kernel_slices_inhomogeneous_periodic} shows the measured spatially nonlocal eddy diffusivity and the MMI-constructed eddy diffusivity. 

The suggested use of MMI is not the only way to mimic the nonlocal behavior of the eddy diffusivity. For example, Hamba \cite{hamba2004nonlocal} first obtained the exact nonlocal eddy diffusivity and then constructed alternative inverse operators by examining their shape against the exact eddy diffusivity kernel shape. The suggested MMI is an alternative that substantially reduces the number of DNS needed by focusing only on a limited number of moments of the nonlocal eddy diffusivity to characterize the kernel shape, while providing a systematic path for determining the coefficients of the inverse operator.

\section{Example MMI models for homogeneous flows} \label{homogeneous flows}
\subsection{Model problem: dispersion by a parallel flow} \label{homogeneous model problem}
As a simple example, consider the dispersion of a passive scalar by a homogeneous, laminar, parallel flow. This problem was first introduced by G. I. Taylor \cite{taylor1953dispersion} and revisited by \cite{mani2021macroscopic} to demonstrate how MFM can be used to obtain the exact nonlocal eddy diffusivity. Moreover, this problem is a simple example in which the scale separation assumption of the Boussinesq approximation does not hold, requiring the consideration of nonlocal effects. Specifically, consider a two-dimensional problem with the governing equation:
\begin{equation}
\frac{\partial c}{\partial t} + \frac{\partial}{\partial x_1}(u_1 c) + \frac{\partial}{\partial x_2}(u_2 c)  = D_M \left(\frac{\partial^2 c}{\partial x_1^2} + \frac{\partial^2 c}{\partial x_2^2} \right),
\end{equation}
where $c(x_1, x_2, t)$ is a passive scalar, $D_M$ is the molecular diffusivity, and $u_j$ is the parallel flow velocity: 
\begin{equation}
    u_1 = U \cos \left(\frac{2\pi}{L_2}x_2 \right), \quad u_2 = 0.
\end{equation}
The domain is $-\infty < x_1 < \infty$ and $0 \leq x_2 < L_2$ with periodic boundary conditions in $x_2$. Nondimensionalizing $x_2$ by $L_2/(2\pi)$, $x_1$ by $UL_2^2/(4\pi^2D_M)$, and $t$ by $L_2^2/(4\pi^2D_M)$ leads to the following nondimensionalized equation:
\begin{equation}
    \frac{\partial c}{\partial t} + \frac{\partial}{\partial x_1}(\cos(x_2)c) = \epsilon^2 \frac{\partial^2 c}{\partial x_1^2} + \frac{\partial^2 c}{\partial x_2^2},
\end{equation}
where $\epsilon = 2 \pi D_M/(L_2U)$ is the only nondimensional parameter. As in \cite{mani2021macroscopic}, for simplification we consider $\epsilon = 0$, corresponding to the limit of large Peclet number, i.e. assume the advective flux is much greater than the diffusive flux in the $x_1$-direction. The governing equation for the homogeneous example problem is
\begin{equation} \label{eq:homogeneous model problem}
\frac{\partial c}{\partial t} +  \frac{\partial}{\partial x_1}(\cos(x_2)c) = \frac{\partial^2 c}{\partial x_2^2}.
\end{equation}
For this problem, averaging is taken over the $x_2$-direction, i.e.
\begin{equation} \label{eq:average in x_2}
    \bar{c}(x_1, t) = \frac{1}{L_2} \int_0^{L_2} c(x_1, x_2, t) dx_2.
\end{equation}
Correspondingly, the mean scalar transport equation for this problem is
\begin{equation} \label{eq:homogeneous mean scalar transport}
    \frac{\partial \bar{c}}{\partial t} + \frac{\partial}{\partial x_1}(\overline{u_1^\prime c^\prime}) = 0,
\end{equation}
where $\overline{u_1^\prime c^\prime} = \overline{\cos(x_2)c^\prime}$ is the unclosed scalar flux. The mean advection, $\partial/\partial x_1(\bar{u}_1\bar{c})$, does not appear in (\ref{eq:homogeneous mean scalar transport}) since $\bar{u}_1 = \overline{\cos({x_2})} = 0$. The mean diffusion, $\partial^2 \bar{c}/\partial x_2^2$, also drops from (\ref{eq:homogeneous mean scalar transport}) due to averaging over $x_2$ and periodic boundary conditions. 

\begin{figure}[t]
    \centering
    \includegraphics[width=0.9\textwidth]{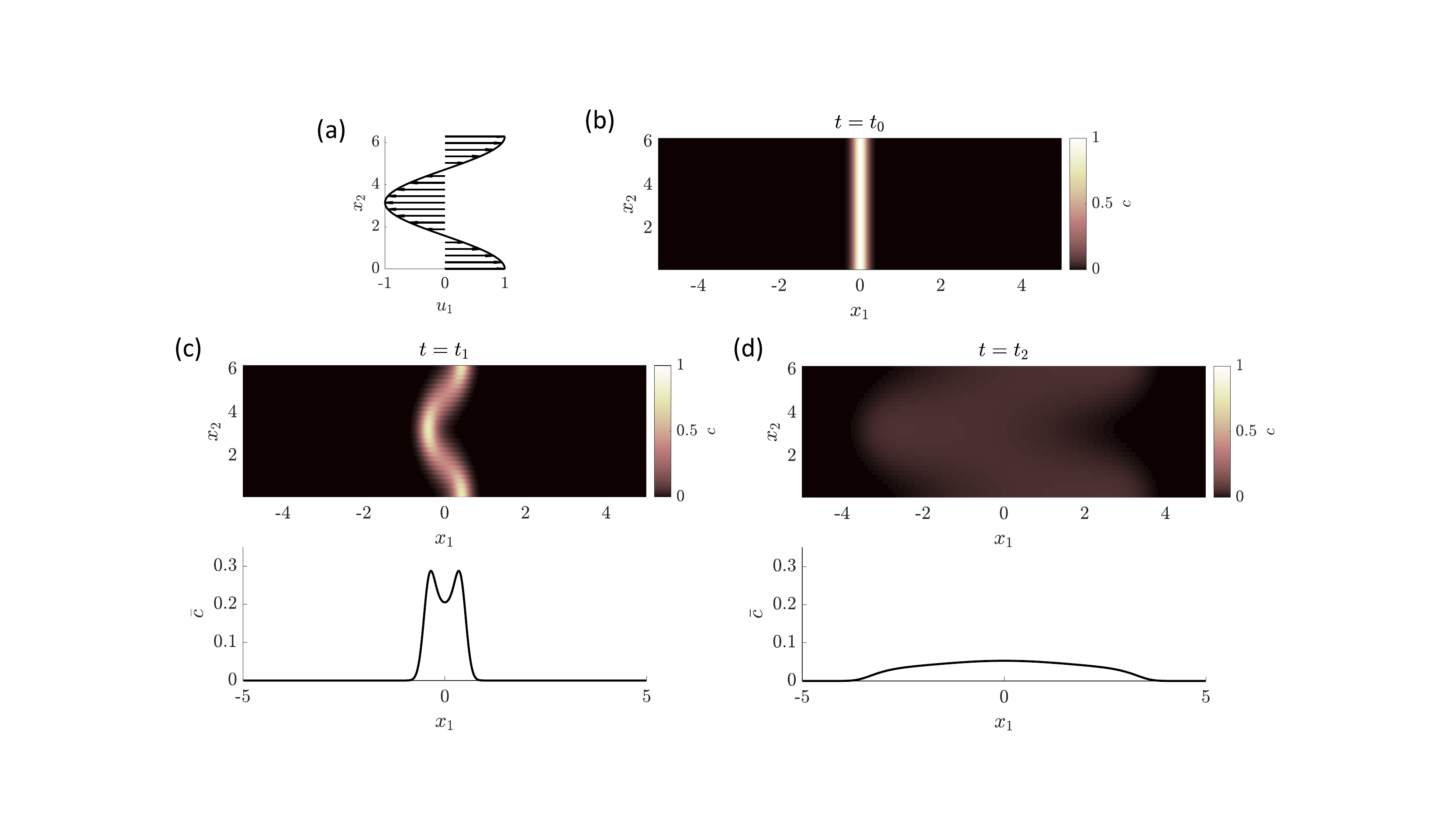}
    \caption{(a) The velocity profile for the homogeneous, parallel flow ($u_1 = \cos(x_2)$, $u_2 = 0$). (b) An initial condition corresponding to the release of a narrow band of passive scalar in the center of the domain ($c(t=0) = \exp(-x_1^2/0.025)$). (c) The dispersed scalar field, $c(x_1, x_2, t)$, and $x_2$-averaged field, $\bar{c}(x_1, t)$, at time, $t_1 = 0.5$. (d) $c(x_1, x_2, t)$ and $\bar{c}(x_1, t)$ at a later time, $t_2 = 4$.}
    \label{fig:homogeneous_2D_DNS} 
\end{figure}

Figure \ref{fig:homogeneous_2D_DNS}a shows the prescribed velocity profile. Figure \ref{fig:homogeneous_2D_DNS}b-d show the initial condition, $c(t=0) = \exp(-x_1^2/0.025)$, and time snapshots of the dispersed scalar field, $c(x_1, x_2, t)$, and averaged field, $\bar{c}(x_1, t)$ solved using DNS. The goal is to predict the complex behavior of $\bar{c}(x_1, t)$ using a one-dimensional partial differential equation. Mani and Park \cite{mani2021macroscopic} also attempted to model this problem by finding the nonlocal eddy diffusivity in Fourier space in both space and time using MFM, analytically fitting the operator in Fourier space, and then transforming it back into physical space. We revisit this problem in order to develop a method that does not need the full nonlocal eddy diffusivity and results in an operator that does not involve the operational square root. 

\subsection{Model comparison} \label{model comparison homogeneous}
We compare some of the models in Section \ref{modeling nonlocal eddy diffusivity} with MMI models for the homogeneous problem. The moments of the nonlocal eddy diffusivity for this problem may be obtained numerically using IMFM or analytically using Taylor's approach for dispersion by a parallel flow \cite{taylor1953dispersion} and its extension by Aris \cite{aris1960dispersion}. Taylor's approach uses the transport equation for the fluctuations, $c'$, and applies the same length/time scale separation assumptions as the Boussinesq approximation to find the leading-order term balance, resulting in the Boussinesq model. Higher-order moments are found by considering perturbative corrections to the leading-order term balance, resulting in models similar to the explicit model in Section \ref{modeling nonlocal eddy diffusivity}. The first few moments of the nonlocal eddy diffusivity are: $D^0 = 1/2$, $D^{1_s} = 0$, $D^{2_s} = 1/32$, and $D^{1_t} = -1/2$. 

\begin{figure}[t]
    \centering
    \includegraphics[width=0.7\textwidth]{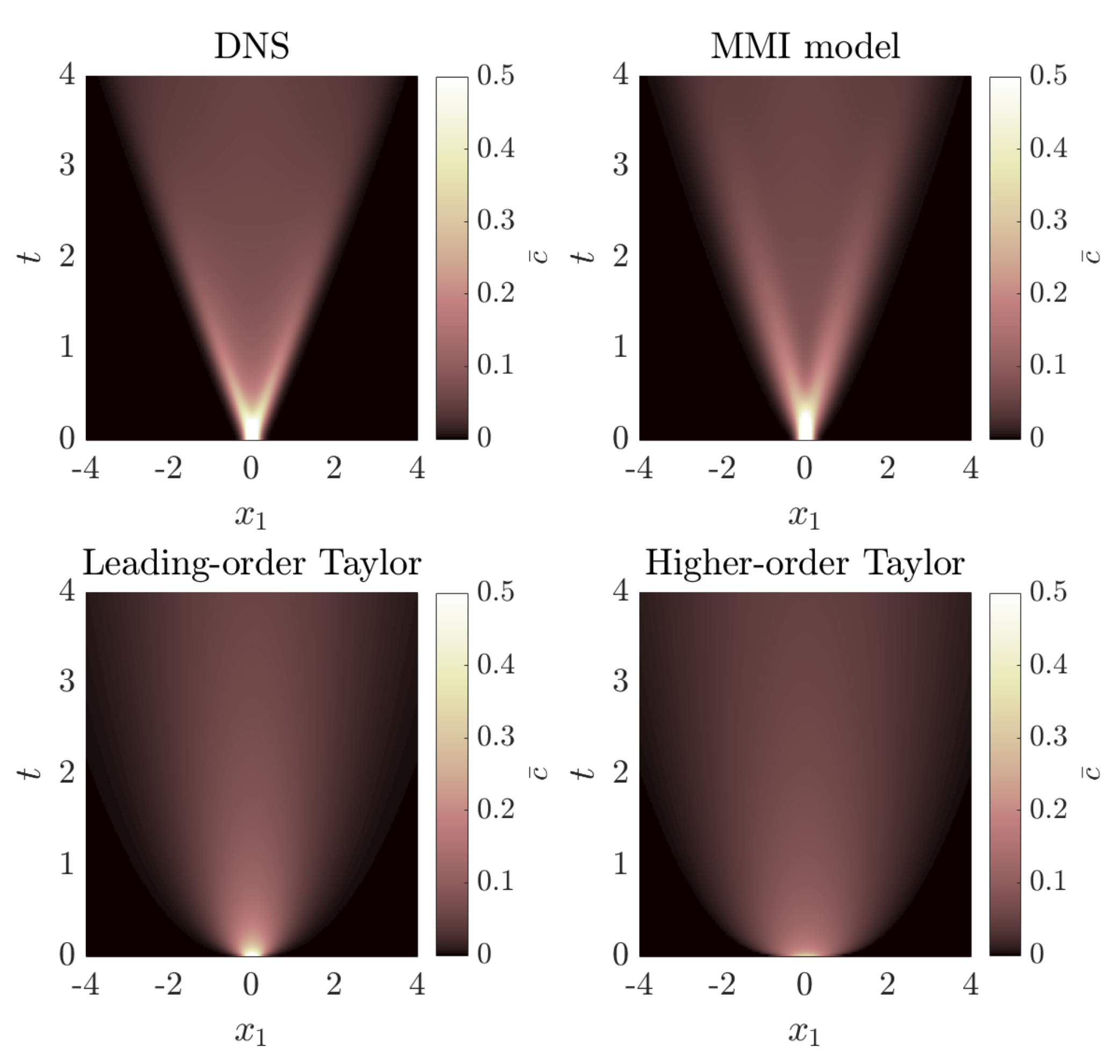}
    \caption{Model comparison of the averaged field, $\bar{c}(x_1, t)$, using the initial condition $c(t=0) = \exp(-x_1^2/0.025)$. The spatiotemporal MMI model in Equation (\ref{eq:spatiotemporal MMI model homogeneous}) closely captures the spread of the averaged field, whereas the leading-order Taylor and higher-order Taylor models overpredict the spread of the averaged field.}
    \label{fig:modelcomparison_homogeneous} 
\end{figure}

The Boussinesq model or leading-order Taylor model is 
\begin{equation}
    \label{eq:leading order Taylor}
    -\overline{u_1'c'} = \frac{1}{2}\frac{\partial \bar{c}}{\partial x_1},
\end{equation}
and the explicit model or higher-order Taylor model is 
\begin{equation}
    \label{eq:higher order Taylor}
    -\overline{u_1'c'} = \frac{1}{2}\frac{\partial \bar{c}}{\partial x_1} + \frac{1}{32} \frac{\partial^3\bar{c}}{\partial x_1^3} - \frac{1}{2}\frac{\partial^2 \bar{c}}{\partial t \partial x_1}.
\end{equation}
As discussed in Mani and Park \cite{mani2021macroscopic}, for this problem, the higher-order Taylor model is modified such that the last term in Equation (\ref{eq:higher order Taylor}) does not appear and the signs of the coefficients are consistent with the dissipation mechanism:
\begin{equation}
    \label{eq:higher order Taylor modified}
    -\overline{u_1'c'} = \frac{1}{2}\frac{\partial \bar{c}}{\partial x_1} - \frac{7}{32} \frac{\partial^3\bar{c}}{\partial x_1^3}.
\end{equation}
The MMI model only incorporating temporal nonlocality and whose coefficients are determined using the procedure in Section \ref{modeling nonlocal eddy diffusivity} is
\begin{equation}
\left[\frac{\partial }{\partial t}+1\right](-\overline{u_1^\prime c^\prime}) = \frac{1}{2}\frac{\partial \bar{c}}{\partial x_1},
\end{equation}
and similarly the MMI model incorporating spatiotemporal nonlocality is 
\begin{equation}
\label{eq:spatiotemporal MMI model homogeneous}
\left[\frac{\partial }{\partial t} + \left (1 - \frac{1}{16}\frac{\partial^2}{\partial x_1^2}\right)\right](-\overline{u_1^\prime c^\prime}) = \frac{1}{2}\frac{\partial \bar{c}}{\partial x_1}.
\end{equation}
Figure \ref{fig:full_kernel_homogeneous} shows the exact measured nonlocal eddy diffusivity, and Figure \ref{fig:MMI_kernel_homogeneous} shows the shape of the nonlocal eddy diffusivity captured by the spatiotemporal MMI model in Equation (\ref{eq:spatiotemporal MMI model homogeneous}). Appendix \ref{spatiotemporal kernel} gives details on how these spatiotemporal eddy diffusivities are computed.

\begin{figure}[t]
     \centering
     \begin{subfigure}[t]{0.45\textwidth}
         \centering
         \includegraphics[width=\textwidth]{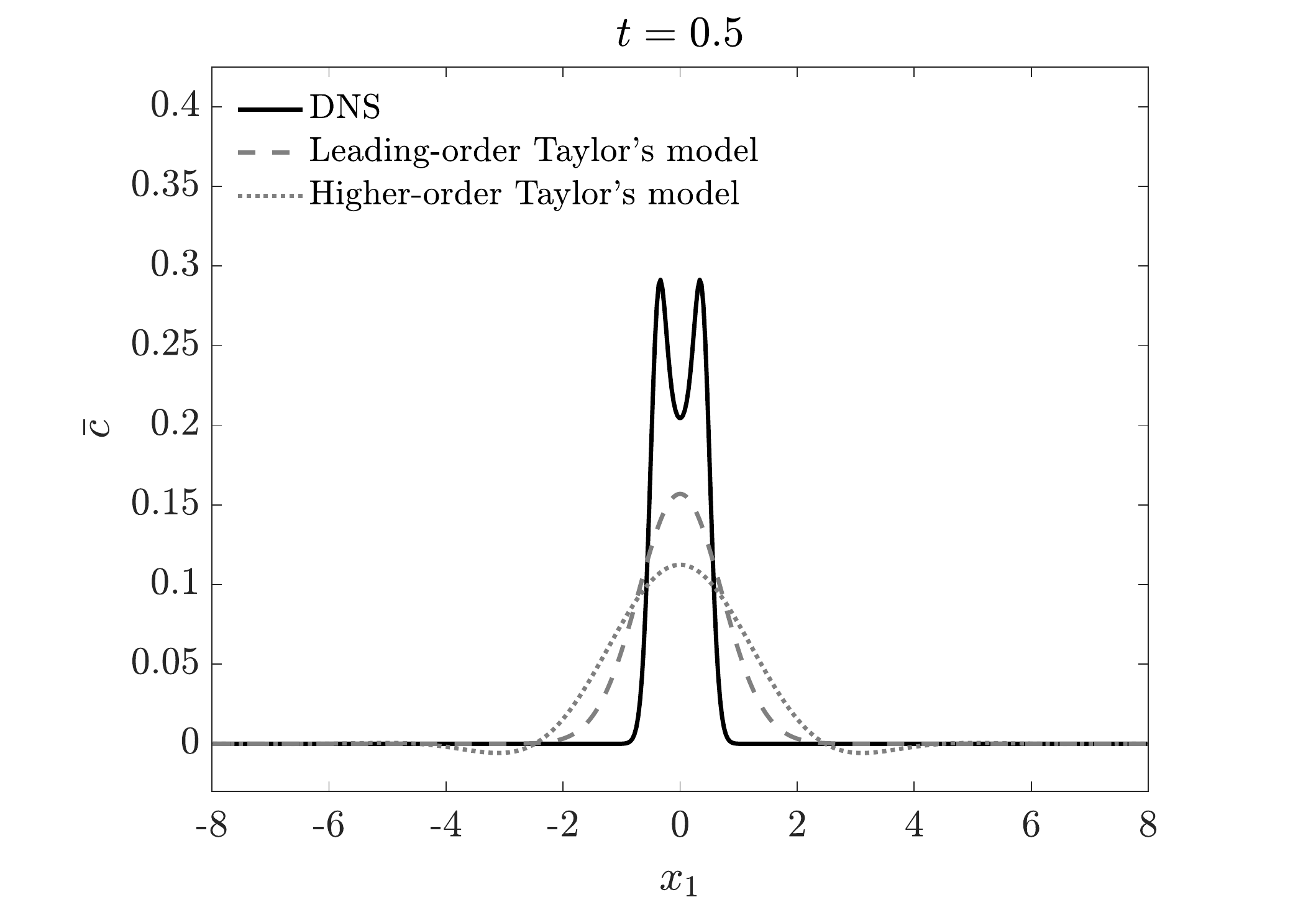}
        \caption{Early-time comparison.}
        \label{fig:modelcomparison_taylor_t05}
     \end{subfigure}
     \begin{subfigure}[t]{0.45\textwidth}
         \centering
         \includegraphics[width=\textwidth]{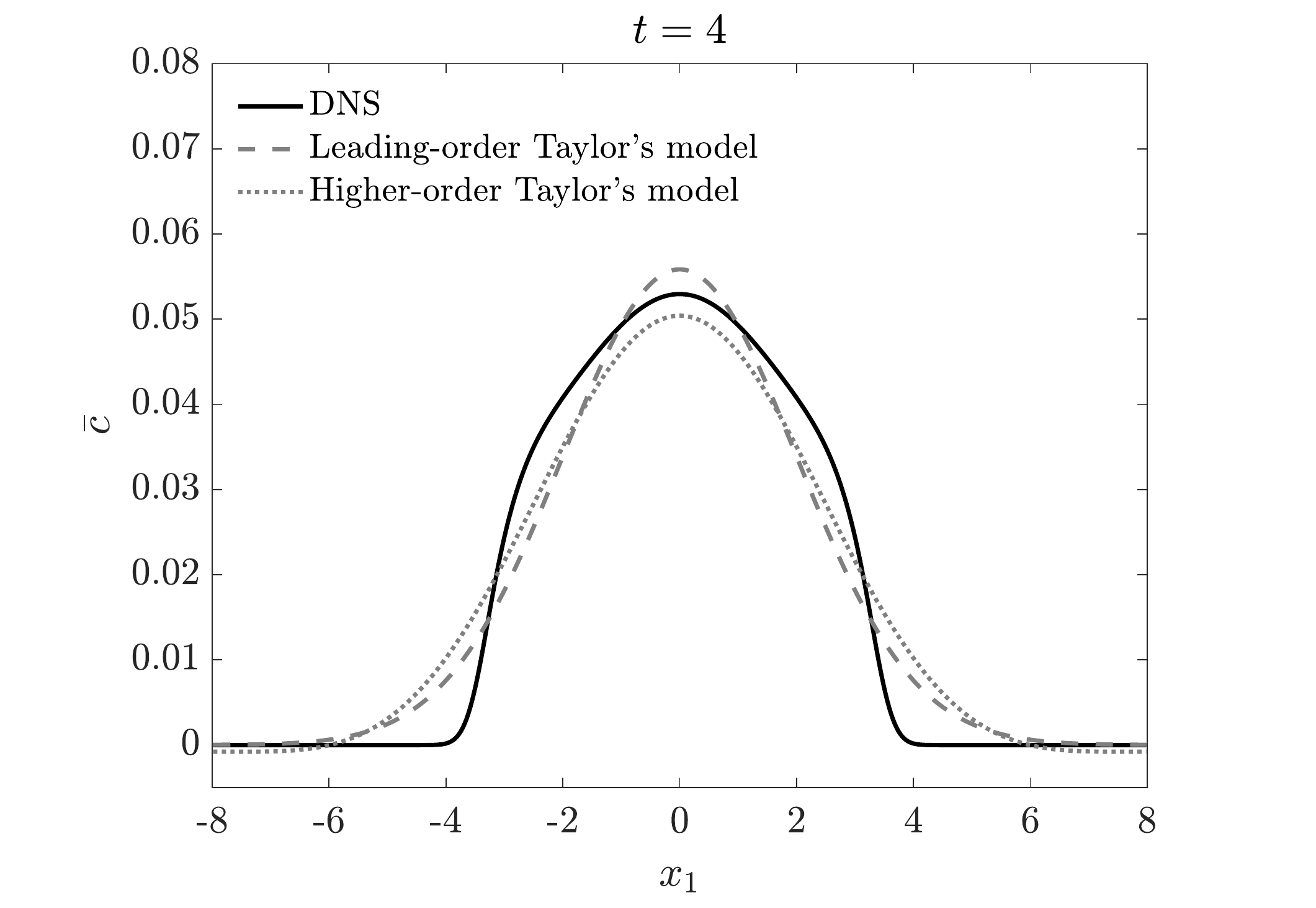}
         \caption{Late-time comparison.}
         \label{fig:modelcomparisontaylor_t4}
     \end{subfigure}
     \caption{Model comparison of leading-order Taylor's model and higher-order Taylor's model.}
     \label{fig:modelcomparison_tslices_taylor}
\end{figure}

Figure \ref{fig:modelcomparison_homogeneous} shows the evolution of the averaged field, $\bar{c}(x_1, t)$. Compared with the DNS solution, the spatiotemporal MMI model closely predicts the spread of the averaged field. The leading-order Taylor model causes the mean field to spread out too quickly, indicating the importance of including nonlocality. The higher-order Taylor model performs even worse, which demonstrates that adding a finite number of higher-order corrections as part of an infinite Taylor series expansion of the eddy diffusivity may not guarantee model improvement. Figure \ref{fig:modelcomparison_tslices_taylor} shows a comparison between the leading-order Taylor's model and higher-order Taylor's model at an early time, $t = 0.5$, and a later time, $t = 4$.  

\begin{figure}[t]
     \centering
     \begin{subfigure}[t]{0.45\textwidth}
         \centering
         \includegraphics[width=\textwidth]{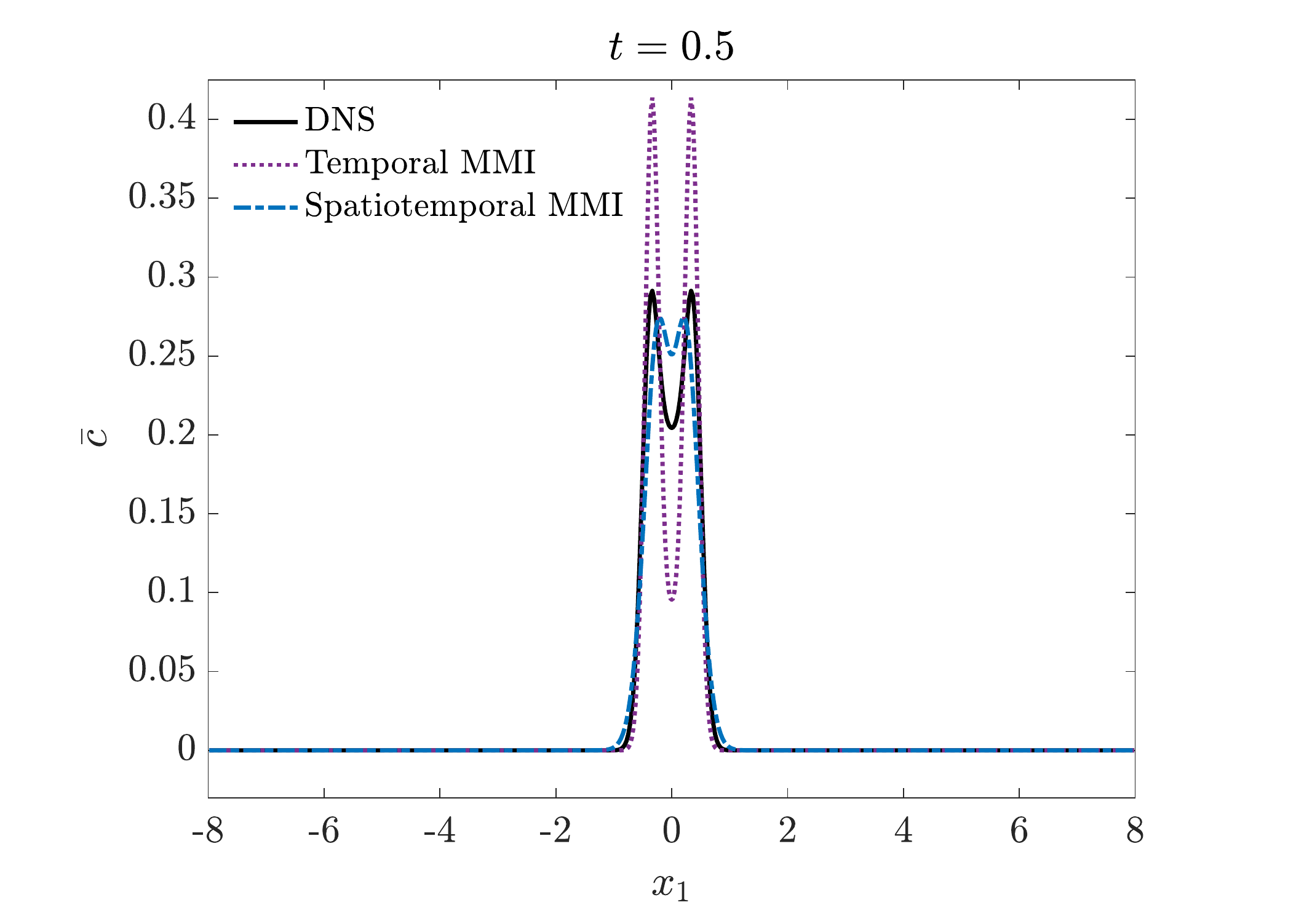}
        \caption{Early-time comparison.}
        \label{fig:modelcomparison_MMI_t05}
     \end{subfigure}
     \begin{subfigure}[t]{0.45\textwidth}
         \centering
         \includegraphics[width=\textwidth]{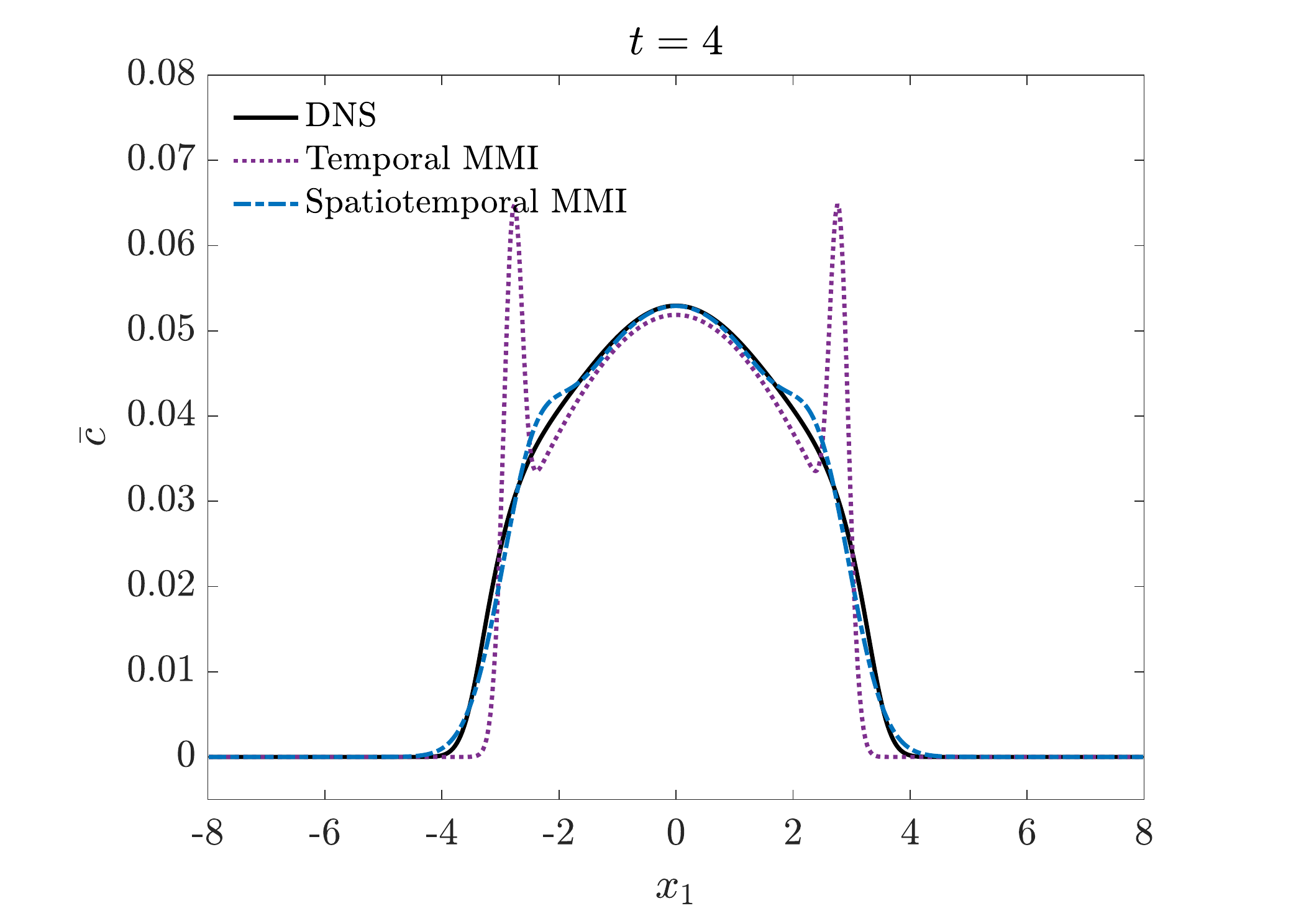}
         \caption{Late-time comparison.}
         \label{fig:modelcomparison_MMI_t4}
     \end{subfigure}
     \caption{Model comparison of the temporal MMI model and spatiotemporal MMI model.}
     \label{fig:modelcomparison_tslices_MMI}
\end{figure}

Figure \ref{fig:modelcomparison_tslices_MMI} shows a comparison between the MMI model incorporating temporal nonlocality and the MMI model incorporating spatiotemporal nonlocality. Both models capture the qualitative shape of the DNS solution better than either leading-order Taylor's model or higher-order Taylor's model at early times. The spatiotemporal MMI model reasonably matches the DNS solution whereas the temporal MMI model shows some overshoot. The spatiotemporal MMI model excellently captures the late-time solution. All four models are expected to perform well at late times in the limit of slowly varying mean field, i.e. where the Boussinesq approximation becomes valid. However, in this case, the spatiotemporal MMI model does well even outside of this limit. Note also that the higher-order Taylor model actually produces a negative solution, whereas the MMI models do not have this issue. 

\subsection{Comparison with other nonlocal eddy diffusivity models}
We now compare the spatiotemporal MMI model in Equation (\ref{eq:spatiotemporal MMI model homogeneous}) with another nonlocal model presented in \cite{mani2021macroscopic} for the same homogeneous problem. Mani and Park \cite{mani2021macroscopic}  approximate the nonlocal eddy diffusivity by fitting an operator to match MFM data in the limits of $k, \omega \rightarrow 0$ and $k, \omega \rightarrow \infty$, where $k$ is the wavenumber corresponding to the Fourier transform in $x_1$-direction and $\omega$ the frequency corresponding to the Fourier transform in time. Transforming back into physical space, the unclosed scalar flux \cite{mani2021macroscopic} is modeled as
\begin{equation}
    \label{eq:MFM inspired model}
    -\frac{\partial}{\partial x_1}\overline{u_1^\prime c^\prime}(x_1, t) = \left[ -\sqrt{\left(\mathcal{I} + \frac{\partial}{\partial t}\right)^2 - \frac{\partial^2}{\partial x_1^2}} + \left( \mathcal{I} + \frac{\partial}{\partial t}\right) \right]\bar{c}(x_1, t),
\end{equation}
where $\mathcal{I}$ is the identity operator. Note that this MFM-inspired, eddy diffusivity operator does not have the cost-saving advantages of using MMI since any numerical implementation of the model in (\ref{eq:MFM inspired model}) would require nonlocal operations in physical space involving full matrices. We show a comparison of this MFM-inspired model with the MMI model to address a more general modeling question: When adding nonlocal corrections to the local model, is it more appropriate to match the limits of large $k$ and $\omega$ or the low-order moments of the nonlocal eddy diffusivity?

To illustrate the effect of matching the limits of large $k$ and $\omega $ on the eddy diffusivity, consider the Taylor series expansion in Equation (\ref{eq:Kramers-Moyal expansion simplified}), simplified for the homogeneous problem in \ref{homogeneous model problem}:
\begin{equation}
    -\overline{u_1^\prime c^\prime} = \left[ D^0 + D^{1_s}\frac{\partial}{\partial x_1} + D^{2_s}\frac{\partial^2}{\partial x_1^2} + \dots + D^{1_t}\frac{\partial}{\partial t} + \cdots\right]\frac{\partial \bar{c}}{\partial x_{1}},
\end{equation}
where the moments of the eddy diffusivity are constants. Taking the Fourier transform leads to:
\begin{equation}
    -\widehat{\overline{u_1^\prime c^\prime}} = \left[ D^0 + ikD^{1_s} - k^2D^{2_s}+ \dots + i\omega D^{1_t} + \cdots\right]ik\hat{\bar{c}}.
\end{equation}
This shows that matching the limit of $k,\omega \rightarrow 0$ is equivalent to matching the zeroth-order moment of the eddy diffusivity, $D^0$. Matching the first and second moments of the eddy diffusivity, as done by MMI models, adds corrections in the limit of small $k$ and $\omega$. This prioritizes better capturing smooth and slowly varying solutions. Whereas, matching large $k$ and $\omega$, as done in \cite{mani2021macroscopic}, is equivalent to matching high-order moments of the nonlocal eddy diffusivity. This would prioritize capturing sharp and quickly-varying features in the solutions. 

\begin{figure}[t]
     \centering
     \begin{subfigure}[t]{0.45\textwidth}
         \centering
         \includegraphics[width=\textwidth]{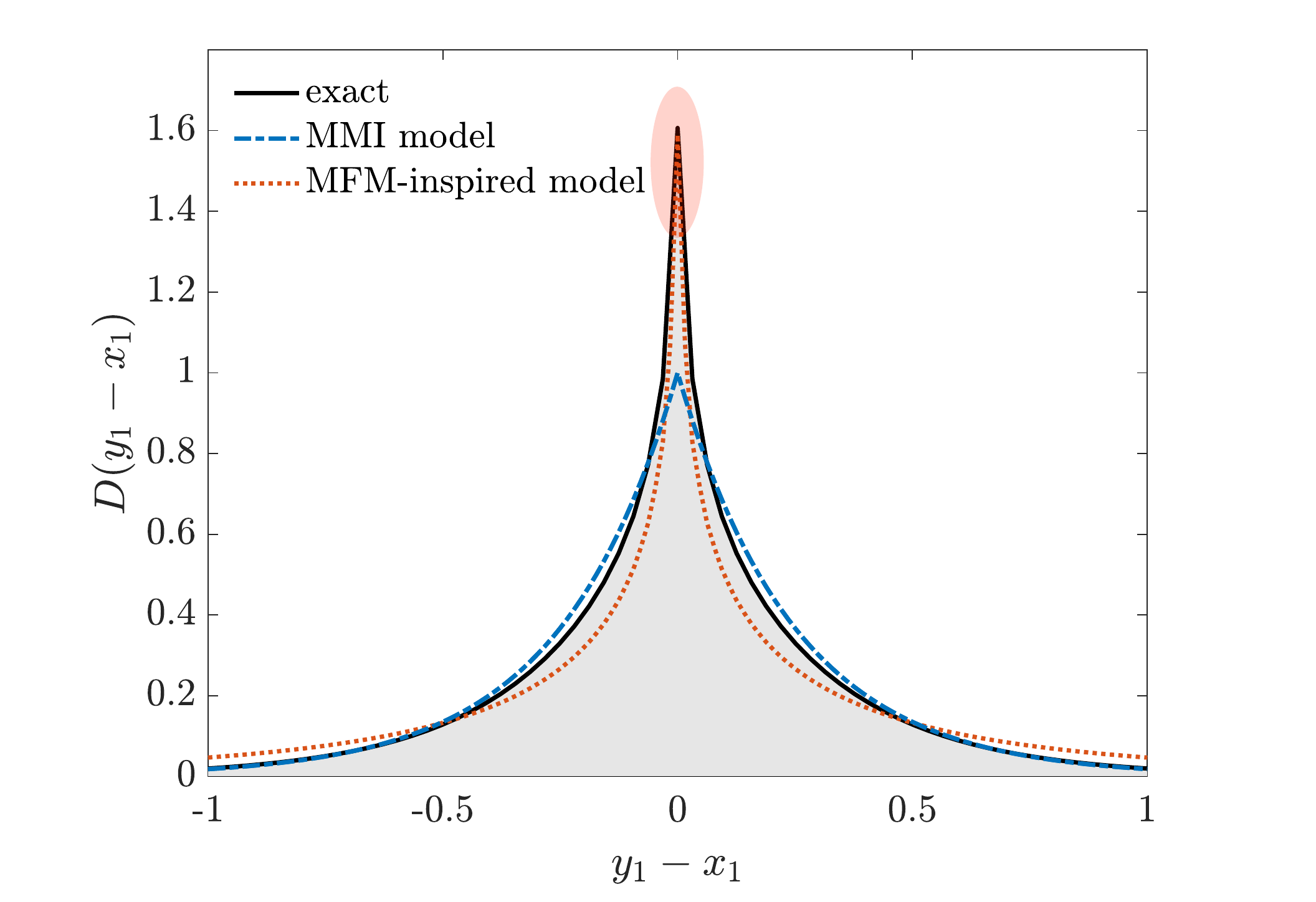}
        \caption{}
        \label{fig:physical spatial kernel}
     \end{subfigure}
     \begin{subfigure}[t]{0.45\textwidth}
         \centering
         \includegraphics[width=\textwidth]{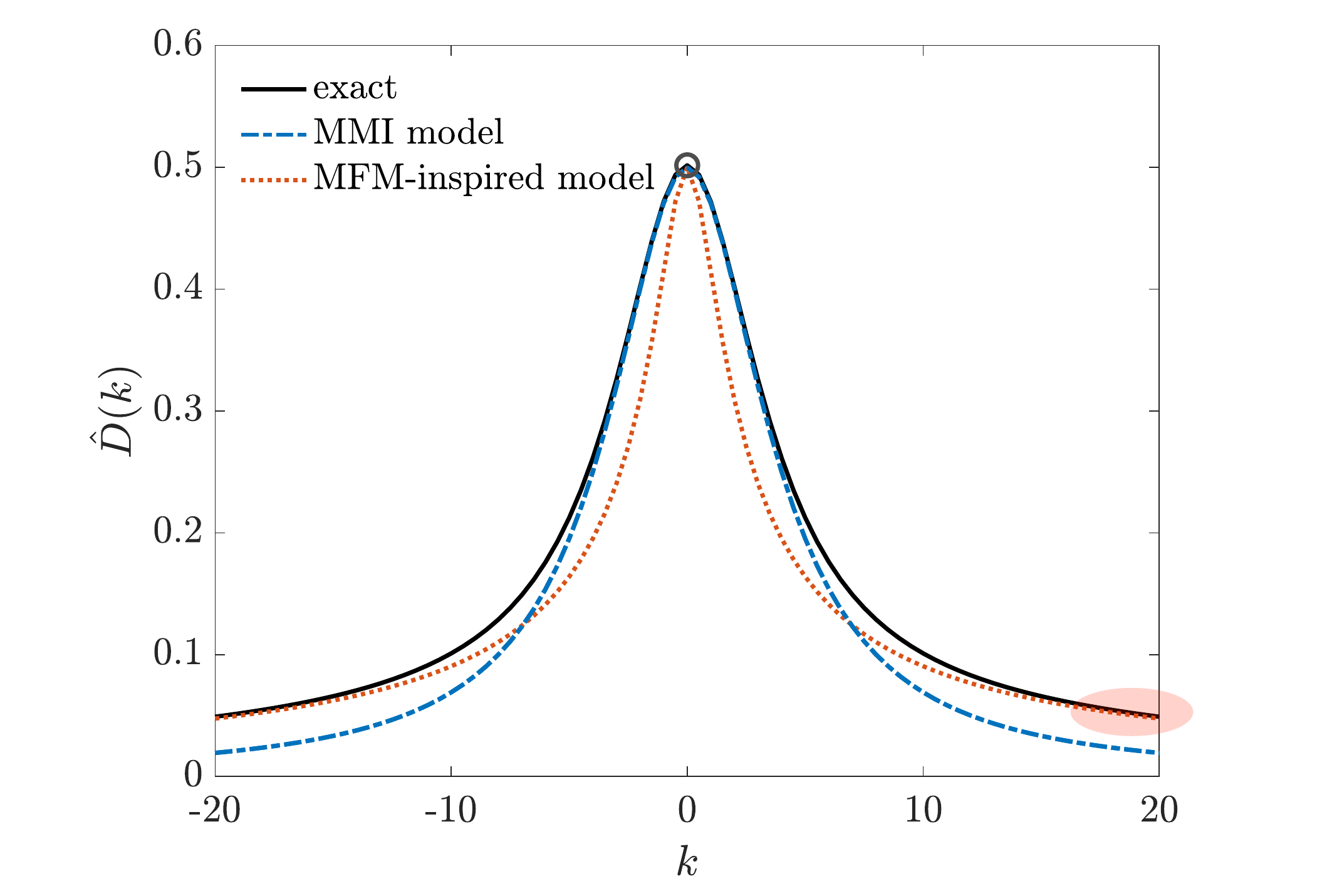}
         \caption{}
         \label{fig:FT spatial kernel}
     \end{subfigure}
     \caption{Comparison of the spatially nonlocal eddy diffusivities (equivalent to the spatiotemporally nonlocal eddy diffusivities integrated over $\tau$) (a) in physical space and (b) in Fourier space.}
\end{figure}

For illustration, Figure \ref{fig:physical spatial kernel} shows a comparison of the exact and modeled spatially nonlocal eddy diffusivities for the homogeneous problem (equivalent to the spatiotemporally nonlocal eddy diffusivities in Figure \ref{fig:full_kernel_homogeneous} and \ref{fig:MMI_kernel_homogeneous} integrated over $\tau$). Figure \ref{fig:FT spatial kernel} shows the Fourier transform of the spatially nonlocal eddy diffusivities. Because the exact and MMI-modeled spatially nonlocal eddy diffusivities were first obtained in physical space and then Fourier transformed, there is some numerical discretization error due to the use of second-order finite differences; however, the truncated plotting window shown in Fourier space is converged. The eddy diffusivity for the MFM-inspired model is obtained in Fourier space analytically by Fourier transforming Equation (\ref{eq:MFM inspired model}) and evaluating at $\omega = 0$ (equivalent to integration over $\tau$):
\begin{equation}
    \label{eq:MFM inspired FT diffusivity}
    \hat{D}(k) = \frac{-\sqrt{1+k^2}+1}{-k^2}.
\end{equation}
The eddy diffusivity of the MFM-inspired model in physical space is then obtained by taking the inverse Fourier transform of (\ref{eq:MFM inspired FT diffusivity}).

In Figure \ref{fig:physical spatial kernel}, the shaded area in gray under the eddy diffusivity, $D^0$, corresponds to $\hat{D}(0)$ in Fourier space in Figure \ref{fig:FT spatial kernel}. The first spatial moment of eddy diffusivity, $D^{1_s}$, is related to the first derivative of the kernel in Fourier space, $\frac{d}{dk} \hat{D}(0)$, and so forth. In Fourier space, the MMI model matches the shape of the eddy diffusivity for small $k$ as shown in Figure \ref{fig:FT spatial kernel}. Whereas, the high wavenumber region, shaded in red in Figure \ref{fig:FT spatial kernel}, corresponds to the peak in physical space at $D(0)$ which involves a sharp feature in Figure \ref{fig:physical spatial kernel}. This peak at $D(0)$ is captured by the MFM-inspired model. In physical space, matching low-order moments better captures the overall shape of the nonlocal eddy diffusivity, whereas matching large $k$ and $\omega$ captures the large wavelength/frequency features of the eddy diffusivity. 

\begin{figure}[t]
     \centering
     \begin{subfigure}[t]{0.45\textwidth}
         \centering
         \includegraphics[width=\textwidth]{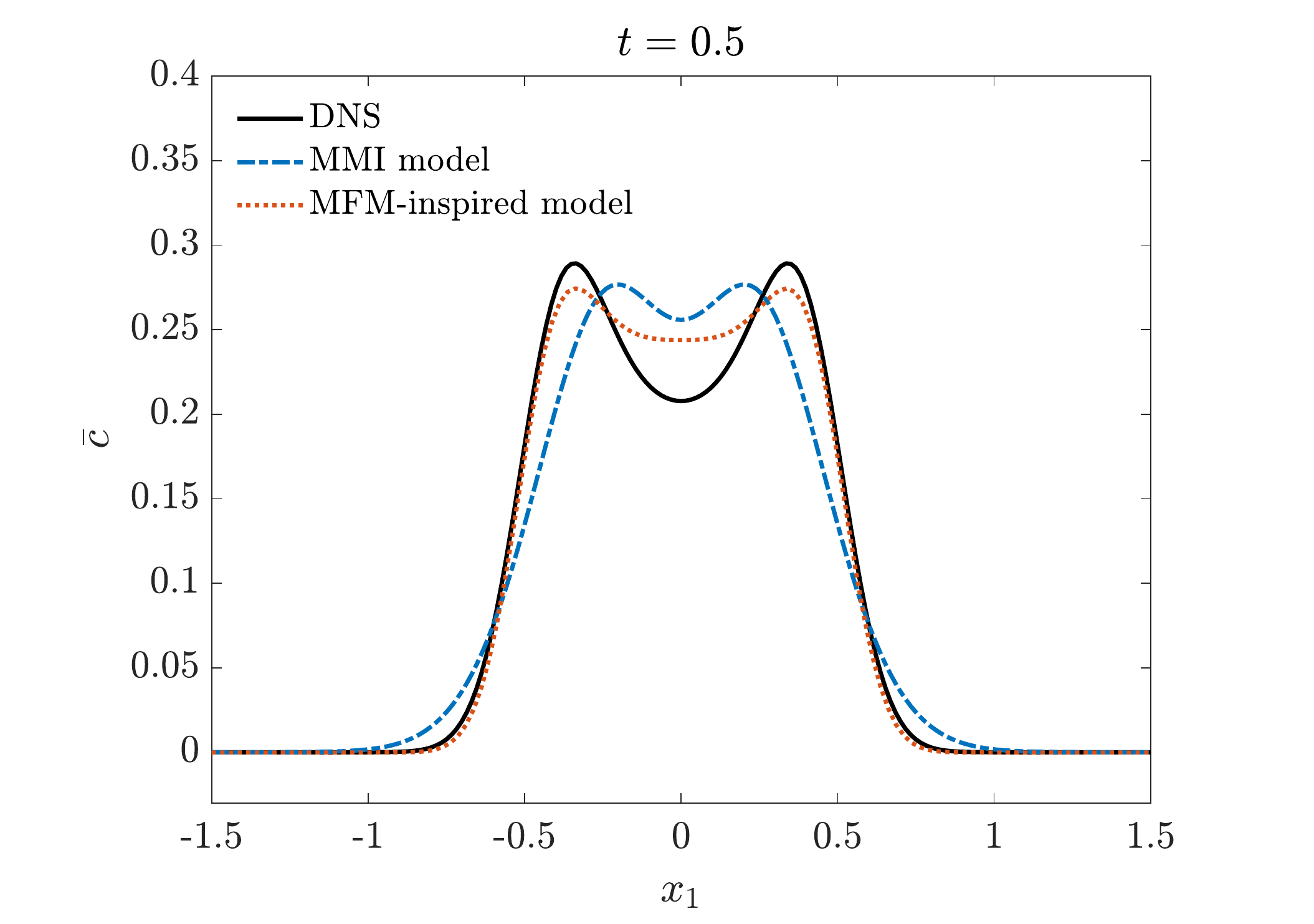}
        \caption{Early time comparison.}
        \label{fig:modelcomparisonMFM_t05}
     \end{subfigure}
     \begin{subfigure}[t]{0.45\textwidth}
         \centering
         \includegraphics[width=\textwidth]{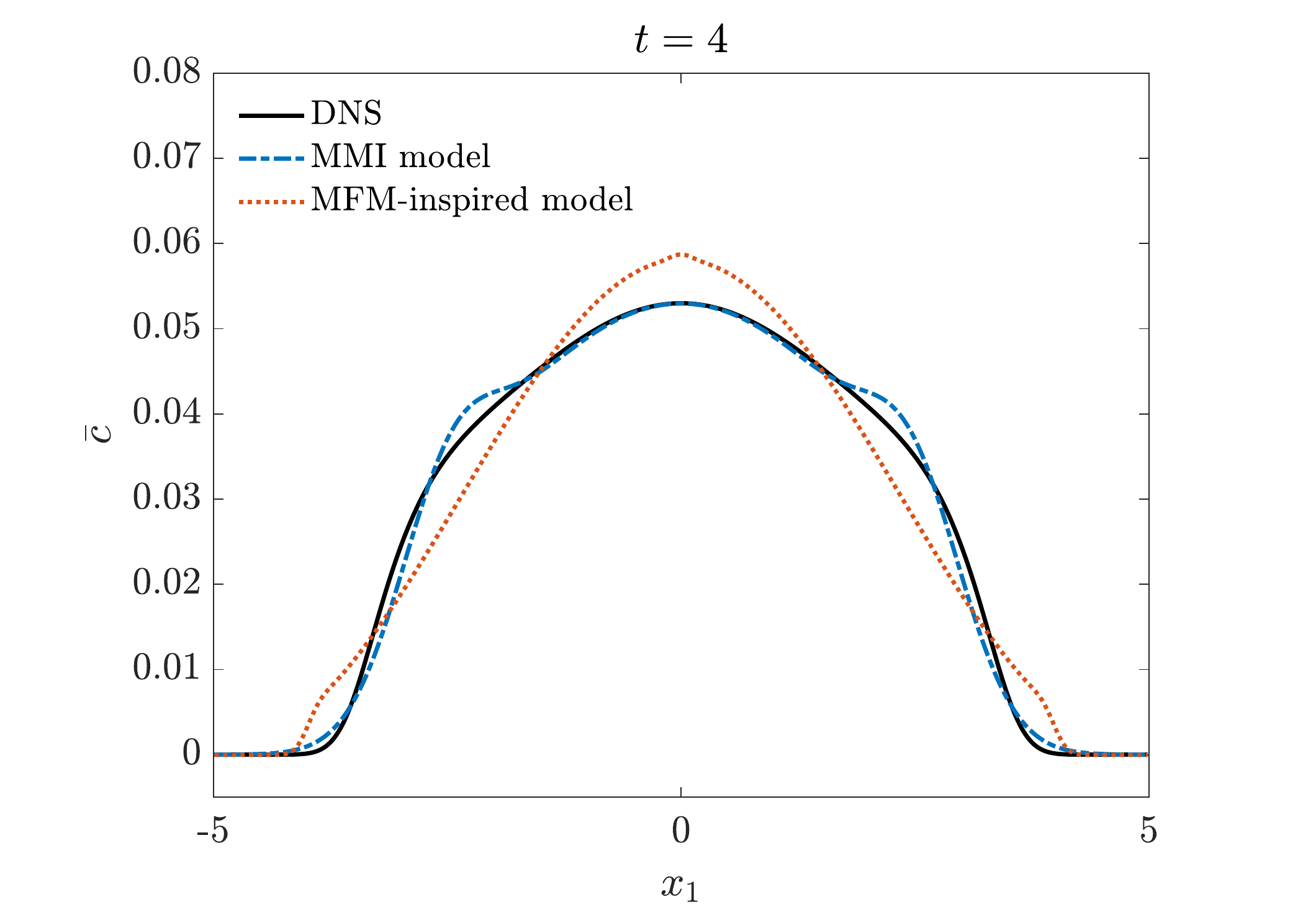}
         \caption{Late time comparison.}
         \label{fig:modelcomparisonMFM_t4}
     \end{subfigure}
     \caption{MMI model and MFM-inspired model comparison at early time, $t = 0.5$, and late time, $t = 4$.}
     \label{fig:modelcomparisonMFM_tslices}
\end{figure}

Figure \ref{fig:modelcomparisonMFM_tslices} shows a solution comparison between the MMI model in Equation (\ref{eq:spatiotemporal MMI model homogeneous}) and the MFM-inspired model at early and late time. At early time, where the solution is dominated by a small-scale feature that quickly disperses, the MFM-inspired model performs better. For late time, where the solution features are smooth and both models are expected to perform well, the MMI model performs better than MFM-inspired model. Appendix \ref{fractional order} shows a comparison with a simple fractional-order closure operator \cite{mehta2019discovering}\cite{song2018universal}\cite{di2021two}, which produces inferior results to both the MMI and MFM-inspired models. This is expected since a simple fractional-order Laplacian cannot capture the limits of large $k$ and $\omega$ or low-order corrections to the leading moment of the eddy diffusivity.
 
Whether a model should match the limits of $k$ and $\omega$ or the low-order moments of the nonlocal eddy diffusivity depends on the problem of interest. If there is a singularity or very sharp feature in the solution, then a model matching the limits of $k$ and $\omega$ may be more appropriate. Otherwise, if the solution is reasonably smooth (as is true for many practical applications), then a model matching the low-order moments is more appropriate.

\section{Example MMI models for inhomogeneous flows} \label{inhomogeneous flows}
 We begin with an inhomogeneous example with periodic boundary conditions, and then discuss wall-bounded flows and the challenges of determining the MMI coefficients in the near-wall region.  

\subsection{Inhomogeneous problem with periodic boundary conditions} \label{inhomogeneous periodic model problem}
\begin{figure}[t]
     \centering
     \begin{subfigure}[t]{0.45\textwidth}
         \centering
         \includegraphics[width=\textwidth]{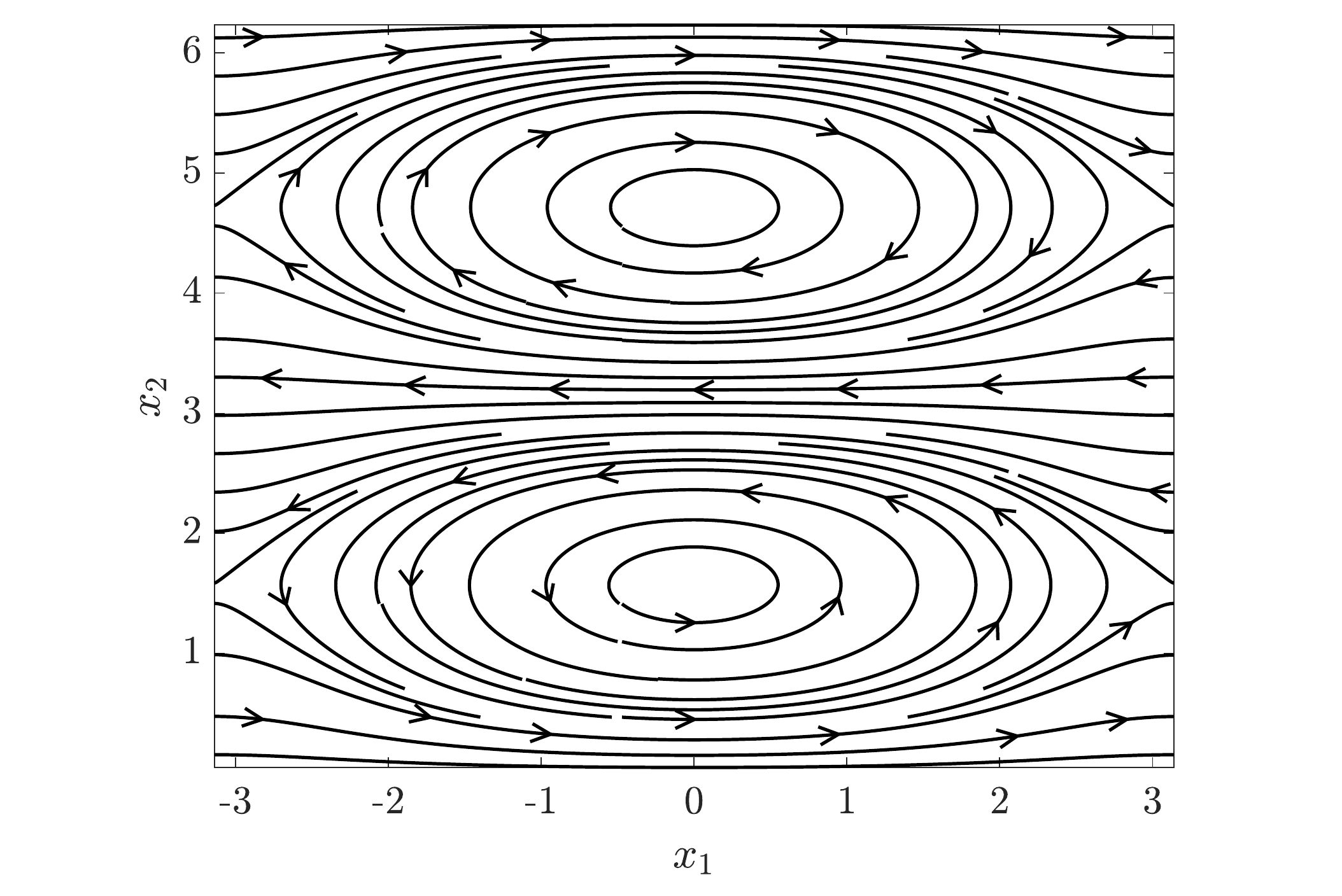}
        \caption{}
        \label{fig:velocity_field_inhomogeneous_periodic}
     \end{subfigure}
     \begin{subfigure}[t]{0.45\textwidth}
         \centering
         \includegraphics[width=\textwidth]{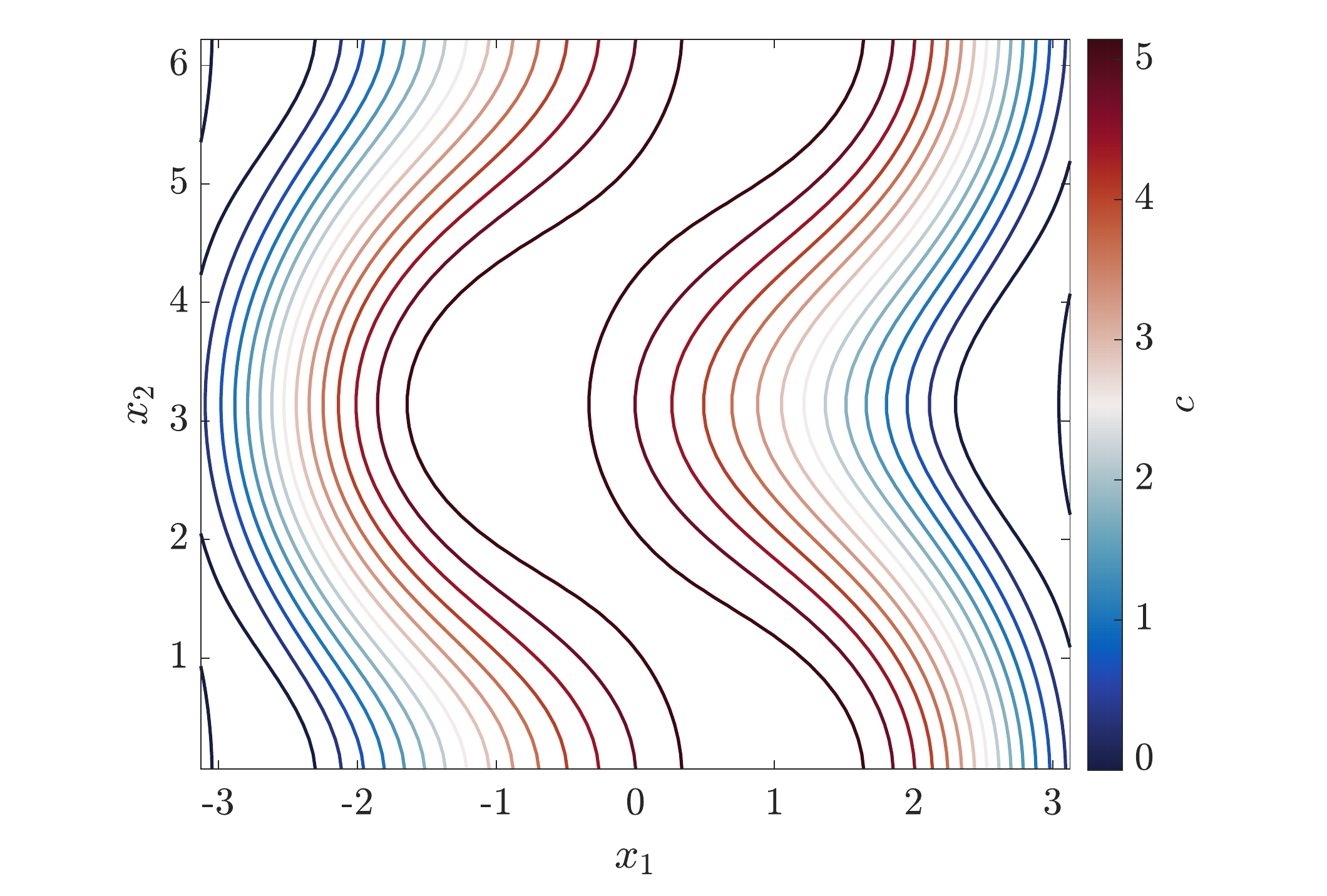}
         \caption{}
         \label{fig:DNS_inhomogeneous_periodic}
     \end{subfigure}
     \caption{(a) Streamlines of the velocity field in Equation (\ref{eq:inhomogeneous periodic velocity}). (b) Contour plot of $c(x_1, x_2)$ from DNS.}
\end{figure}

Consider a two-dimensional domain corresponding to the cross-section of a channel with periodic boundary conditions at the left and right walls at $x_1 = \pm \pi$, and a no flux condition, $\partial c/\partial x_2 = 0$, at the top and bottom walls at $x_2 = 0, 2\pi$. The flow consists of two vortices given by the velocity field:
\begin{equation} \label{eq:inhomogeneous periodic velocity}
u_1 = \frac{1}{2}[2 + \cos(x_1)]\cos(x_2), u_2 = \frac{1}{2}\sin(x_1)\sin(x_2).
\end{equation}
Streamlines of the velocity field are shown in Figure \ref{fig:velocity_field_inhomogeneous_periodic}. The steady, governing equation is
\begin{equation} \label{eq:inhomogeneous governing equation}
    \frac{\partial}{\partial x_1}(u_1 c) + \frac{\partial}{\partial x_2}(u_2 c) = \epsilon^2 \frac{\partial^2 c}{\partial x_1^2} + \frac{\partial^2 c}{\partial x_2^2} + f,
\end{equation}
where $f$ is an external source function. The parameter, $\epsilon^2$, results from directional nondimensionalization as in Section \ref{homogeneous model problem}. For this example problem, we consider $\epsilon^2 = 0.05$ and $f$ to be an oscillatory source function given by $f = \cos(x_1)$.

Figure \ref{fig:DNS_inhomogeneous_periodic} shows contours of $c(x_1, x_2)$ from DNS. The problem is discretized using second-order central difference on a uniform staggered mesh with $N_1 = 200$ grid points in $x_1$ and $N_2 = 50$ grid points in $x_2$. Due to the boundary conditions of the problem, $c(x_1, x_2)$ can arbitrarily be shifted by a constant. Hence, there is an additional constraint such that at the first point in $x_1$, the average of $c(x_1, x_2)$ over $x_2$ is zero. 

As in Section \ref{homogeneous model problem}, averaging is defined in the $x_2$-direction as
\begin{equation} \label{eq:inhomogeneous averaging}
    \bar{c}(x_1) = \frac{1}{L_2}\int_0^{L_2} c(x_1, x_2) dx_2,
\end{equation}
where $L_2 = 2\pi$. The corresponding mean scalar transport equation for this problem is 
\begin{equation} \label{eq:inhomogeneous mean scalar transport}
    \frac{d}{dx_1}\overline{u_1^\prime c^\prime} = \epsilon^2\frac{d^2\bar{c}}{dx_1^2} + \bar{f}.
\end{equation}

The MMI model matching up to the second-order spatial moment of the nonlocal eddy diffusivity is:
\begin{equation}
    \label{eq:spatial MMI model again}
    \left[1 + a_1(x_1)\frac{d}{d x_1} + a_2(x_1)\frac{d^2}{d x_1^2} \right](-\overline{u_1^\prime c'}) = a_0(x_1)\frac{d \bar{c}}{d x_1}, 
\end{equation}
where the procedure for determining the coefficients is described in Section \ref{modeling nonlocal eddy diffusivity}. Figure \ref{fig:MMI_coeff_inhomogeneous_periodic} shows the coefficients for the MMI model, and Figure \ref{fig:kernel_slices_inhomogeneous_periodic} shows cross sections of the exact nonlocal eddy diffusivity obtained using  MFM and the modeled eddy diffusivity. The MMI model closely captures the double-sided exponential shape of the exact eddy diffusivity including the slight asymmetry at some $x_1$-locations. Appendix \ref{MFM decomposition} provides an appropriate IMFM formulation for obtaining moments of the eddy diffusivity for problems in which the periodic boundary conditions are incompatible with the IMFM required $\bar{c}$.
Appendix \ref{MFM decomposition} also provides details for obtaining the exact eddy diffusivity for periodic problems. 

\begin{figure}[t]
    \centering
    \includegraphics[width=0.5\textwidth]{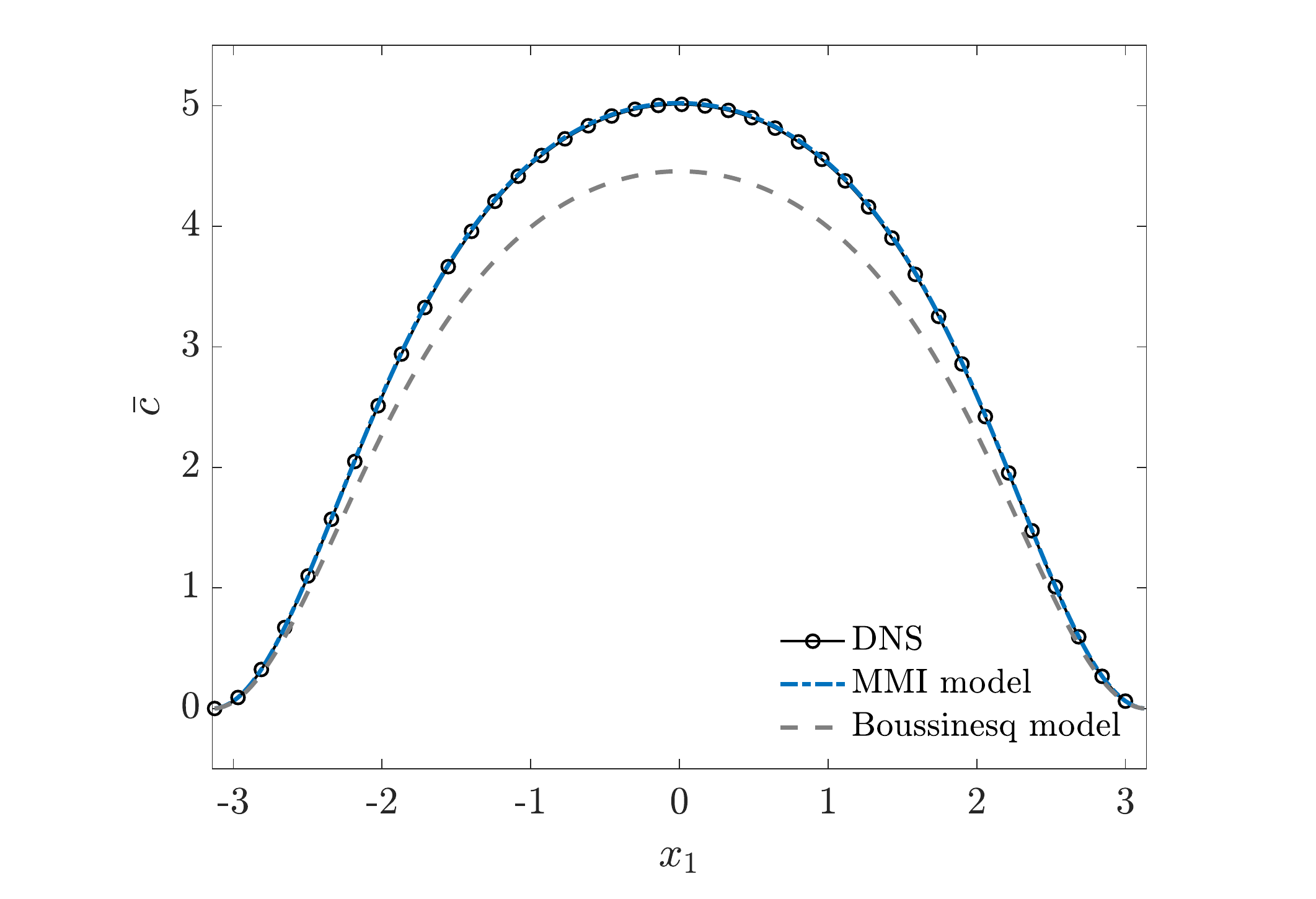}
    \caption{Model comparison for the inhomogeneous problem with periodic boundary conditions. The MMI model is almost indistinguishable from the DNS solution.}
    \label{fig:cbar_inhomogeneous_periodic} 
\end{figure}

Figure \ref{fig:cbar_inhomogeneous_periodic} shows a comparison between the MMI model and DNS solution for the inhomogeneous problem with periodic boundary conditions. The DNS solution corresponds to averaging the full solution in Figure \ref{fig:DNS_inhomogeneous_periodic} over the $x_2$-direction. The Boussinesq model given by
\begin{equation} \label{eq:Boussinesq model inhomogeneous}
    -\overline{u_1'c'} = D^0(x) \frac{\partial \bar{c}}{\partial x_1}
\end{equation}
is also shown for comparison. The Boussinesq model greatly underpredicts the solution while the MMI model solution is almost indistinguishable from the DNS solution.

\subsection{Wall-bounded inhomogeneous flows}
\label{inhomogeneous wall-bounded model problem}
\begin{figure}[t]
     \centering
     \begin{subfigure}[t]{0.45\textwidth}
         \centering
         \includegraphics[width=\textwidth]{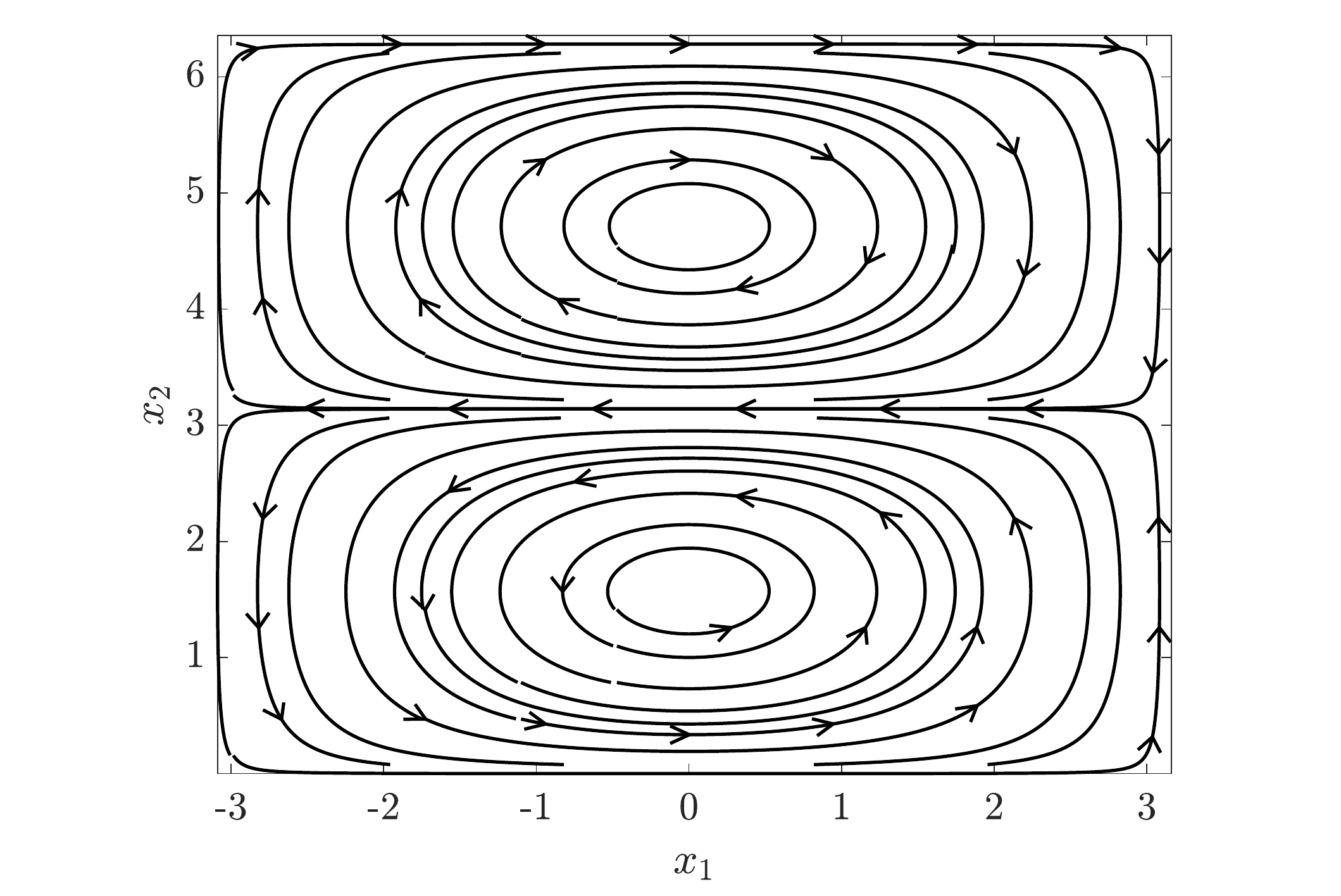}
        \caption{}
        \label{fig:velocity_field_inhomogeneous_wall}
     \end{subfigure}
     \begin{subfigure}[t]{0.45\textwidth}
         \centering
         \includegraphics[width=\textwidth]{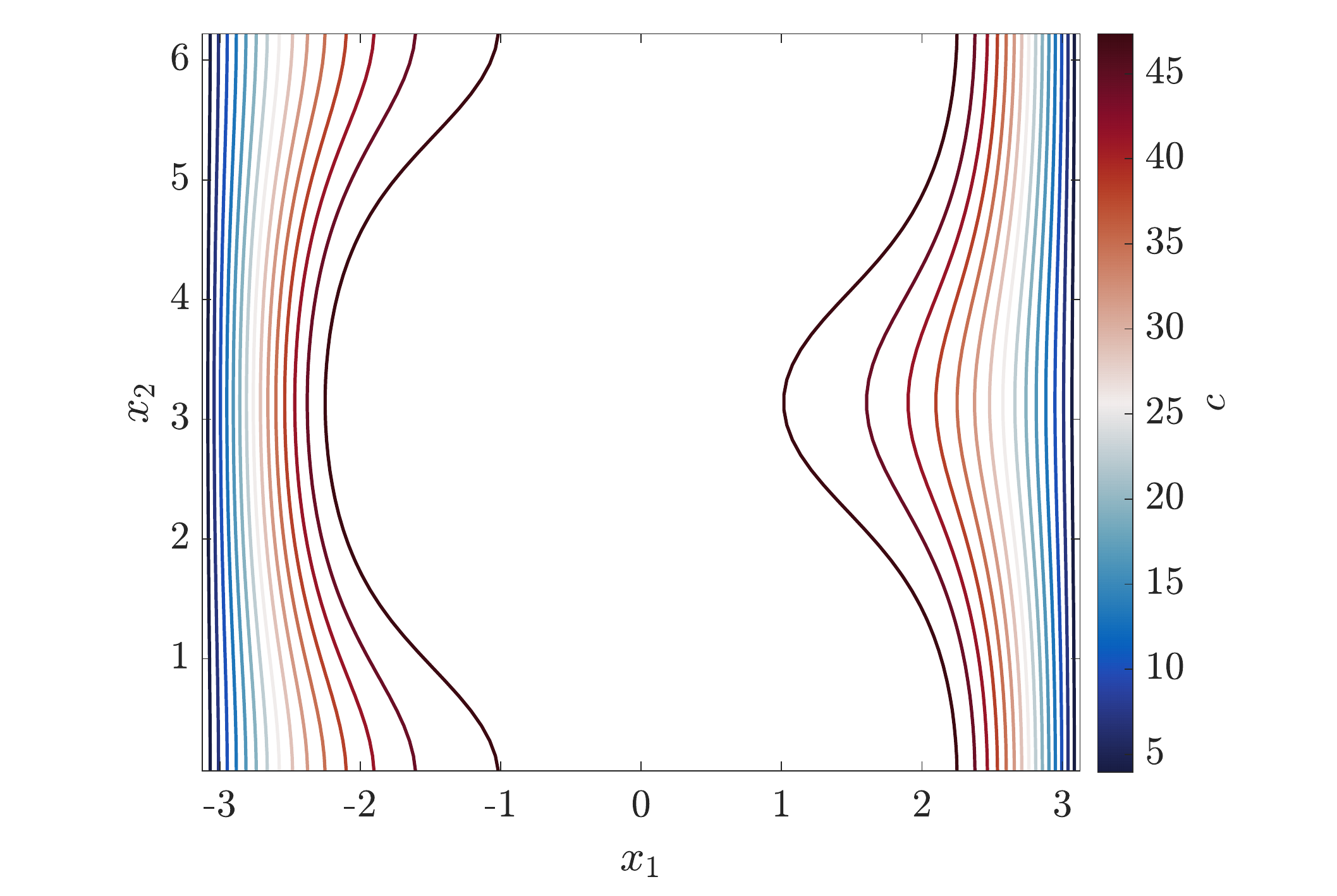}
         \caption{}
         \label{fig:DNS_inhomogeneous_wall}
     \end{subfigure}
     \caption{(a) Streamlines of the velocity field in Equation (\ref{eq:inhomogeneous wall velocity}). (b) Contour plot of $c(x_1, x_2)$ from DNS.}
\end{figure}

As an example application of the MMI model to wall-bounded inhomogeneous flows, we consider the model problem of Mani and Park \cite{mani2021macroscopic} which uses the same two-dimensional channel geometry as in Section \ref{inhomogeneous periodic model problem}, but replaces the periodic boundary conditions with solid walls and Dirichlet boundary conditions $c(x_1 = \pm \pi)= 0$. To satisfy the no-slip and no-penetration conditions at the solid wall, the velocity field is modified to be:
\begin{equation} \label{eq:inhomogeneous wall velocity}
    u_1 = [1+\cos(x_1)]\cos(x_2), u_2 = \sin(x_1)\sin(x_2).
\end{equation}
Streamlines of the velocity field are shown in Figure \ref{fig:velocity_field_inhomogeneous_wall}. The governing equation is given by Equation (\ref{eq:inhomogeneous governing equation}) with $\epsilon^2 = 0.05$ as before. The source function, $f$, is specified to be a constant, $f = 1$. Contours of $c(x_1, x_2)$ from DNS with grid resolution $N_1 = 200$ and $N_2 = 50$ are shown in Figure \ref{fig:DNS_inhomogeneous_wall}. Averaging is defined over $x_2$ by Equation (\ref{eq:inhomogeneous averaging}), and the corresponding mean scalar equation is given by Equation (\ref{eq:inhomogeneous mean scalar transport}) as before. 

\begin{figure}[t]
     \centering
     \begin{subfigure}[t]{0.45\textwidth}
         \centering
         \includegraphics[width=\textwidth]{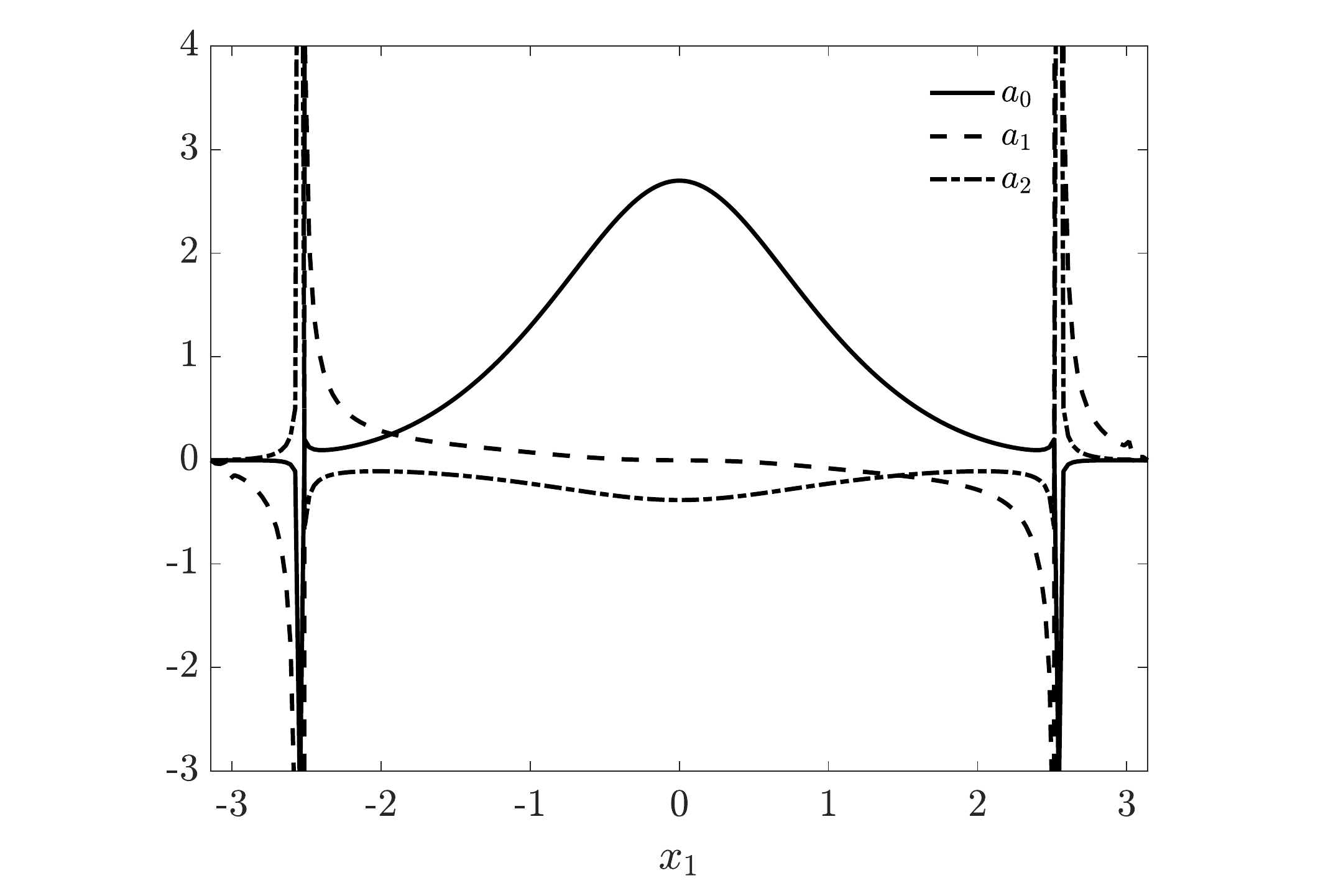}
        \caption{}
        \label{fig:MMI_coeff_inhomogeneous_wall}
     \end{subfigure}
     \begin{subfigure}[t]{0.45\textwidth}
         \centering
         \includegraphics[width=\textwidth]{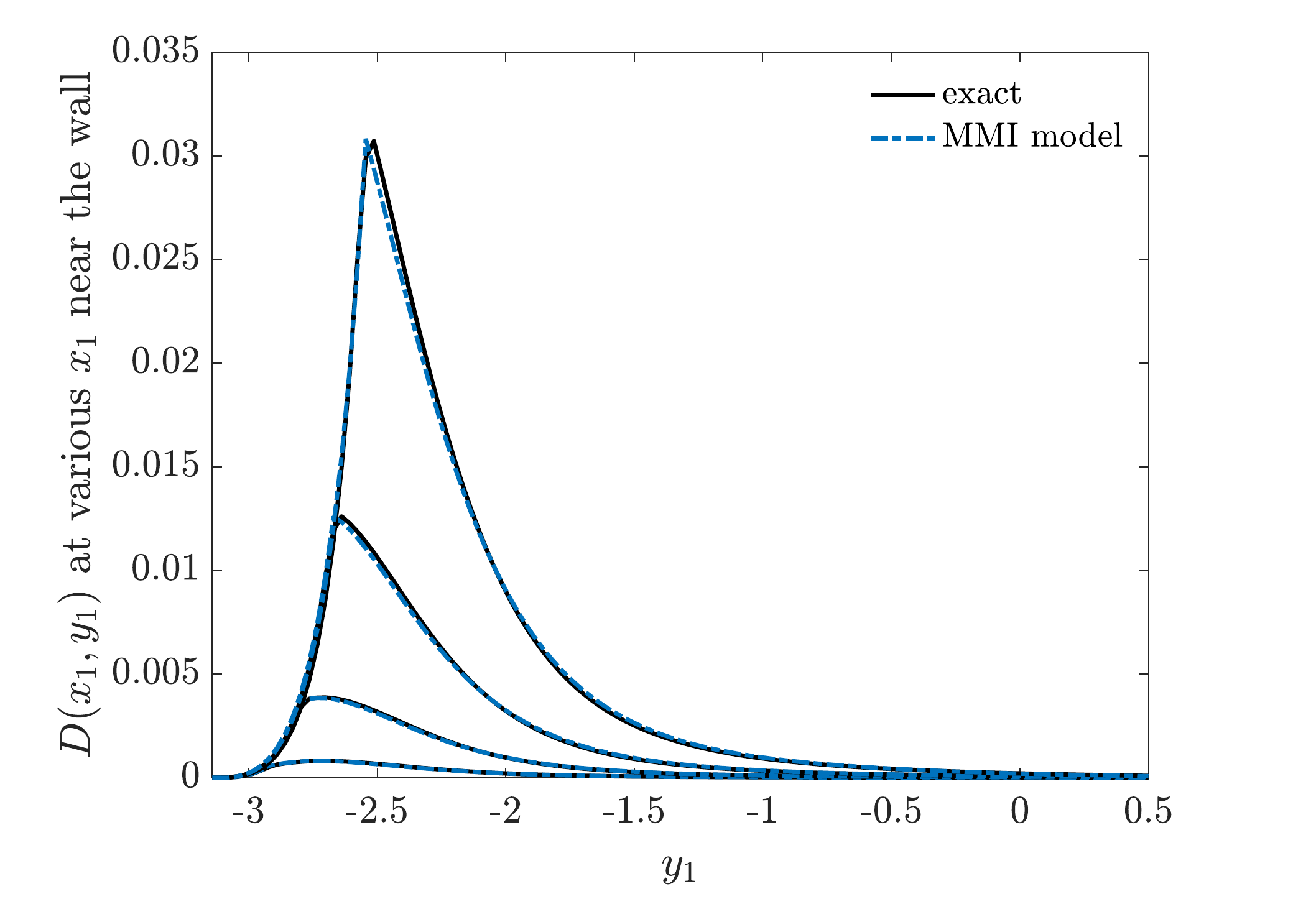}
         \caption{}
         \label{fig:kernel_slices_inhomogeneous_wall}
     \end{subfigure}
     \caption{(a) MMI coefficients for Equation (\ref{eq:spatial MMI model again}) for the wall-bounded inhomogeneous problem. (b) The exact and modeled nonlocal eddy diffusivity shown for various $x_1$ approaching the wall ($x_1 = -2.922, -2.796, -2.670, -2.545$).}
\end{figure}

\begin{figure}[t]
    \centering
    \includegraphics[width=0.5\textwidth]{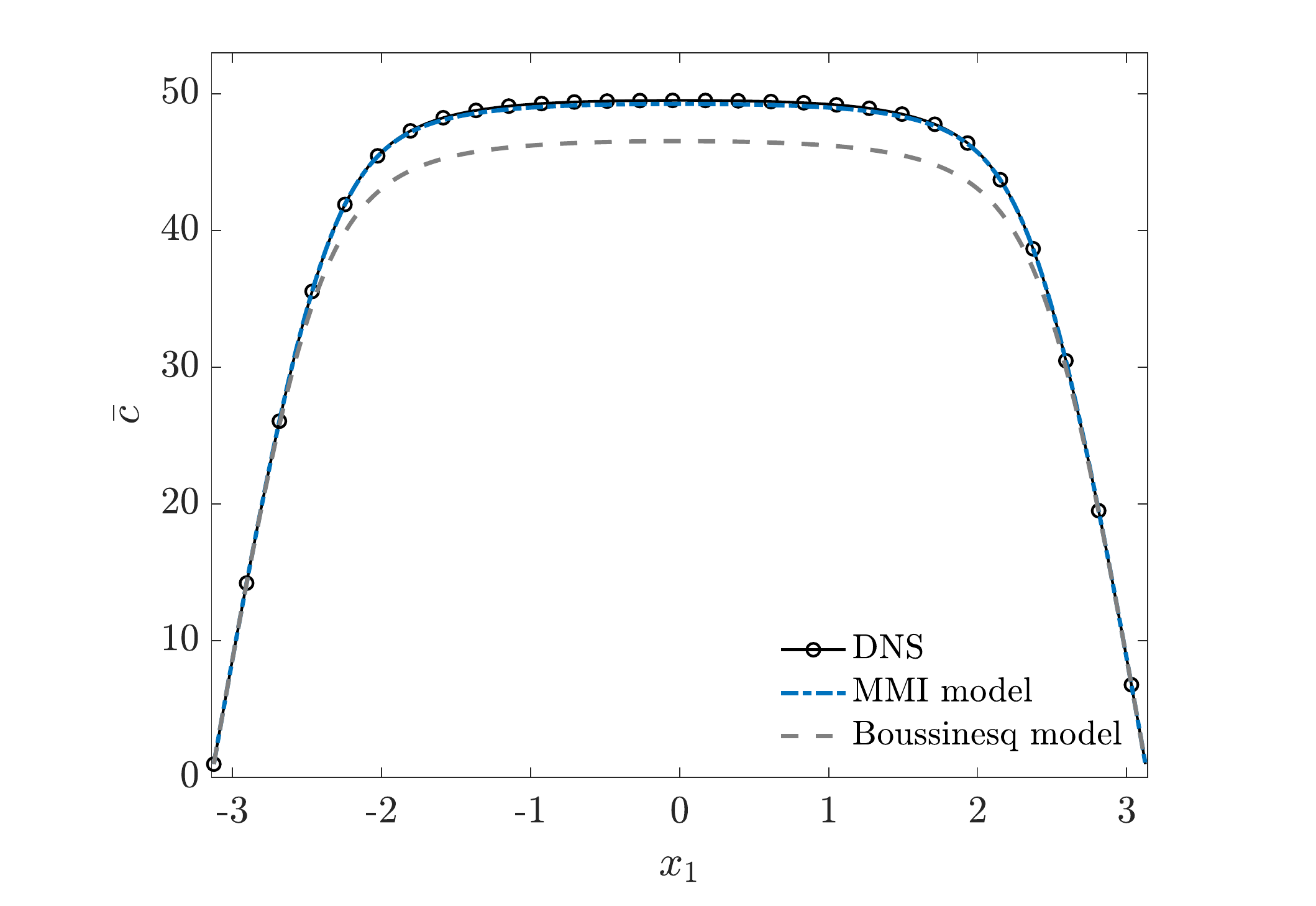}
    \caption{Model comparison for the wall-bounded inhomogeneous problem. The MMI model is almost indistinguishable from the DNS solution.}
    \label{fig:cbar_inhomogeneous_wall} 
\end{figure}

The steady MMI model is given in Equation (\ref{eq:spatial MMI model again}), and the coefficients are found via the procedure described in Section \ref{modeling nonlocal eddy diffusivity}. Figure \ref{fig:MMI_coeff_inhomogeneous_wall} shows the MMI coefficients for the wall-bounded inhomogeneous flow. The coefficients are well-behaved in the center of the domain; however, near the wall, there is a sharp spike in the MMI coefficients at $|x_1| \approx 2.5$. The location of the singularity remains fixed under mesh refinement, indicating that the issue is not due to the numerics. Despite the ill-behaved coefficients,
the resulting $\bar{c}(x_1)$ from the MMI model is still very accurate, as shown in Figure \ref{fig:cbar_inhomogeneous_wall}, and greatly outperforms the Boussinesq model given in Equation (\ref{eq:Boussinesq model inhomogeneous}).

To gain an understanding of why this singularity in the coefficients occurs, Figure \ref{fig:kernel_slices_inhomogeneous_wall} shows the exact eddy diffusivity obtained using MFM at various $x_1$-locations approaching the wall. While the modeled eddy diffusivity closely follows the exact eddy diffusivity, both eddy diffusivities deviate from a double-sided exponential shape, gradually becoming smoother and smaller in magnitude near the wall as the flow also diminishes.

The transition in kernel behavior due to the presence of the wall causes the singularity seen in the MMI coefficients. The ill-behaved coefficients are a result of the specific MMI model form in Equation (\ref{eq:spatial MMI model again}). In the procedure for determining the coefficients, the determinant of the pointwise matrix formed by the linear system in Equations (\ref{eq:alpha0 eqn})-(\ref{eq:alpha2 eqn}) crosses zero at the location of the singularity.

However, near the wall, viscous effects dominate. For example, the molecular diffusivity, $\epsilon^2$, is larger than the leading-order eddy diffusivity, $D^0$, until $|x_1| \approx 2.35$, which includes the region of singularity at $|x_1| \approx 2.5$. The MMI model form is unimportant near the wall, and the resulting $\bar{c}(x_1)$ is still very accurate as shown in Figure \ref{fig:cbar_inhomogeneous_wall}.

\subsubsection*{Coefficient regularization}
We provide a coefficient regularization technique to remedy the singularity in the MMI coefficients in Figure \ref{fig:MMI_coeff_inhomogeneous_wall}. Because the molecular term dominates near the wall, a portion of the molecular diffusion flux with small parameter $\sigma$ is added when determining the MMI coefficients: 
\begin{equation} \label{eq:inhomogeneous MMI regularization fitting}
    \left[1 + a_1(x_1)\frac{d}{d x_1} + a_2(x_1)\frac{d^2}{d x_1^2} \right]\left(-\overline{u_1^\prime c'} + \sigma \frac{d \bar{c}}{d x_1}\right) = a_0(x_1)\frac{d \bar{c}}{d x_1},
\end{equation}
where $\sigma$ is a constant smaller than the molecular diffusivity, $\epsilon^2$. With this choice, away from the wall, the added regularization is negligible. Near the wall, the added regularization dominates and prevents the singularity in the coefficients by preventing the effective kernel shape from diminishing. In other words, as $-\overline{u_1^\prime c^\prime}$ goes to zero near the wall, the additional $\sigma d \bar{c}/d x_1$ term adds a Dirac delta function to the kernel that keeps it from diminishing. 

Once the coefficients are determined using Equation (\ref{eq:inhomogeneous MMI regularization fitting}), the closure model for the scalar flux is:
\begin{equation} \label{eq:inhomogeneous MMI regularization intermediate}
    -\overline{u_1^\prime c'} = \left[1 + a_1(x_1)\frac{d}{d x_1} + a_2(x_1)\frac{d^2}{d x_1^2} \right]^{-1} a_0(x_1)\frac{d \bar{c}}{d x_1} - \sigma \frac{d\bar{c}}{dx_1},
\end{equation}
where the portion of the molecular diffusion flux that was added for regularization when determining the coefficients is now subtracted. 

However, for model implementation we found the following closure form to be more robust: 
\begin{equation} \label{eq:inhomogeneous MMI regularization model}
    \left[1 + a_1(x_1)\frac{d}{d x_1} + a_2(x_1)\frac{d^2}{d x_1^2} \right](-\overline{u_1^\prime c'}) = (a_0(x_1) - \sigma) \frac{d \bar{c}}{d x_1},
\end{equation}
where the coefficients are still determined using Equation (\ref{eq:inhomogeneous MMI regularization fitting}).
With this model implementation, even choices of $\sigma$ larger than $\epsilon^2$ (up to $\sigma \approx 3 
\epsilon^2$) still produced very accurate solutions with up to around one percent maximum error. In the procedure for determining  the coefficients, due to the presence of the MMI operator acting on the regularization term in Equation (\ref{eq:inhomogeneous MMI regularization fitting}), the added regularization is not purely local. The model form in Equation (\ref{eq:inhomogeneous MMI regularization model}) mimics an added nonlocal effect of the regularization term. Equation (\ref{eq:inhomogeneous MMI regularization model}) also still ensures that the implemented MMI model matches the zeroth moment, i.e. case when $\bar{c} = x_1$. The presence of $\sigma$ will affect the higher-order moments of the eddy diffusivity; for example, substituting $\bar{c} = x_1^2/2$ into Equation (\ref{eq:inhomogeneous MMI regularization model}) results in an extra $\sigma(x_1 + a_1)$ when compared with Equation (\ref{eq:alpha1 eqn}). However, as long as $\sigma$ is small, this error in the higher-order moments is also small. 

\begin{figure}[t]
     \centering
     \begin{subfigure}[t]{0.45\textwidth}
         \centering
         \includegraphics[width=\textwidth]{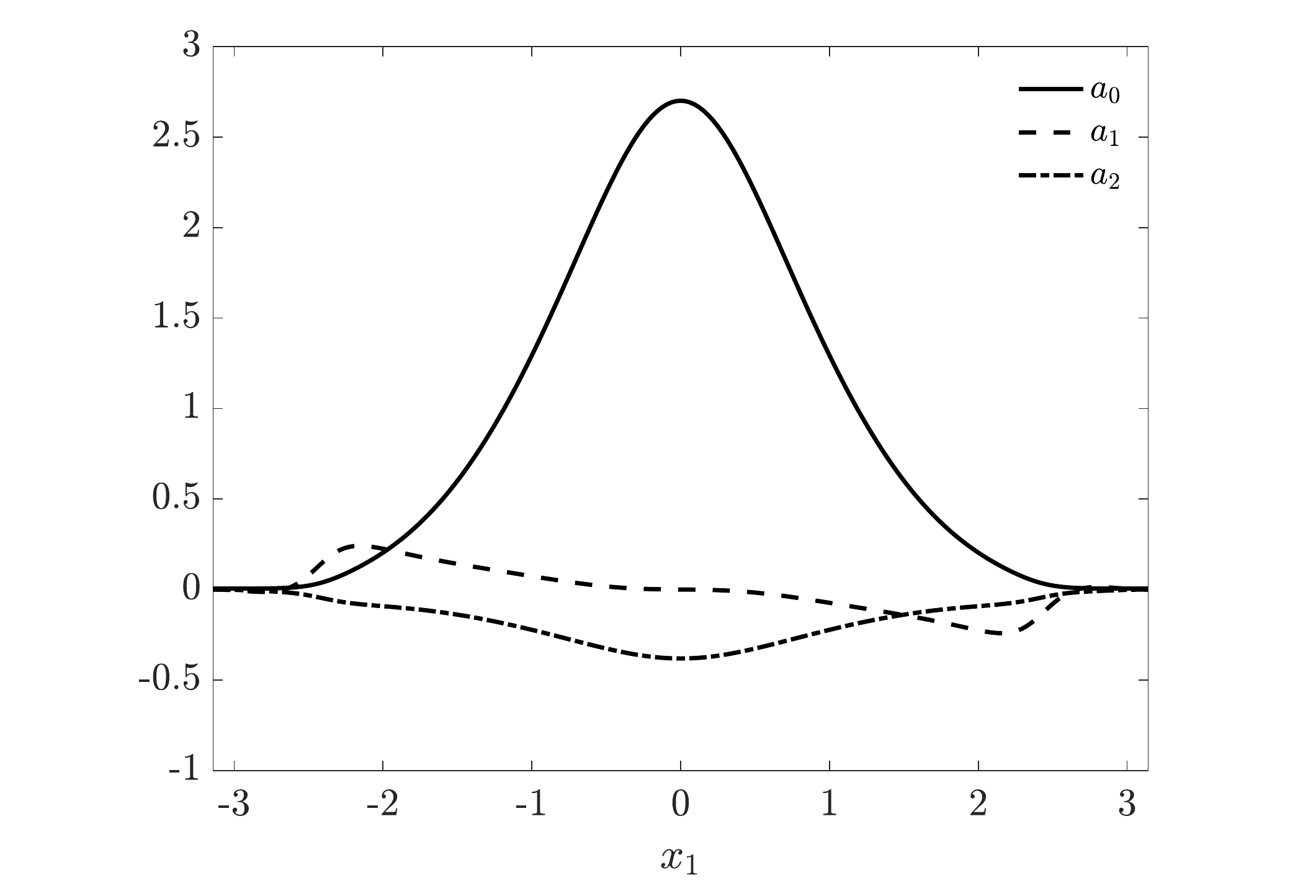}
        \caption{}
        \label{fig:MMI_regularization_coeff}
     \end{subfigure}
     \begin{subfigure}[t]{0.45\textwidth}
         \centering
         \includegraphics[width=\textwidth]{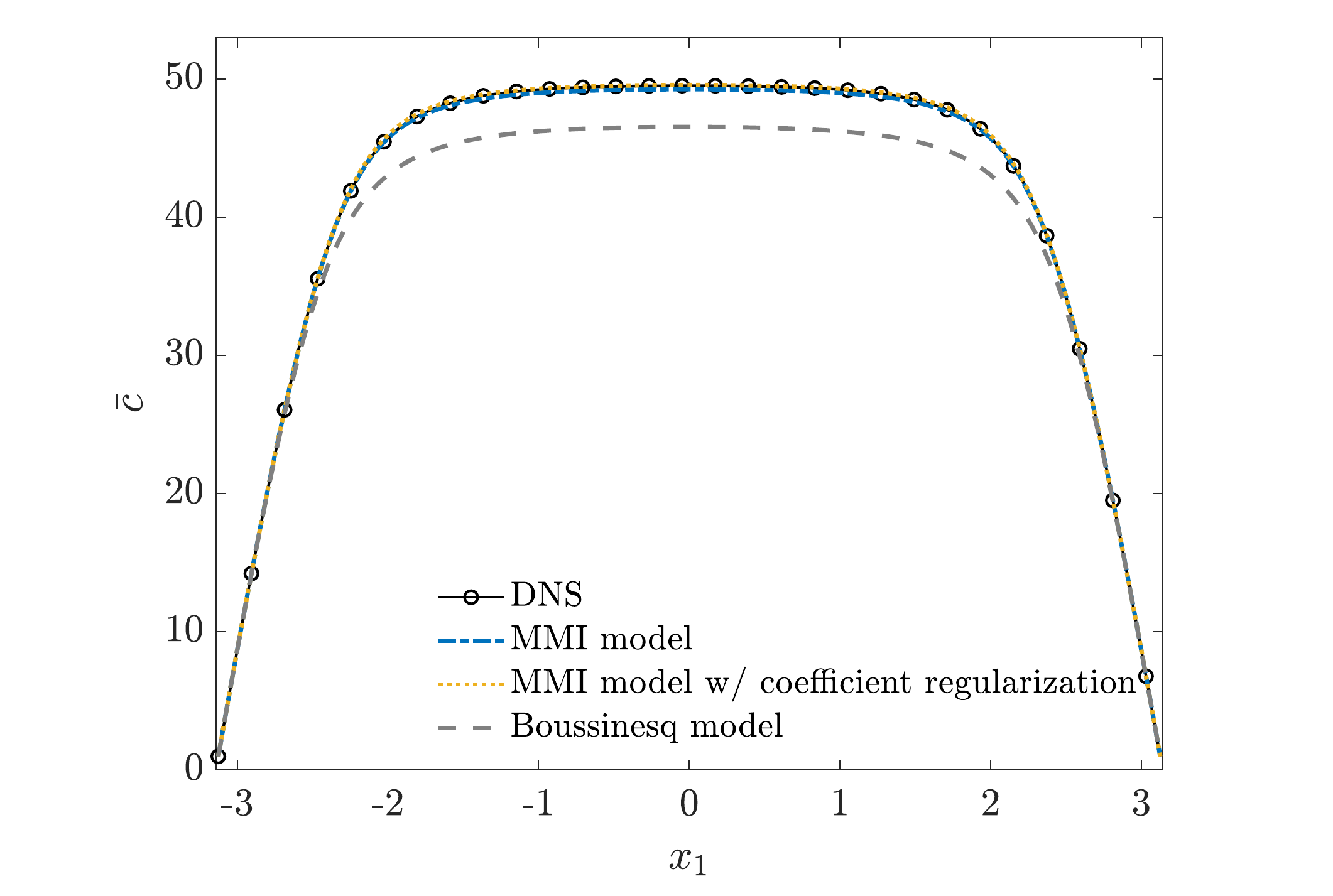}
         \caption{}
         \label{fig:MMI_regularization_cbar}
     \end{subfigure}
     \caption{(a) MMI coefficients using the regularization technique in Equation (\ref{eq:inhomogeneous MMI regularization fitting}) for the wall-bounded inhomogeneous problem. (b) Model comparison for the wall-bounded inhomogeneous problem.}
\end{figure}

Figure \ref{fig:MMI_regularization_coeff} shows the MMI coefficients for the wall-bounded inhomogeneous problem with coefficient regularization. The coefficient, $\sigma$, is chosen to be $0.1 \epsilon^2$ where $\epsilon^2$ is the nondimensionalized molecular diffusivity in the $x_1$-direction. With regularization, the coefficients are now well-behaved; however, this is not yet systematic with regards to choice of $\sigma$.

Figure \ref{fig:MMI_regularization_cbar} shows a comparison between the MMI model with coefficient regularization in Equation (\ref{eq:inhomogeneous MMI regularization model}) and the DNS solution. The MMI model with coefficient regularization performs even slightly better the original MMI model in Equation (\ref{eq:spatial MMI model again}) although this depends on the choice of $\sigma$. While the coefficient regularization technique performs well for this wall-bounded inhomogeneous problem, this does not address the original issue with the potential model form error. The MMI model form in Equation (\ref{eq:spatial MMI model again}) admits a variety of exponential kernel shapes, but there are some shapes it cannot capture, and the model form may need to be modified. Alternative formulations for the MMI model are shown in Appendix \ref{other MMI formulas}, but this is still an area of ongoing investigation. 

\section{Conclusion}
A nonlocal eddy diffusivity can greatly improve modeling of mean scalar transport when the Boussinesq approximation is invalid. However, both obtaining and implementing full nonlocal eddy diffusivities are too expensive for most practical applications, and thus we introduce matched moment inverse (MMI) operators to model the nonlocal eddy diffusivity. These operators offer leading-order corrections to capture nonlocal effects in the eddy diffusivity by systematically matching the moments of the true nonlocal eddy diffusivity, while avoiding the known divergence issues of explicit models that truncate the Kramers-Moyal expansion. The resulting model is in the form of a differential equation rather than a computationally- or memory- intensive nonlocal integral. Moreover, these models only require information about the moments of the nonlocal eddy diffusivity, which can be obtained using one simulation per moment, rather than the full nonlocal eddy diffusivity, which would require as many simulations as degrees of freedom in the averaged space. Furthermore, we show that MMI models are expected to perform well as long as the mean scalar field is reasonably smooth; for fields with sharp features, one may want to consider other operators that match the large and small wavenumber limits instead. 

In this work, we demonstrate the application of MMI models to homogeneous and inhomogeneous example problems. We discuss challenges in inhomogeneous wall-bounded flows where the scalar flux goes to zero near the wall, and show a coefficient regularization technique although this is not yet systematic. 

\section{Acknowledgements}
This work was supported by the Boeing Company and the Office of Naval Research under grant N00013-20-1-2718. J. L. was additionally supported by the Burt and Deedee McMurtry Stanford Graduate Fellowship. 

\appendix
\section{Comparison of methods for obtaining the nonlocal eddy diffusivity} \label{Hamba comparison}
 This appendix is motivated by the observation of two apparently different derivations for the exact expression of the nonlocal eddy diffusivity. Using a Green's function solution, Hamba \cite{hamba1995analysis}\cite{hamba2004nonlocal} derived an exact expression for the nonlocal eddy diffusivity, while Mani and Park \cite{mani2021macroscopic} presented a derivation based on macroscopic forcing of a linear system. In this appendix, we demonstrate that when IMFM is used to obtain the full nonlocal eddy diffusivity (as opposed to using IMFM to obtain moments as used extensively in this work), its formulation is compatible with the earlier work by Hamba. In Hamba's approach, the transport equation for the scalar fluctuation, $c'$, is given by
\begin{equation}
    \label{eq:scalar fluctuation eqn Hamba}
    \frac{\partial c'}{\partial t} + \frac{\partial}{\partial x_j}(u_jc' - \overline{u_j'c'}) - D_M \frac{\partial^2 c'}{\partial x_j\partial x_j} = -u_j'\frac{\partial \bar{c}}{\partial x_j}.
\end{equation}
The mean scalar gradient on the right-hand-side is considered a source term for $c'$. Using the Green's function, $g_i^\prime(\mathbf{x},\mathbf{y},t,\tau)$, satisfying
\begin{equation} 
    \label{eq:Green's function Hamba}
    \frac{\partial g_i^\prime}{\partial t} + \frac{\partial}{\partial x_j}(u_j g_i^\prime - \overline{u_j'g_i^\prime}) - D_M \frac{\partial^2 g_i^\prime}{\partial x_j\partial x_j} = u_i'\delta(\mathbf{x}-\mathbf{y})\delta(t-\tau),
\end{equation}
the solution to Equation (\ref{eq:scalar fluctuation eqn Hamba}) is
\begin{equation}
    c'(\mathbf{x},t) = -\int_{\mathbf{y},\tau} g_i^\prime(\mathbf{x}, \mathbf{y},t,\tau)\frac{\partial\bar{c}}{\partial x_i}\big|_{\mathbf{y},\tau}d\mathbf{y}d\tau.
\end{equation}
Hence, the scalar flux can be written as 
\begin{equation}
    \label{eq:scalar flux Hamba}
    -\overline{u_j'c'}(\mathbf{x},t) = \int_{\mathbf{y},\tau} \overline{u_j'(\mathbf{x},t)g_i^\prime(\mathbf{x}, \mathbf{y},t,\tau)}\frac{\partial\bar{c}}{\partial x_i}\big|_{\mathbf{y},\tau}d\mathbf{y}d\tau,
\end{equation}
where 
\begin{equation}
    D_{ji}(\mathbf{x},\mathbf{y},t,\tau) = \overline{u_j'(\mathbf{x},t)g_i^\prime(\mathbf{x},\mathbf{y},t,\tau)}
\end{equation}
is the nonlocal eddy diffusivity, and Equation (\ref{eq:scalar flux Hamba}) is identical to Equation (\ref{eq:general eddy diffusivity}).

To compare the approach presented by Mani and Park \cite{mani2021macroscopic}, which arrived at the same nonlocal eddy diffusivity, consider the forced scalar transport equation:
\begin{equation}
    \label{eq:forced scalar transport}
    \frac{\partial c}{\partial t} + \frac{\partial}{\partial x_j}(u_jc) - D_M \frac{\partial^2 c}{\partial x_j \partial x_j} = s.
\end{equation}
To obtain the full nonlocal eddy diffusivity, one can specify the macroscopic forcing, $s = \bar{s}$, to maintain the mean scalar gradient as a Dirac delta function, and then postprocess $-\overline{u_j^\prime c
^\prime}$. To allow further comparison with Hamba's approach, the forced mean scalar transport equation is
\begin{equation}
    \label{eq:forced mean scalar transport}
    \frac{\partial \bar{c}}{\partial t} + \frac{\partial}{\partial x_j}(\bar{u}_j\bar{c}) + \frac{\partial}{\partial x_j}(\overline{u_j'c'}) - D_M \frac{\partial^2 \bar{c}}{\partial x_j\partial x_j} = \bar{s}.
\end{equation}
Substituting $\bar{s} = s$ from Equation (\ref{eq:forced mean scalar transport}) into Equation (\ref{eq:forced scalar transport}) leads to the scalar fluctuation equation:
\begin{equation}
    \frac{\partial c^\prime}{\partial t} + \frac{\partial}{\partial x_j}(u_j c^\prime) + u_j^\prime \frac{\partial \bar{c}}{\partial x_j} - \frac{\partial}{\partial x_j}(\overline{u_j^\prime c^\prime}) - D_M \frac{\partial^2 c^\prime}{\partial x_j \partial x_j} = 0,
\end{equation}
which is identical to Hamba's equation in (\ref{eq:scalar fluctuation eqn Hamba}). 

Additionally, to compare the MFM forcing with the implied forcing in Hamba's approach, we rearrange (\ref{eq:forced mean scalar transport}): 
\begin{equation}
    -\frac{\partial}{\partial x_j}(\overline{u_j'c'}) = \frac{\partial \bar{c}}{\partial t} + \frac{\partial}{\partial x_j}(\bar{u}_j\bar{c}) - D_M\frac{\partial^2\bar{c}}{\partial x_j\partial x_j} - \bar{s},
\end{equation}
and observe that the corresponding term in the Green's function equation (Equation (\ref{eq:Green's function Hamba})), $-\partial/\partial x_j(\overline{u_j'g_i'})$, is an implied forcing term that contains the macroscopic forcing. An analogous comparison between Hamba \cite{hamba2005nonlocal} and Mani and Park \cite{mani2021macroscopic} can also be made for momentum transport and Reynolds stress closures.

\section{Obtaining the spatiotemporal eddy diffusivity for the homogeneous model problem}\label{spatiotemporal kernel} 
The spatiotemporal eddy diffusivity shown in Figure \ref{fig:full_kernel_homogeneous}  may be obtained using the Green's-function-based approach of Hamba \cite{hamba2004nonlocal} or similarly, IMFM as detailed here. For the homogeneous model problem in Section \ref{homogeneous model problem}, the unclosed scalar flux can be written as
\begin{equation}
-\overline{u_1^\prime c^\prime}(x_1, t) = \int_0^t \int_{-\infty}^{\infty} D(y_1-x_1, \tau-t)\frac{\partial \bar{c}}{\partial x_1}\big|_{y_1, \tau} dy_1 d\tau,
\end{equation}
where $\bar{()}$ denotes averaging in the $x_2$-direction, and $D(y_1-x_1, \tau-t)$ is the spatiotemporal eddy diffusivity. Using IMFM and prescribing $\partial \bar{c}/\partial x_1$ as a Dirac delta function in both space and time:
\begin{equation}
-\overline{u_1^\prime c^\prime}(x_1, t) = \int_0^t \int_{-\infty}^{\infty} D(y_1-x_1, \tau-t) \delta(y_1, \tau) dy_1 d\tau,
\end{equation}
then by the sifting property of the delta function:
\begin{equation}
    -\overline{u_1^\prime c^\prime}(x_1, t) = D(-x_1, -t).
\end{equation}
The spatiotemporal eddy diffusivity can be obtained by postprocessing $-\overline{u_1^\prime c^\prime}(x_1,t)$ since the flow is homogeneous.

\section{Comparison with a fractional-order operator} \label{fractional order}

Several recent works \cite{mehta2019discovering}\cite{song2018universal}\cite{di2021two} examine using fractional-order operators for nonlocal models. A simple model with a fractional-order Laplacian for the homogeneous problem in Section \ref{homogeneous model problem} is:
\begin{equation}\label{eq:fractional model}
    \frac{\partial \bar{c}}{\partial t} = \frac{1}{2} \left(\frac{\partial^2}{\partial x_1^2}\right)^{\alpha/2}\bar{c},
\end{equation}
where $0<\alpha<2$, and the coefficient in front of the fractional-order Laplacian is chosen such that when $\alpha = 2$, the model reduces to the leading-order Taylor model (Boussinesq model) in Equation (\ref{eq:leading order Taylor}). Equation (\ref{eq:fractional model}) can be solved by Fourier transforming in $x_1$:
\begin{equation}\label{eq:fractional model FT}
   \frac{\partial \bar{\hat{c}}}{\partial t} = \frac{1}{2}\left(-\left(k^2\right)^{\alpha/2}\right)\bar{\hat{c}},
\end{equation}
where $k$ is the corresponding wavenumber in $x_1$, and time-advancing in Fourier space.  

\begin{figure}[t]
    \centering
    \includegraphics[width=0.45\textwidth]{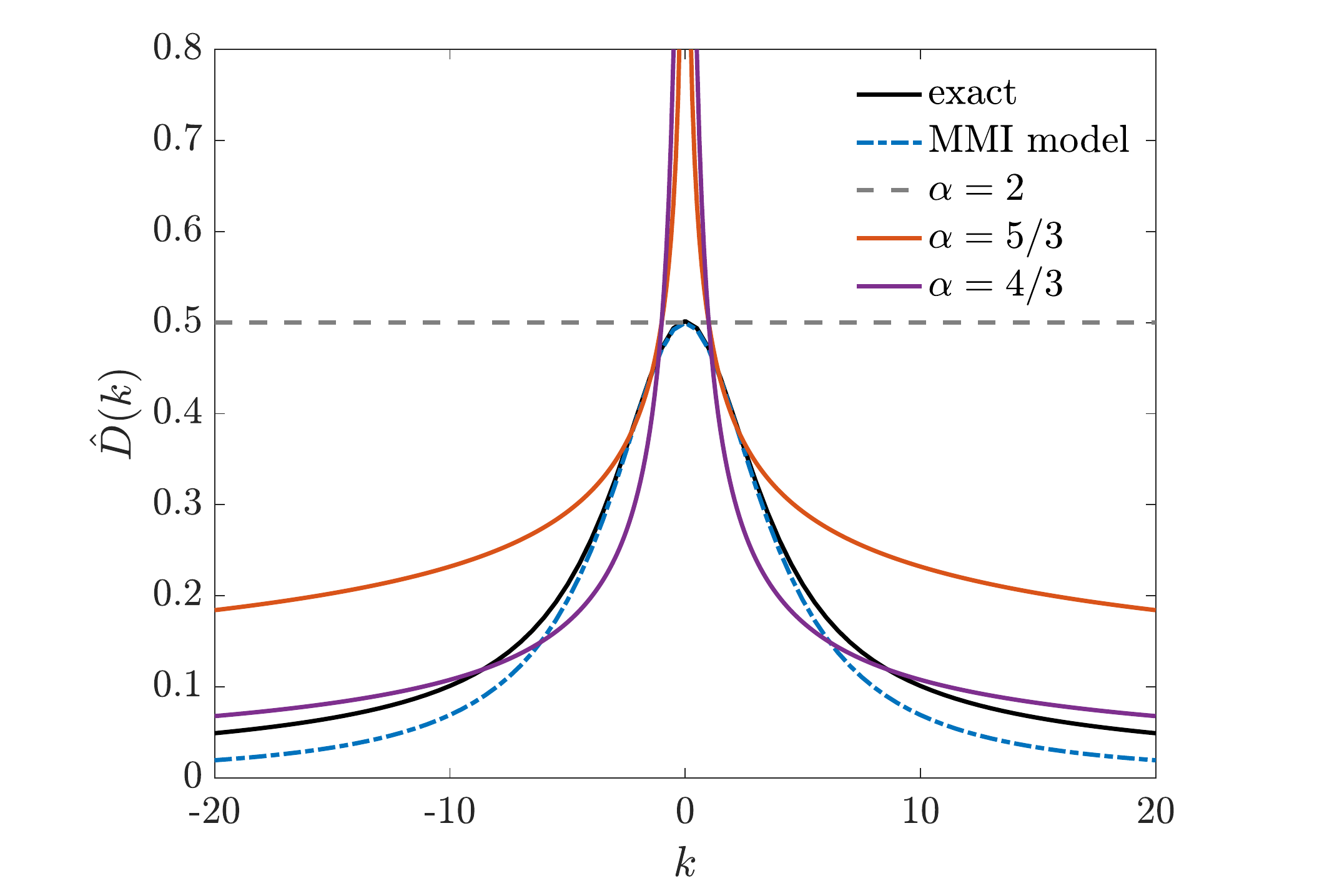}
    \caption{Nonlocal eddy diffusivity of the simple fractional-order model shown in Fourier space for various choices of $\alpha$ compared with the exact and MMI-modeled nonlocal eddy diffusivity.}
    \label{fig:fractional_kernel_fourier} 
\end{figure}

To obtain the nonlocal eddy diffusivity, recall that the right-hand-side of Equation (\ref{eq:fractional model FT}) is a model for the derivative of the unclosed scalar flux:
\begin{equation}
    ik(-\overline{\widehat{u_1^\prime c^\prime}}) = \frac{1}{2}\left(-\left(k^2\right)^{\alpha/2}\right)\bar{\hat{c}}. 
\end{equation}
Rearranging, 
\begin{equation}
    -\overline{\widehat{u_1^\prime c^\prime}} = \frac{1}{2}\left(k^2\right)^{\alpha/2-1}(ik\bar{\hat{c}}),
\end{equation}
where the gradient of $\bar{c}$ in Fourier space is $ik\bar{\hat{c}}$, and correspondingly the nonlocal eddy diffusivity in Fourier space is:
\begin{equation} \label{eq:fractional model eddy diffusivity}
    \hat{D}(k) = \frac{1}{2}\left(k^2\right)^{\alpha/2-1}.
\end{equation}
Figure \ref{fig:fractional_kernel_fourier} shows the nonlocal eddy diffusivity in Fourier space for several $\alpha$ in comparison with the exact and MMI-modeled nonlocal eddy diffusivity.

\begin{figure}[t]
     \centering
     \begin{subfigure}[t]{0.45\textwidth}
         \centering
         \includegraphics[width=\textwidth]{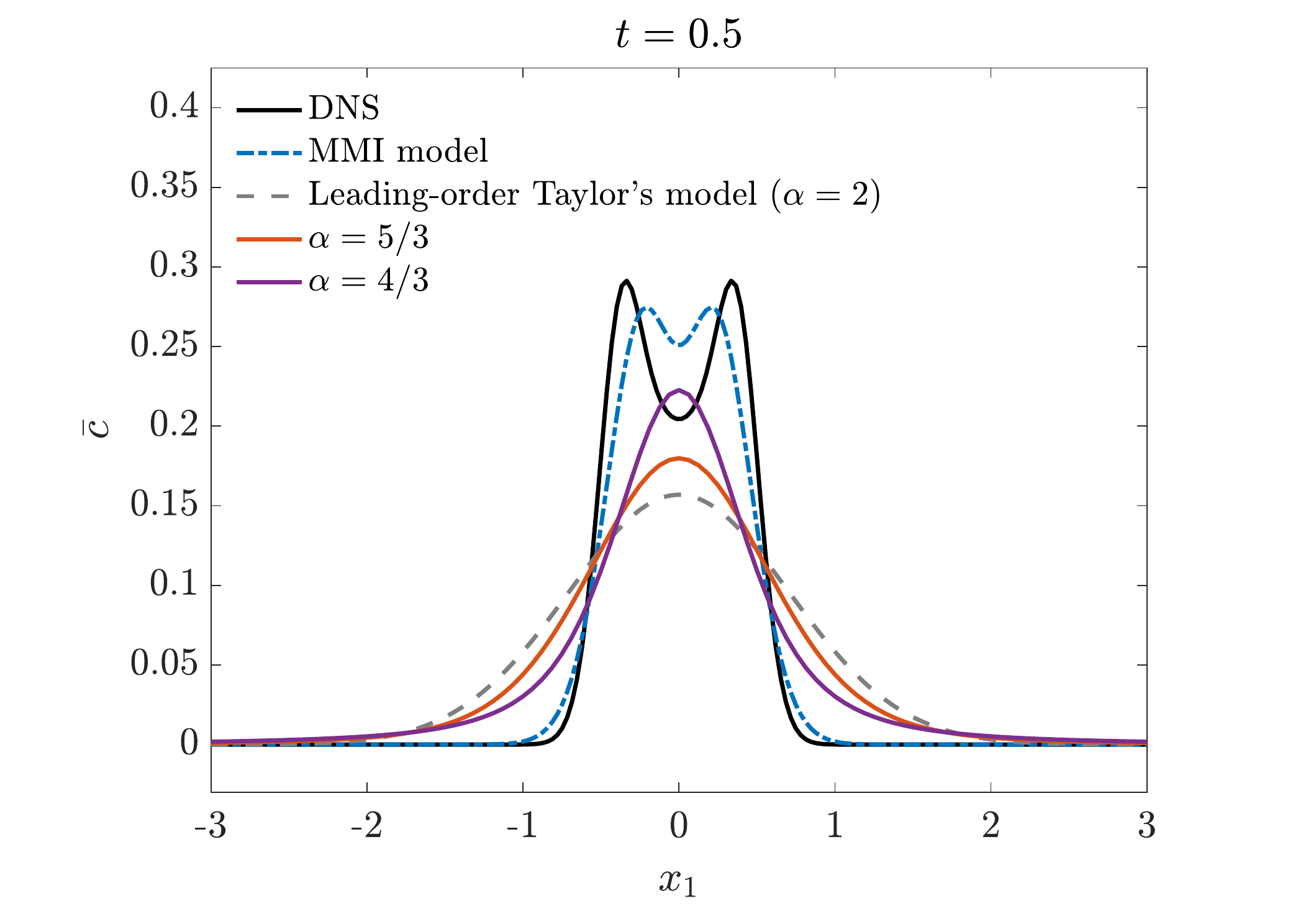}
        \caption{Early time comparison.}
        \label{fig:fractional_modelcomparison_t05}
     \end{subfigure}
     \begin{subfigure}[t]{0.45\textwidth}
         \centering
         \includegraphics[width=\textwidth]{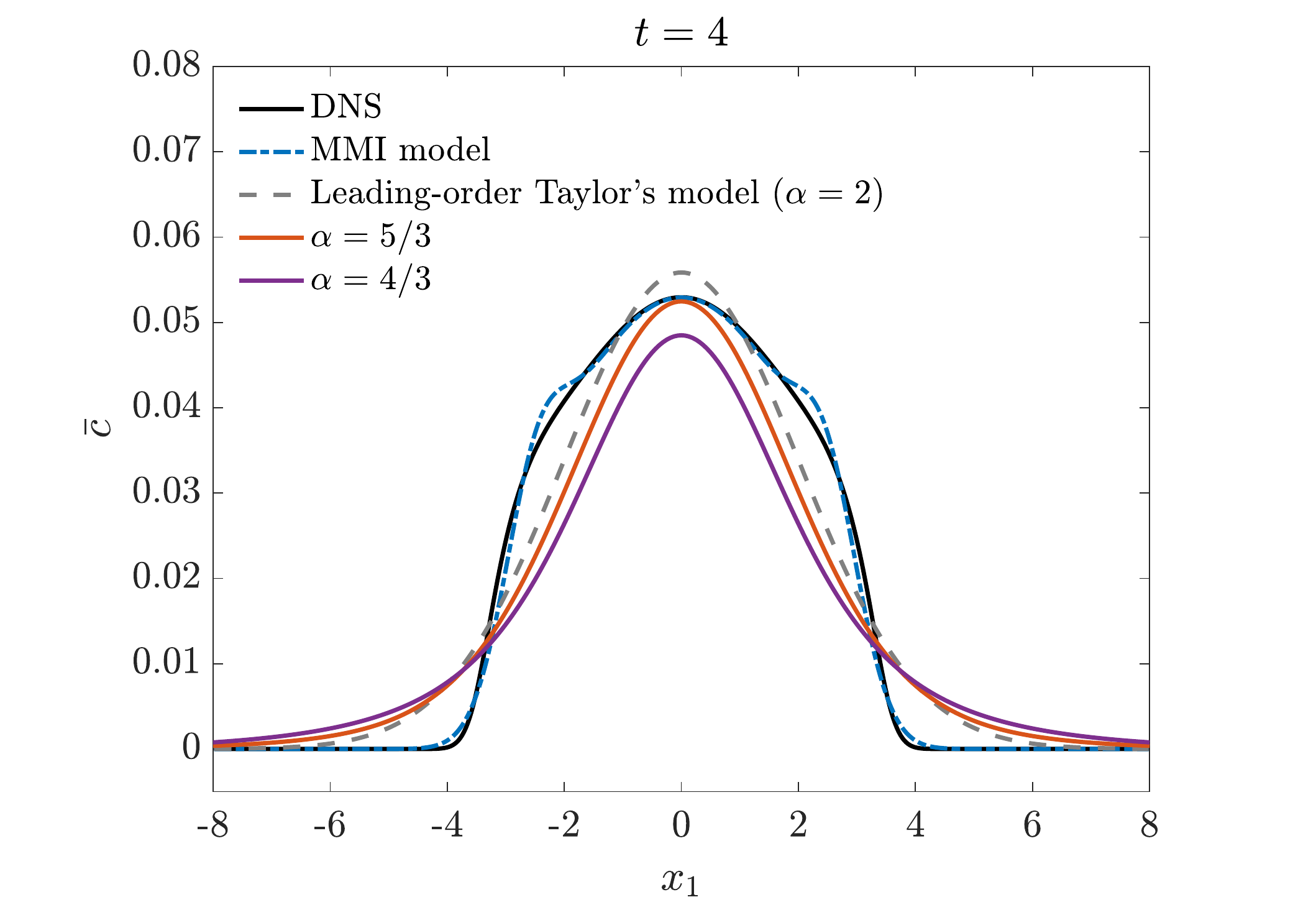}
         \caption{Late time comparison.}
         \label{fig:fractional_modelcomparison_t4}
     \end{subfigure}
     \caption{Fractional-order model comparison for the homogeneous problem in Section \ref{homogeneous model problem} at early time, $t = 0.5$, and late time, $t = 4$.}
     \label{fig:fractional_modelcomparison_tslices}
\end{figure}

Figure \ref{fig:fractional_modelcomparison_tslices} shows an early time ($t = 0.5)$ and late time ($t = 4$) comparison of the spatiotemporal MMI model in Equation (\ref{eq:spatiotemporal MMI model homogeneous}) and the fractional-order model for several choices of $\alpha$. For early time, the simple fractional-order model with a constant $\alpha$ does not capture the double-peaked feature in the DNS solution of $\bar{c}(x_1)$, whereas the MMI model does. For late time, the fractional-order model overpredicts the spread of $\bar{c}(x_1)$. As shown in Figure \ref{fig:fractional_kernel_fourier}, the nonlocal eddy diffusivity of the fractional-order model is larger than the exact nonlocal eddy diffusivity in both of the limits of large $k$ and small $k$. Thus, the fractional-order model solution disperses too quickly for both early time, where narrow (large wavenumber) features are present, and late time, where very smooth (small wavenumber) features are present.

This result may be remedied by a more sophisticated fractional-order model with a variable $\alpha$ but this is not considered here. 

\section{MFM for periodic problems} 
\label{MFM decomposition}
\subsection{Obtaining moments of the nonlocal eddy diffusivity}
\label{inverse MFM decomposition}
The $\bar{c}$ required for IMFM, e.g. $\bar{c} = x_1$ for the zeroth moment, may be incompatible with the periodic boundary conditions of the problem as for the inhomogeneous model problem in Section \ref{inhomogeneous periodic model problem}. We decompose $c(x_1,x_2) = \bar{c}(x_1) + c^\prime(x_1,x_2)$ where $\bar{c}(x_1)$ may be nonperiodic but $c^\prime(x_1,x_2)$ is periodic.

Moreover, following a  similar line of reasoning as for the input-output relationship between $-\overline{u_1^\prime c^\prime}(x_1)$ and $\bar{c}(x_1)$, $c^\prime(x_1, x_2)$ can be Taylor series expanded as 
\begin{equation}
\label{eq:IMFM decomposition 1}
    c^\prime(x_1, x_2) = c_0(x_1, x_2) \frac{\partial \bar{c}}{\partial x_1} + c_1(x_1, x_2) \frac{\partial^2 \bar{c}}{\partial x_1^2} + c_2(x_1, x_2)\frac{\partial^3 \bar{c}}{\partial x_1^3} + \dots,
\end{equation}
where $c_0(x_1,x_2)$, $c_1(x_1,x_2)$, etc. are to be determined. Once determined, to get to the desired moments, multiply Equation (\ref{eq:IMFM decomposition 1}) by $-u_1^\prime(x_1,x_2)$ and average over $x_2$,
\begin{equation}
\label{eq:IMFM decomposition 2}
    -\overline{u_1^\prime c^\prime}(x_1) = -\overline{u_1^\prime c_0}(x_1) \frac{\partial \bar{c}}{\partial x_1} - \overline{u_1^\prime c_1}(x_1) \frac{\partial^2 \bar{c}}{\partial x_1^2} - \overline{u_1^\prime c_2}\frac{\partial^3 \bar{c}}{\partial x_1^3} - \dots,
\end{equation}
which leads to $D^0(x_1) = - \overline{u_1^\prime c_0}(x_1)$, $D^{1_s}(x_1) = - \overline{u_1^\prime c_1}(x_1)$, $D^{2_s}(x_1) = - \overline{u_1^\prime c_2}(x_1)$, etc. 
 
For example, to obtain the zeroth moment of the nonlocal eddy diffusivity for the inhomogeneous model problem in Section \ref{inhomogeneous periodic model problem}, substitute $\bar{c} = x_1$ into Equation (\ref{eq:IMFM decomposition 1}) to get $c^\prime(x_1,x_2) = c_0(x_1,x_2)$, and substitute $c(x_1,x_2) = x_1 + c_0(x_1,x_2)$ into the governing equation for the inhomogeneous model problem (Equation (\ref{eq:inhomogeneous governing equation})):
\begin{equation}
\label{eq:IMFM decomposition g0}
    u_1 + u_1\frac{\partial c_0}{\partial x_1} + u_2\frac{\partial c_0}{\partial x_2} = \epsilon^2\frac{\partial^2 c_0}{\partial x_1^2} + \frac{\partial^2 c_0}{\partial x_2^2} + s(x_1),
\end{equation}
where $s(x_1)$ is the IMFM forcing required to enforce $\overline{c^\prime}(x_1) = \overline{c_0}(x_1) = 0$. One can then solve for $c_0(x_1,x_2)$, and obtain the zeroth moment by forming $D^0(x_1) = - \overline{u_1^\prime c_0}(x_1)$. Using Equation (\ref{eq:IMFM decomposition g0}) for $c_0(x_1,x_2)$ rather than the full governing equation for $c(x_1,x_2)$ bypasses the issue of needing to explicitly enforce $\bar{c}$ with periodic boundary conditions. 

Similarly, to obtain the first spatial moment of the nonlocal eddy diffusivity, substitute $c(x_1,x_2) = x_1^2/2 + c_0(x_1,x_2)x_1 + c_1(x_1,x_2)$ into the governing equation for the inhomgeneous model problem, and subtract $x_1$ times Equation (\ref{eq:IMFM decomposition g0}):
\begin{equation}
\label{eq:IMFM decomposition g1}
    u_1c_0 + u_1\frac{\partial c_1}{\partial x_1} + u_2\frac{\partial c_1}{\partial x_2} = \epsilon^2 + 2\epsilon^2\frac{\partial c_0}{\partial x_1} + \epsilon^2\frac{\partial^2 c_1}{\partial x_1^2} + \frac{\partial^2 c_1}{\partial x_2^2} + s(x_1),
\end{equation}
where $s(x_1)$ is the forcing required to enforce $\overline{c_1}(x_1)=0$. One can then solve for $c_1(x_1,x_2)$, and obtain the first spatial moment by forming $D^{1_s}(x_1) = - \overline{u_1^\prime c_1}(x_1)$. Note Equation (\ref{eq:IMFM decomposition g1}) relies on having $c_0(x_1,x_2)$ from Equation (\ref{eq:IMFM decomposition g0}).

One can obtain the second spatial moment of the nonlocal eddy diffusivity from solving the equation for $c_2(x_1,x_2)$, and so forth. As with IMFM where obtaining the second moment relies on having the zeroth and first moments, the equation for $c_2(x_1,x_2)$ relies on having $c_0(x_1,x_2)$ and $c_1(x_1,x_2)$. However, this decomposition does not raise the cost of obtaining the moments, still requiring one simulation per moment. We leave the extension of this formulation to unsteady and chaotic flows as future work. 

\subsection{Obtaining the full nonlocal eddy diffusivity}
This section provides details for obtaining the nonlocal eddy diffusivity for the inhomogeneous problem with periodic boundary conditions in Section \ref{inhomogeneous periodic model problem}. Either the Green's-function-based approach of Hamba \cite{hamba2004nonlocal} or MFM can be used to obtain the nonlocal eddy diffusivity. However, due to the small number of degrees of freedom in this problem, we obtain the nonlocal eddy diffusivity by directly inverting the discretized advection-diffusion operator and projecting it into the averaged space as detailed in Section \ref{intro to MFM}. Due to the periodic boundary conditions, some additional treatment is needed, which is detailed here. In other words, the averaged operator, $[\bar{\mathcal{L}}]$, can be obtained via Equation (\ref{eq:linalg Lbar}) and can be further written as
\begin{equation} \label{eq:Lbar periodic}
    [\bar{\mathcal{L}}] = -[d/dx_1]([D] + \epsilon^2 [\mathcal{I}])[d/dx_1],
\end{equation}
where $[D]$ is the desired nonlocal eddy diffusivity matrix and $[\mathcal{I}]$ is the identity matrix. However, due to the periodic boundary conditions, $[d/dx_1]$ is uninvertible, and thus one cannot simply solve for $[D]$ using Equation (\ref{eq:Lbar periodic}). 

Rather, the decomposition in Section \ref{inverse MFM decomposition} is applied, and let
\begin{equation}
    [c^\prime] = [c_d][d\bar{c}/dx_1],
\end{equation}
which is equivalent to Equation (\ref{eq:IMFM decomposition 1}) before it is Taylor-series expanded. Then, 
\begin{equation}
    [-\overline{u_1^\prime c^\prime}] = -[P][u_1^\prime][c^\prime] = -[P][u_1^\prime][c_d][d\bar{c}/dx_1] = [D][d\bar{c}/dx_1],
\end{equation}
 where $[P]$ is the projection (i.e. averaging) matrix, and thus the nonlocal eddy diffusivity matrix is
\begin{equation}
    \label{eq:D matrix eqn}
    [D] = -[P][u_1^\prime][c_d].
\end{equation}

To obtain $[c_d]$, substitute $c = \bar{c} + c^\prime$ into the governing equation for the inhomogeneous model problem (Equation (\ref{eq:inhomogeneous governing equation})) with the MFM forcing, $s(x_1)$:
\begin{equation}
    \label{eq:inhomogeneous decomposed eqn}
    u_1\frac{\partial c^\prime}{\partial x_1} + u_2\frac{\partial c^\prime}{\partial x_2} - \epsilon^2\frac{\partial^2 c^\prime}{\partial x_1^2} - \frac{\partial^2 c^\prime}{\partial x_2^2} = -u_1\frac{\partial \bar{c}}{\partial x_1} + \epsilon^2\frac{\partial^2 \bar{c}}{\partial x_1^2} + s(x_1),
\end{equation}
where the role of $s(x_1)$ is to enforce the condition $\overline{c^\prime}(x_1) = 0$. 
In matrix-operator form, Equation (\ref{eq:inhomogeneous decomposed eqn}) is written as
\begin{equation} \label{eq:linalg decomposition 1}
    [\mathcal{L}][c^\prime] = [\bar{\mathcal{L}}_1][\partial \bar{c}/\partial x_1] + [s],
\end{equation}
where $[\bar{\mathcal{L}}_1] = -[u_1] + \epsilon^2[\partial/\partial x_1]$. Substituting $[s] = [E][\bar{s}]$ into Equation (\ref{eq:linalg decomposition 1}), and forming a matrix system to simultaneously solve for $[c^\prime]$ and $[\bar{s}]$ such that $[P][c^\prime] = 0$ leads to:
\begin{equation}
    \left[
    \begin{array}{c|c}
        \mathcal{L} & -E\\
        \hline
        P & 0
    \end{array}
    \right] \left[ 
    \begin{array}{c}
         c^\prime \\
         \bar{s} 
    \end{array}
    \right] = \left[
    \begin{array}{c}
        \bar{\mathcal{L}}_1 \\
        0
    \end{array}
    \right] 
    \left[\frac{\partial \bar{c}}{\partial x_1}
    \right].
\end{equation}
Rearranging, 
\begin{equation}
    \left[ 
    \begin{array}{c}
         c^\prime \\
         \bar{s} 
    \end{array}
    \right] = \left[
    \begin{array}{c|c}
        \mathcal{L} & -E\\
        \hline
        P & 0
    \end{array}
    \right]^{-1} \left[
    \begin{array}{c}
        \bar{\mathcal{L}}_1 \\
        0
    \end{array}
    \right] 
    \left[\frac{\partial \bar{c}}{\partial x_1}
    \right] = \left[
    \begin{array}{c}
        c_d \\
        *
    \end{array}
    \right] 
    \left[\frac{\partial \bar{c}}{\partial x_1}
    \right],
\end{equation}
allows one to obtain $[c_d]$ and subsequently the nonlocal eddy diffusivity, $[D]$, using Equation (\ref{eq:D matrix eqn}). 

\section{An alternative MMI formulation}
\label{other MMI formulas}
For the inhomogeneous problems in Section \ref{inhomogeneous flows}, an alternative to the steady MMI model in Equation (\ref{eq:spatial MMI model again}) is
\begin{equation} \label{eq:inhomogeneous MMI alt}
    \left[1 + a_1(x_1) \frac{d}{dx_1} + a_2(x_1)\frac{d^2}{dx_1^2} \right]\left(\frac{-\overline{u_1^\prime c^\prime}}{a_0(x_1)}\right) = \frac{d\bar{c}}{dx_1}.
\end{equation}
By choosing $a_0(x_1) = D^0(x_1)$, the MMI formulation in Equation (\ref{eq:inhomogeneous MMI alt}) matches the zeroth moment of the exact nonlocal eddy diffusivity (i.e. for $\bar{c} = x_1$, the model recovers $-\overline{u_1^\prime c^\prime}|_{\bar{c}=x_1} = D^0$). The remaining coefficients, $a_1(x_1)$ and $a_2(x_1)$, can be determined by matching the other low-order moments via specifying $\bar{c} = x_1^2/2$ and $\bar{c} = x_1^3/6$ as done in Section \ref{modeling nonlocal eddy diffusivity}. This alternative formulation has one fewer coefficient to solve for than the original MMI formulation, but may have singularity issues if the zeroth moment of the eddy diffusivity goes to zero, for example near a wall. 

\begin{figure}[t]
     \centering
     \begin{subfigure}[t]{0.45\textwidth}
         \centering
         \includegraphics[width=\textwidth]{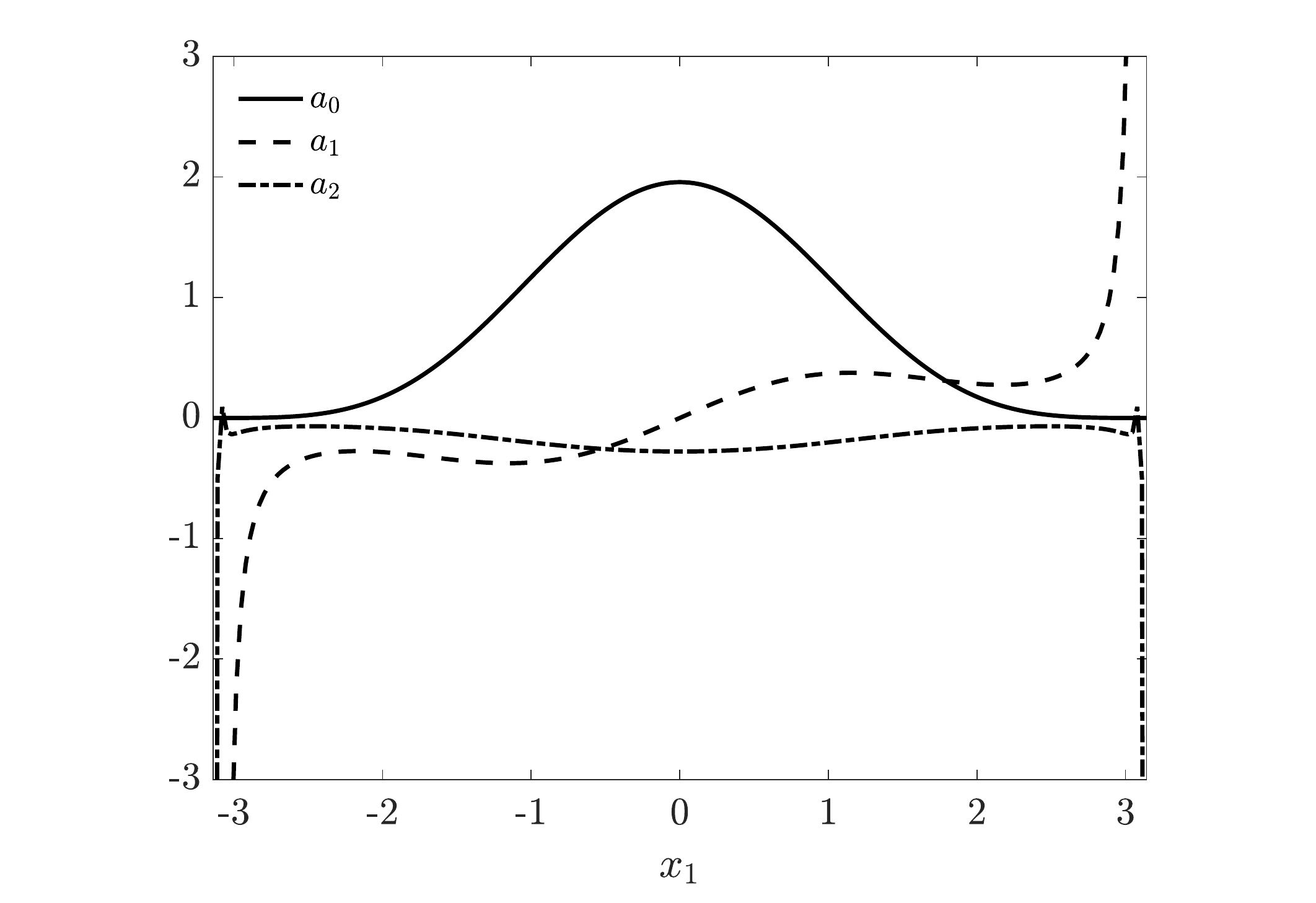}
        \caption{}
        \label{fig:MMI_coeff_alt}
     \end{subfigure}
     \begin{subfigure}[t]{0.45\textwidth}
         \centering
         \includegraphics[width=\textwidth]{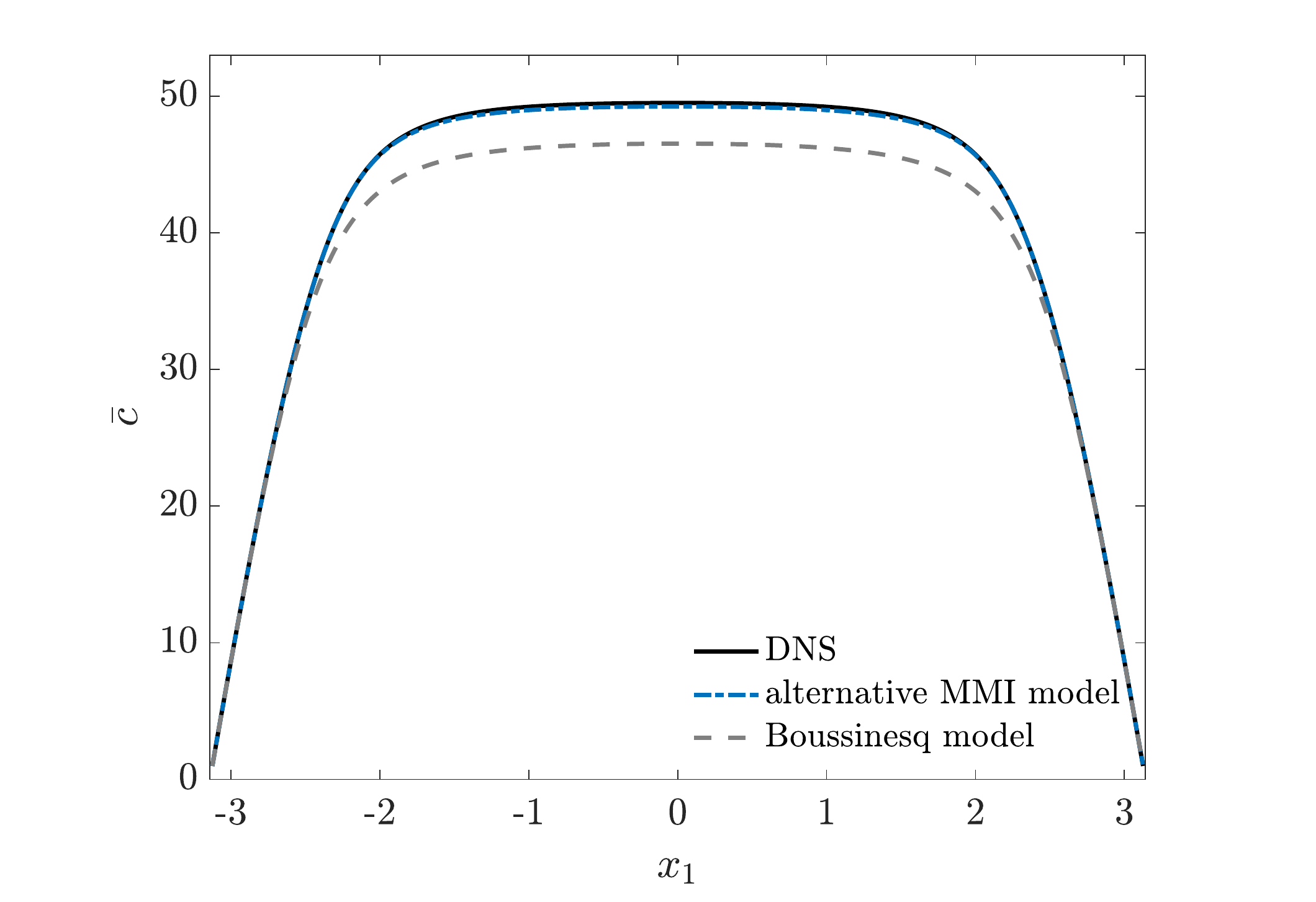}
         \caption{}
         \label{fig:MMI_cbar_alt}
     \end{subfigure}
     \caption{(a) Coefficents of the alternative MMI formulation in Equation (\ref{eq:inhomogeneous MMI alt}) for the wall-bounded inhomogeneous problem in Section \ref{inhomogeneous wall-bounded model problem}. (b) Model comparison for the wall-bounded inhomogeneous problem.}
\end{figure}

Figure \ref{fig:MMI_coeff_alt} shows the coefficients of the alternative MMI formulation for the wall-bounded inhomogeneous model problem in Section \ref{inhomogeneous wall-bounded model problem}, and Figure \ref{fig:MMI_cbar_alt} shows the resulting solution, $\bar{c}(x_1)$, of the alternative MMI formulation closely matching the DNS solution. Although the solution of the alternative MMI formulation is very similar to that of the original MMI formulation in Figure \ref{fig:cbar_inhomogeneous_wall}, the coefficients show some differences particularly in $a_1(x_1)$ and the location of the singularities. At the wall, both $\overline{u_1^\prime c^\prime}(x_1)$ and $D^0(x_1)$ go to zero, leading to a zero divided by zero and numerical issues in determining the coefficients at the wall. However, at the wall, viscous effects also dominate and the eddy diffusivity model is unimportant leading to a well-behaved solution. 

\begin{figure}[t]
     \centering
     \begin{subfigure}[t]{0.45\textwidth}
         \centering
         \includegraphics[width=\textwidth]{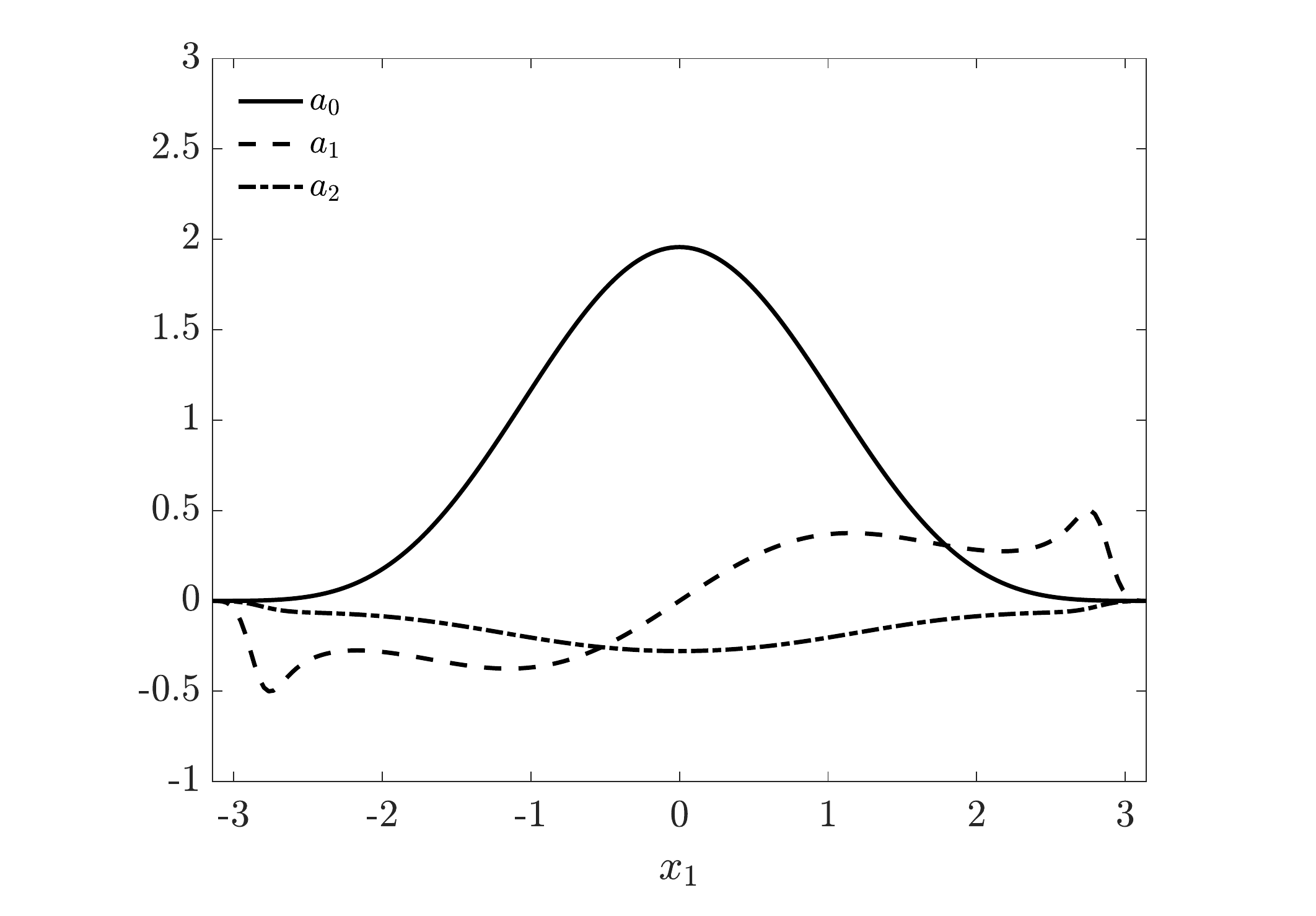}
        \caption{}
        \label{fig:MMI_regularization_coeff_alt}
     \end{subfigure}
     \begin{subfigure}[t]{0.45\textwidth}
         \centering
         \includegraphics[width=\textwidth]{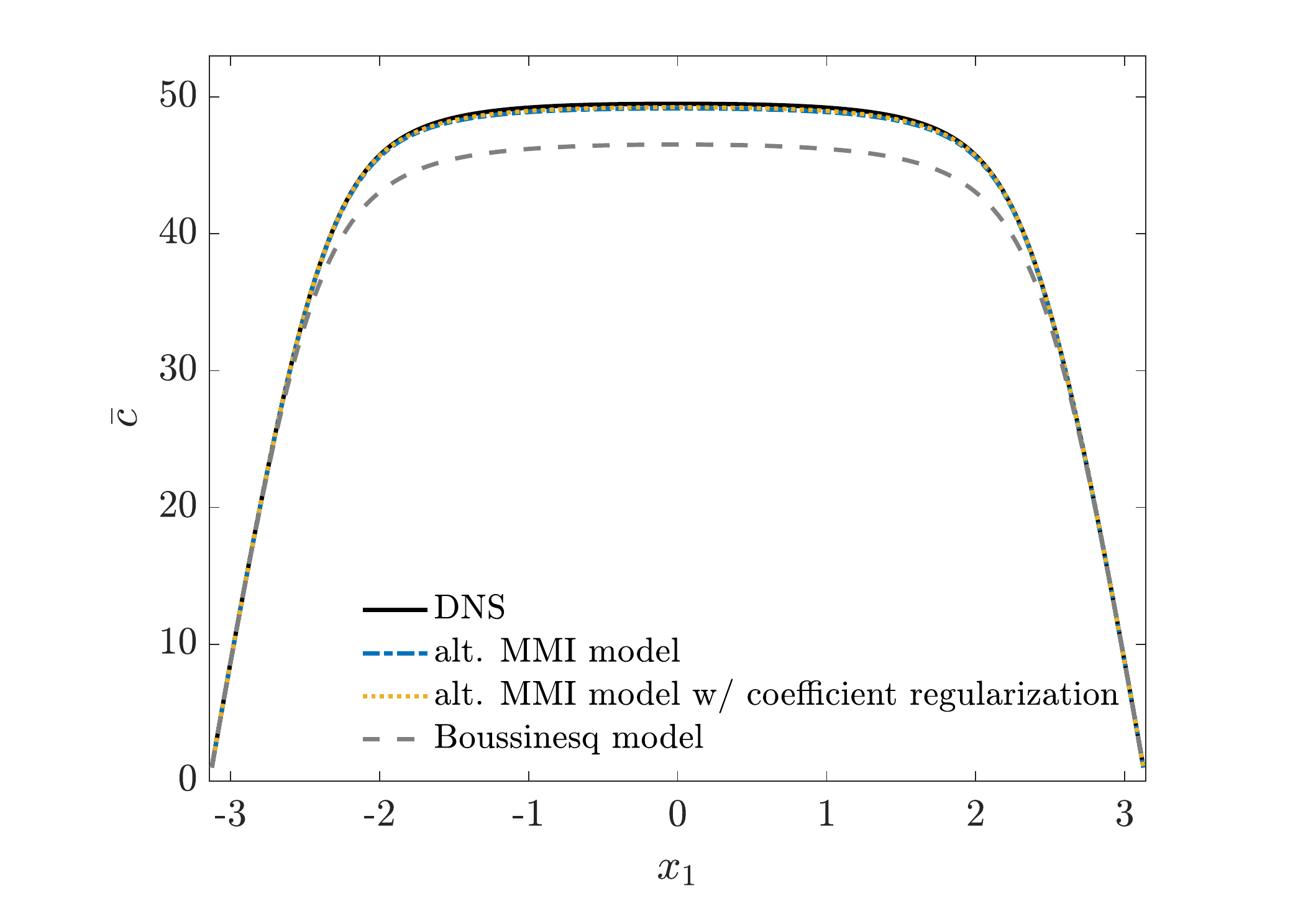}
         \caption{}
         \label{fig:MMI_regularization_cbar_alt}
     \end{subfigure}
     \caption{(a) Coefficients for the alternative MMI formulation with coefficient regularization and $\sigma = 0.01\epsilon^2$.  (b) Model comparison for the wall-bounded inhomogeneous problem in Section \ref{inhomogeneous wall-bounded model problem}.}
\end{figure}

A coefficient regularization technique similar to the one shown in Section \ref{inhomogeneous wall-bounded model problem} may be used. A small parameter $\sigma$ is introduced for determining the MMI coefficients:
\begin{equation} \label{eq:inhomogeneous MMI regularization alt}
    \left[1 + a_1(x_1) \frac{d}{dx_1} + a_2(x_1)\frac{d^2}{dx_1^2} \right]\left(\frac{-\overline{u_1^\prime c^\prime} + \sigma \frac{d\bar{c}}{dx_1}}{D^0 + \sigma}\right) = \frac{d\bar{c}}{dx_1}.
\end{equation}
The $\sigma$ parameter is added to both the numerator and denominator in order to match the zeroth moment (i.e. for $\bar{c} = x_1$, the model recovers $-\overline{u_1^\prime c^\prime}|_{\bar{c}=x_1} + \sigma = D^0 + \sigma$). Equation (\ref{eq:inhomogeneous MMI regularization alt}) is used purely for determining the model coefficients; for ease of implementation, the final model is still Equation (\ref{eq:inhomogeneous MMI alt}) with $a_0(x_1) = D^0(x_1)$. As with the previous coefficient regularization technique in Section \ref{inhomogeneous wall-bounded model problem}, this introduces a small amount of error in matching the first- and second-order moments, but the tradeoff is better-behaved coefficients. Figure \ref{fig:MMI_regularization_coeff_alt} shows the coefficients for the alternative MMI formulation using coefficient regularization with $\sigma = 0.01\epsilon^2$, and Figure \ref{fig:MMI_regularization_cbar_alt} shows a comparison of the model solution with DNS. While the alternative MMI model with coefficient regularization performs slightly better than without coefficient regularization, the choice of $\sigma$ is not yet systematic.

\newpage
\bibliographystyle{ieeetr}
\bibliography{main}

\end{document}